\documentclass[12pt]{article}

\usepackage{graphicx}
\usepackage{float}
\usepackage{amsmath}
\usepackage{verbatim}
\usepackage{fancyhdr}
\usepackage{amsfonts}
\usepackage{floatpag}
\usepackage{authblk}
\usepackage{color}   
\usepackage{lmodern}
\usepackage{geometry}    
\usepackage{multicol}
\usepackage{slashed}
\usepackage{lscape}
\usepackage{listings}
\usepackage{caption}
\usepackage{fancyvrb}
\usepackage{multirow}
\usepackage{cite}
\usepackage{textcomp}
\usepackage{xcolor}

\usepackage[colorlinks=true
,urlcolor=blue 
,anchorcolor=blue
,citecolor=blue
,filecolor=blue
,linkcolor=blue
,menucolor=blue
,pagecolor=blue
,linktocpage=true
]{hyperref}

\newcommand{\beq}{\begin{equation}}
\newcommand{\eeq}{\end{equation}}   

\DeclareMathOperator{\ggHttH}{ggH+ttH}
\DeclareMathOperator{\VBFVH}{VBF+VH}
\DeclareMathOperator{\ggH}{ggH}
\DeclareMathOperator{\ttH}{ttH}
\DeclareMathOperator{\VH}{VH}
\DeclareMathOperator{\ZH}{ZH}
\DeclareMathOperator{\WH}{WH}
\DeclareMathOperator{\VBF}{VBF}

\DeclareMathOperator{\SM}{SM}
\DeclareMathOperator{\gaga}{\gamma\gamma}
\DeclareMathOperator{\tautau}{\tau\tau}
\DeclareMathOperator{\BR}{\mathcal{B}}
\DeclareMathOperator{\Ree}{Re}
\DeclareMathOperator{\Imm}{Im}

\DeclareMathOperator{\invisible}{invisible}

\DeclareMathOperator{\Lilith}{\texttt{Lilith}}
\DeclareMathOperator{\XML}{\texttt{XML}}
\DeclareMathOperator{\stau}{\tilde{\tau}}

\newcommand{\lilith}{\texttt{Lilith}}

\makeatletter
\let \@sverbatim \@verbatim
\def \@verbatim {\@sverbatim \verbatimplus}
{\catcode`'=13 \gdef \verbatimplus{\catcode`'=13 \chardef '=13 }} 
\makeatother

\begin{document}

\thispagestyle{empty}  \setcounter{page}{0}  

\begin{flushright}
LPSC15027\\
CTPU-15-01\\
February 2015\\
\end{flushright}

\vskip 1.4 true cm

\begin{center}
{\LARGE \bf{Lilith: a tool for constraining new physics \\[0.2cm] from Higgs measurements}}\\[1.9cm]
\textsc{J\'er\'emy Bernon}$^{1}$ \textsc{and B\'eranger Dumont}$^{2,1}$%
\vspace{0.5cm}\\[9pt]\smallskip{$^1$\small \textsl{\textit{Laboratoire de Physique Subatomique et de Cosmologie, }}}\linebreak
{\small \textsl{\textit{Universit\'e Grenoble-Alpes, CNRS/IN2P3,}}}\linebreak
{\small \textsl{\textit{53 Avenue des Martyrs, F-38026 Grenoble, France}.}}
\vspace{0.4cm}\\[2pt]\smallskip{$^2$\small \textsl{\textit{Center for Theoretical Physics of the Universe, Institute for Basic Science (IBS), }}}\linebreak
{\small \textsl{\textit{Daejeon 305-811, Republic of Korea}.}}\\[1.3cm]\textbf{Abstract}\smallskip
\end{center}

\begin{quote}
\noindent 

The properties of the observed Higgs boson with mass around 125~GeV can be affected in a variety of ways by new physics beyond the Standard Model (SM). The wealth of experimental results, targeting the different combinations for the production and decay of a Higgs boson, makes it a non-trivial task to assess the compatibility of a non-SM-like Higgs boson with all available results.
In this paper we present {\tt Lilith}, a new public tool for constraining new physics from signal strength measurements performed at the LHC and the Tevatron.
{\tt Lilith} is a {\tt Python} library that can also be used in {\tt C} and {\tt C++}/{\tt ROOT} programs.
The Higgs likelihood is based on experimental results stored in an easily extensible {\tt XML} database, and is evaluated from the user input, given in {\tt XML} format in terms of reduced couplings or signal strengths.
The results of {\tt Lilith} can be used to constrain a wide class of new physics scenarios.

\let   \thefootnote    \relax
\footnotetext{$^{1}\;$bernon@lpsc.in2p3.fr} 
\footnotetext{$^{2}\;$dum33@ibs.re.kr}
\end{quote}

\newpage

\hypersetup{linkcolor=black}
\tableofcontents

\newpage
\hypersetup{linkcolor=blue}

\section{Introduction}

The discovery of a Higgs boson with properties compatible with those of the SM and mass around 125~GeV at CERN's Large Hadron Collider (LHC)~\cite{Aad:2012tfa, Chatrchyan:2012ufa} was a major breakthrough. 
Indeed, the Higgs boson was the last elementary particle predicted by the SM remaining to be observed.
But, more importantly, the Higgs field has a key role in the SM as it triggers the breaking of the electroweak symmetry and gives masses to the elementary particles.
Precision measurements of the properties of the observed boson are of utmost importance to assess its role in the breaking of the electroweak symmetry. They could reveal a more complicated Higgs sector, indicating the presence of more elementary scalars or compositeness of the observed particle, and could also shed light on a large variety of beyond-the-SM (BSM) particles that couple to the Higgs boson.
Conversely, precision measurements can be used to rule out new physics scenarios affecting the properties of the Higgs boson.

That the mass of the observed Higgs boson is about 125~GeV is a fortunate coincidence as many decay modes of the SM Higgs boson are accessible with a modest integrated luminosity at the LHC~\cite{Heinemeyer:2013tqa}. Hence, complementary information on the properties of the Higgs boson were already obtained from the measurements performed during Run~I of the LHC at 7--8~TeV center-of-mass energy~\cite{ATLAS-CONF-2014-009,Khachatryan:2014jba}.
A large variety of models of new physics (both effective and explicit ones) can be constrained from the measurements presented in terms of signal strengths. These results were used in a large number of phenomenological studies in the past three years (see Refs.~\cite{Falkowski:2013dza,Giardino:2013bma,Ellis:2013lra,Djouadi:2013qya,Dumont:2013wma,Belanger:2013xza,Chpoi:2013wga,Lopez-Val:2013yba,Banerjee:2013apa,Belyaev:2013ida,Cheung:2013rva,Cao:2013cfa,Wang:2013sha,Dumont:2013npa,Fan:2014txa,Belanger:2014roa,Bechtle:2014ewa,Ellis:2014dva,Dumont:2014wha,Cheung:2014noa,Bernon:2014vta,deBlas:2014ula,Ciuchini:2014dea,Bergstrom:2014vla,Robens:2015gla,Cheung:2015uia} for a sample of recent studies based on the full data collected at Run~I).

However, it is not straightforward to put constraints on new physics from the measured signal strengths.
Indeed, a large number of analyses have already been performed by the ATLAS and CMS collaborations. They usually include several event categories, and present signal strength results in different ways. Extracting all necessary information from the figures of the various publications is a tedious and lengthy task. Moreover, as the full statistical models used by the experimental collaborations are not public, a number of assumptions need to be made for constructing a likelihood. The validity of these approximations should be assessed from a comparison with the results provided by ATLAS and CMS.

In order to put constraints on new physics from the LHC Higgs results, many groups have been developing private codes.
Moreover, recently a public tool, {\tt HiggsSignals}~\cite{Bechtle:2013xfa}, became available.
{\tt HiggsSignals} is a {\tt FORTRAN} code that uses the signal strengths for individual measurements, taking into account the associated efficiencies.
In this paper, we present a new public tool, {\tt Lilith}.\footnote{Lilith is a mythological figure often associated with a female demon. It also stands for ``\underline{li}ght \underline{li}kelihood fi\underline{t} for the \underline{H}iggs''.} {\tt Lilith} is a library written in {\tt Python}, that can easily be used in any {\tt Python} script as well as in {\tt C} and {\tt C++}/{\tt ROOT} codes, and for which we also provide a command-line interface.
It follows a different approach than {\tt HiggsSignals} in that it uses as a primary input results in which the fundamental production and decay modes are unfolded from experimental categories. 
The experimental results are stored in {\tt XML} files, making it easy to modify and extend. The user input can be given in terms of reduced couplings or signal strengths for one or multiple Higgs states, and is also specified in an {\tt XML} format.

In Section~\ref{sec:expresults}, we present the signal strength framework used to encode deviations from the SM at the LHC, as well as the experimental results that we use as input in {\tt Lilith}. The parametrization of new physics effects on the observed Higgs boson, as well as derivation of signal strengths, are presented in Section~\ref{sec:npparam}. All technical details on how to use {\tt Lilith} and the {\tt XML} formats that we use are then given in Section~\ref{sec:manual}. Constraints derived from {\tt Lilith} are validated in Section~\ref{sec:validation}, and two concrete examples of its capabilities are given in Section~\ref{sec:examples}.
Finally, prospects for Run~II of the LHC are discussed in Section~\ref{sec:prospects}, and conclusions are given in Section~\ref{conclusions}.

\section{From experimental results to likelihood functions}
\label{sec:expresults}

\subsection{Signal strength measurements}
\label{sec:mumeasurements}

Thanks to the excellent operation of the LHC and to the wealth of accessible final states for a 125~GeV SM-like Higgs boson, the properties of the observed Higgs boson have been measured with unforeseeable precision by the ATLAS and CMS collaborations already during Run~I of the LHC at 7--8~TeV center-of-mass energy~\cite{ATLAS-CONF-2014-009,Khachatryan:2014jba}. LHC searches are targeting the different combinations for the production and decay modes of a Higgs boson.
The SM Higgs boson has five main
production mechanisms at a hadron collider: gluon fusion ($\ggH$), vector-boson fusion ($\VBF$), associated production with an electroweak gauge boson ($\WH$ and $\ZH$, collectively denoted as $\VH$) and associated production with a pair of top quarks ($\ttH$).\footnote{Current searches do not constrain the associated production with a pair of bottom quarks, whose SM cross section is small and which is plagued with the very large QCD background.} Observation of these production modes constrains the couplings of the Higgs to vector bosons ($\VBF$, $\VH$) and to third-generation quarks ($\ggH$, $\ttH$).
The main decay modes accessible at the LHC are $H \to \gamma\gamma$, $H \to Z Z^* \to 4\ell$, $H \to W W^* \to 2\ell2\nu$, $H \to b\bar b$ and $H \to \tau\tau$ (with $\ell \equiv e,\mu$). They can provide complementary information on the couplings of the Higgs to vector bosons (from the decay into $ZZ^*$, $WW^*$, and $\gamma\gamma$) and to third-generation fermions (from the decay into $b\bar b$, $\tau\tau$, and $\gamma\gamma$). Being loop-induced processes, $gg \to H$ and $H \to \gamma\gamma$ also have sensitivity to BSM colored particles and BSM electrically-charged particles, respectively. 

The results of the Higgs searches at the LHC are given in terms of signal strengths, $\mu$, which scale the number of signal events expected for the SM Higgs, $n_s$. For a given set of selection criteria, the expected number of events is therefore $\mu \cdot n_s + n_b$, where $n_b$ is the expected number of background events, so that $\mu = 0$ corresponds to the no-Higgs scenario and $\mu = 1$ to a SM-like Higgs. Equivalently, signal strengths can be expressed as
\begin{align}
\mu = \frac{\sigma \times A \times \varepsilon}{[\sigma \times A \times \varepsilon]^{\rm SM}} \,,
\label{eq:signalstr1}
\end{align}
where $A \times \varepsilon$ is the product of the acceptance and of the efficiency of the selection criteria.
Two assumptions can subsequently be made: first, the
signal is a sum of processes that exist for a 125~GeV SM Higgs boson, {\it i.e.}\ $\sigma = \sum_{X,Y}\sigma(X)\BR(H \to Y)$ for the various production modes $X\in(\ggH$, $\VBF$, $\VH$, $\ttH)$ and decay modes $Y\in(\gaga$, $ZZ^*$, $WW^*$, $b\bar{b}$, $\tautau$, $\ldots)$. Second, the acceptance times efficiency is identical to the SM one for all processes, that is $(A \times \varepsilon)_{X,Y} = [(A \times \varepsilon)_{X,Y}]^{\rm SM}$ for every $X$ and $Y$. 
These conditions require in particular that no new production mechanism (such as $pp \to A \to ZH$, where $A$ is a CP-odd Higgs boson) exist, and that the structure of the couplings of the Higgs boson to SM particules is as in the SM.
Under these conditions, signal strengths read
\beq
\mu = \frac{\sum_{X,Y} (A \times \varepsilon)_{X,Y} \sigma(X)\BR(H\to Y)}{\sum_{X,Y} (A \times \varepsilon)_{X,Y} \sigma^{\rm SM}(X)\BR^{\rm SM}(H\to Y)}
= \sum_{X,Y} {\rm eff}_{X,Y} \frac{\sigma(X)\BR(H\to Y)}{\sigma^{\rm SM}(X)\BR^{\rm SM}(H\to Y)} \,,
\label{eq:signalstr2}
\eeq
where the ${\rm eff}_{X,Y}$ are ``reduced efficiencies'', corresponding to the relative contribution of each combination for the production and decay of a Higgs boson to the signal.
These can be estimated from the $A \times \varepsilon$ obtained in a Monte Carlo simulation of individual processes.
In the case of an inclusive search targeting a given decay mode $Y$ ({\it i.e.} \ $\forall X, (A \times \varepsilon)_{X,Y} = (A \times \varepsilon)_Y$), ${\rm eff}_Y$ is equal to the ratio of SM cross sections, $\sigma^{\rm SM}_X / (\sum_X \sigma^{\rm SM}_X)$.

The signal strength framework used by the ATLAS and CMS collaborations is based on the general form of Eq.~\eqref{eq:signalstr2}, hence on the assumption that new physics results only in the scaling of SM Higgs processes. This makes it possible to combine the information from various Higgs searches and assess the compatibility of given scalings of SM production and/or decay processes from a global fit to the Higgs data.
This framework is very powerful as it can be used to constrain a wide variety of new physics models (some examples can be found in Ref.~\cite{ATLAS-CONF-2014-010}).
This is the approach that we will follow in {\tt Lilith}.
However, in order to derive constraints on new physics, one first needs to construct a likelihood function from the signal strength information given in the experimental publications. In particular, combining the results from several Higgs searches is non-trivial and deserves scrutiny. 

\subsection{Event categories versus unfolded production and decay modes}
\label{sec:mupresentation}

\begin{figure}[t]
\centering   \includegraphics[scale=0.35]{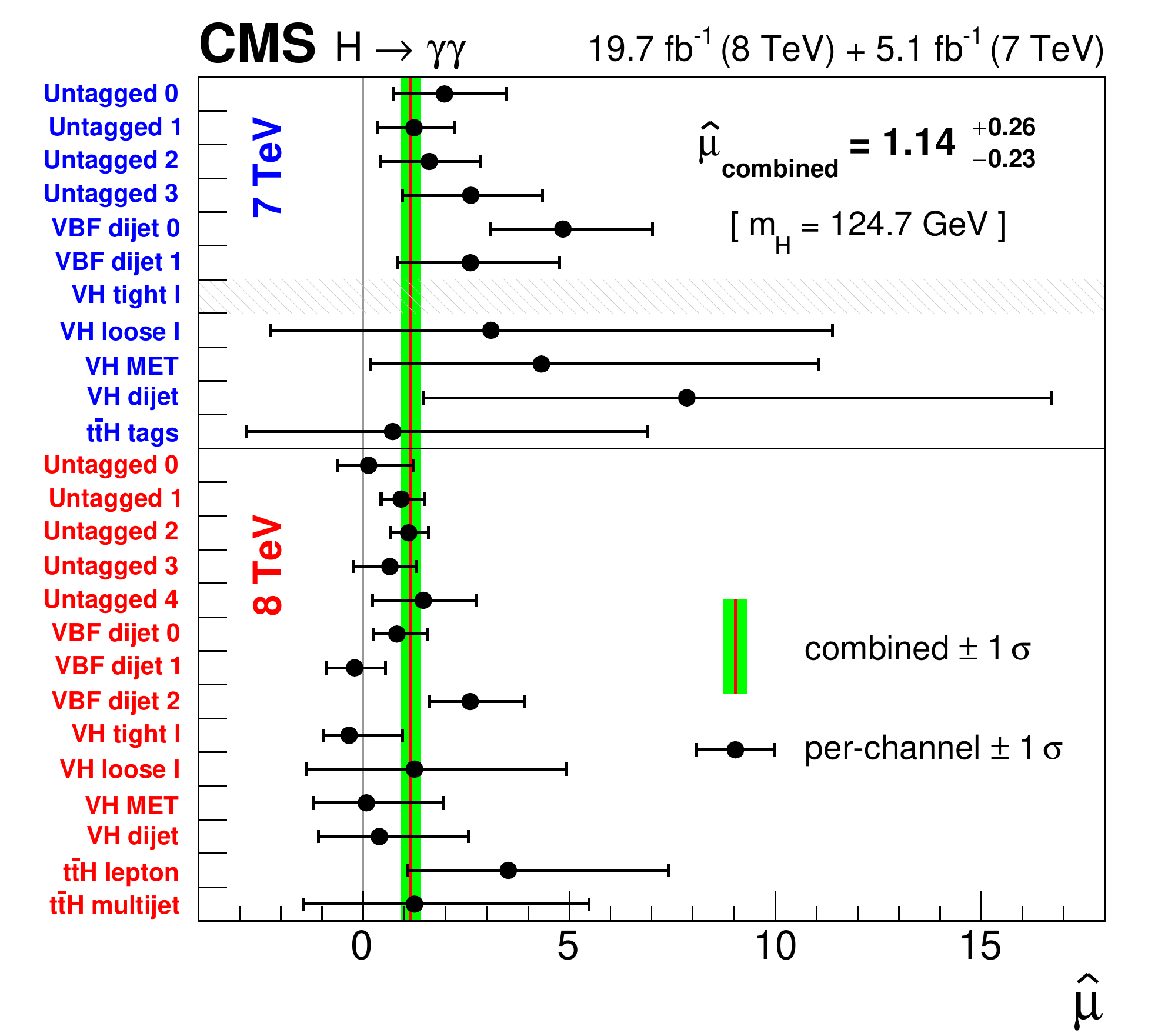} \includegraphics[scale=0.43]{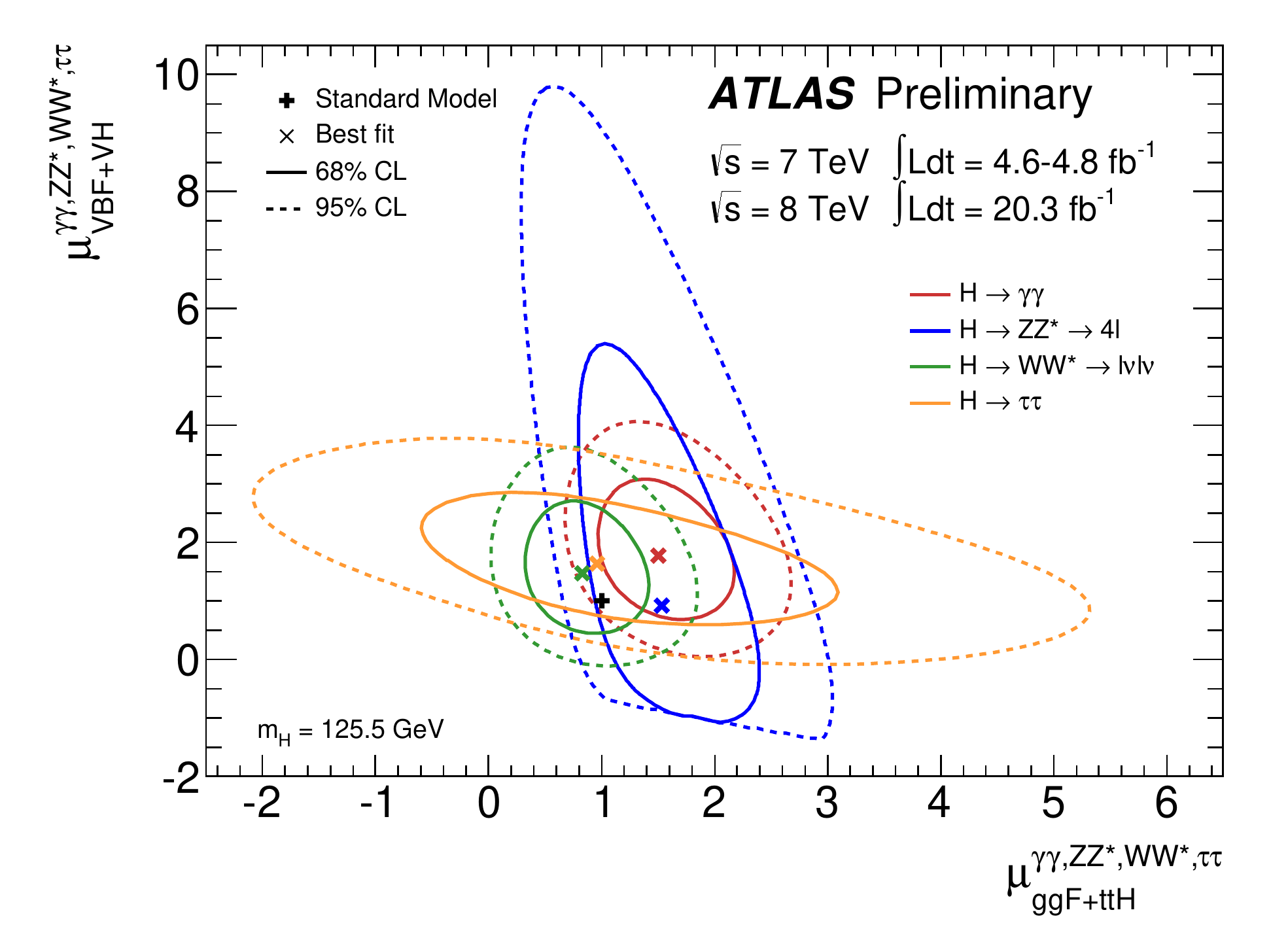}
\caption{Signal strength measurements by the ATLAS and CMS collaborations.
On the left panel, results of the CMS search $H\to\gamma\gamma$~\cite{Khachatryan:2014ira} category per category.
On the right panel, 2-dimensional ATLAS results in which the fundamental production modes are unfolded from experimental categories, in the plane $(\mu({\rm ggH+ttH}, Y), \mu({\rm VBF+VH}, Y))$ for $Y\in(\gaga$, $ZZ^*$, $WW^*$, $\tau\tau)$~\cite{ATLAS-CONF-2014-009}.}
\label{fig:subcat2D}
\end{figure}

The searches for the Higgs boson performed by the ATLAS and CMS collaborations are divided into individual analyses usually focusing on a single decay mode. Within each analysis several event categories are then considered. Among other reasons, these are designed to optimize the sensitivity to the different production mechanisms of the SM Higgs boson (hence, they have different reduced efficiencies ${\rm eff}_{X,Y}$).
In order to put constraints on new physics from the results in a given event category, one needs to extract the measurement of the signal strength and the relevant ${\rm eff}_{X,Y}$ information from the experimental publication. For example, results of the CMS $H\to\gamma\gamma$ analysis~\cite{Khachatryan:2014ira}, in terms of signal strengths for all categories, are shown on the left panel of Fig.~\ref{fig:subcat2D}. With the addition of the reduced efficiencies ${\rm eff}_{X,\gamma\gamma}$, also given in Ref.~\cite{Khachatryan:2014ira}, combinations of $\sigma(X)\BR(H\to \gamma\gamma)$ can be constrained.

However, several problems arise when constructing a likelihood. First of all, as can be seen on the left panel of Fig.~\ref{fig:subcat2D}, only two pieces of information are given: the best fit to the data, that will be denoted as $\hat\mu$ in the following, and the 68\%~confidence level (CL) interval or $1 \sigma$ interval. The full likelihood function category per category is never provided by the experimental collaborations. Assuming that the measurements are approximately Gaussian, it is however possible to reconstruct a simple likelihood, $L(\mu)$, from this information.
In that case, $-2\log L(\mu)$ follows a $\chi^2$ law. From the boundaries of the 68\%~CL interval, left and right uncertainties at 68\%~CL, $\Delta\mu^-$ and $\Delta\mu^+$, with respect to the best fit point can be derived. The likelihood can then be defined as
\begin{equation}
\label{eq:Lchi2}
- 2 \log L(\mu) = \left\{ \begin{array}{l}
\left(\frac{\mu-\hat\mu}{\Delta\mu^-} \right)^2 \ {\rm if}\ \mu<\hat\mu \,, \\
\ \\
\left(\frac{\mu-\hat\mu}{\Delta\mu^+} \right)^2 \ {\rm if}\ \mu>\hat\mu \,,
\end{array}
\right.
\end{equation}
with $\Delta\mu^- = \Delta\mu^+$ in the Gaussian regime.
While this is often a valid approximation to the likelihood, it should be pointed out that signal strength measurements are not necessarily Gaussian, depending in particular on the size of the event sample.

Barring this limitation, Eq.~\eqref{eq:Lchi2} can be used to constrain new physics. However, it requires that at least the 68\%~CL interval and the relevant reduced efficiencies ${\rm eff}_{X,Y}$ are provided by the experimental collaboration for every individual category. This is very often, but not always, the case. 
Categories are sometimes defined without giving the corresponding signal efficiencies (as in, {\it e.g.}, the CMS ttH analysis~\cite{Khachatryan:2014qaa}), and/or the result is given for a (set of) combined signal strength(s) but not in terms of signal strengths category per category (as in the ATLAS $ZZ^*$ and $\tau\tau$ analyses~\cite{Aad:2014eva,Aad:2015vsa} and in the CMS ttH analysis~\cite{Khachatryan:2014qaa}). 
Such combined $\mu$ should in general not be used because they have been obtained under the assumption of SM-like production or decay of the Higgs boson. 
 Whenever the ${\rm eff}_{X,Y}$ are not given in the experimental publications it is in principe possible to obtain estimates from a reproduction of the selection criteria applied on signal samples generated by Monte Carlo simulation. However, this turns out to be a very difficult or impossible task. Indeed, searches for the Higgs boson typically rely on complex search strategies to optimize the sensitivity, such as multivariate analysis techniques that are impossible to reproduce in practice with the information currently available. Whenever the information on reduced efficiencies is not available, we are left to guesswork, with a natural default choice being that ${\rm eff}_X$ is equal to the ratio of SM cross sections, $\sigma^{\rm SM}_X / (\sum_X \sigma^{\rm SM}_X)$, which would correspond to a fully inclusive search.

Constraining new physics from a single LHC Higgs category can already be a non-trivial task and come with some uncertainty because the full information is not provided category per category. However, more severe complications typically arise when using several categories/searches at the same time, as is needed for a global fit to the Higgs data. The simplest solution is to define the full likelihood as the product of individual likelihoods,
\beq
L(\boldsymbol{\mu}) = \prod_{i=1}^{n} L(\mu_i) \quad \Rightarrow \quad \chi^2(\boldsymbol{\mu}) = \sum_{i=1}^{n} \chi^2(\mu_i) = \sum_{i=1}^{n} \left(\frac{\mu_i - \hat\mu_i}{\Delta \mu_i}\right)^2 \,. \label{eq:likesimpleprod}
\eeq
However, this assumes that all measurements are completely independent. We know that this is not the case as the various individual measurements share common systematic uncertainties. They are divided into two categories: the shared experimental uncertainties, coming from the presence of the same final state objects and from the estimation of the luminosity, and the shared theoretical uncertainties, dominated by the contributions from identical production and/or decay modes to the expected Higgs signal in different categories~\cite{CMS-NOTE-2011-005}. The estimation of the experimental uncertainties in ATLAS should be largely independent from the one in CMS, hence these correlations can be treated separately for measurements performed by one collaboration or the other. Conversely, theoretical uncertainties are estimated in the same way in ATLAS and CMS and should be correlated between all measurements.

In the case where all measurements are well within the Gaussian regime, it is possible to take these correlations into account in a simple way, defining our likelihood as
\beq
-2 \log L(\boldsymbol{\mu}) = \chi^2(\boldsymbol{\mu}) = (\boldsymbol{\mu} - \hat{\boldsymbol{\mu}})^T C^{-1} (\boldsymbol{\mu} - \hat{\boldsymbol{\mu}}) \,, \label{eq:gausscovmatrix}
\eeq
where $C^{-1}$ is the inverse of the $n \times n$ covariance matrix,
with $C_{ij} = {\rm cov}[\hat\mu_i,\hat\mu_j]$ (leading to $C_{ii} = \sigma_i^2$). However, the off-diagonal elements of this matrix are not given by the experimental collaborations and are very difficult to estimate from outside the collaboration. This remarkably simple and compact expression for the likelihood (a $n \times n$ matrix) is only valid under the Gaussian approximation; beyond that the expression and the communication of the likelihood become more complicated.

An alternative way for constraining new physics from the experimental results is to consider results in which the fundamental production and decay modes are unfolded from experimental categories. These so-called ``signal strengths in the theory plane'' are defined as
\begin{equation}
	\mu(X,Y) \equiv \frac{\sigma(X)\BR(H\to Y)}{\sigma^{\rm SM}(X)\BR^{\rm SM}(H\to Y)} \,,
	\label{eq:signalstr3}
\end{equation}
where as before $X$ labels the production mode and $Y$ the decay mode of the Higgs boson.
These quantities can be estimated from a fit to the results in several event categories; as the ${\rm eff}_{X,Y}$ will differ from measurement to measurement, complementary information on the various $(X,Y)$ couples can be obtained and break the degeneracies.
The resulting signal strengths are directly comparable to the predictions in a given new physics model.
They have first been used in phenomenological studies in Refs.~\cite{Cacciapaglia:2012wb,Belanger:2012gc}.

It has become a common 
practice of the ATLAS and CMS collaborations to present such results in 2-dimensional likelihood planes for every decay mode.
In that case, the five production modes of the SM Higgs boson are usually combined to form just two effective $X$ modes, VBF + VH and ggH + ttH. The likelihood is then shown in the $(\mu({\rm ggH+ttH}, Y), \mu({\rm VBF+VH}, Y))$ plane. The ATLAS results in this 2-dimensional plane for $Y\in(\gaga$, $ZZ^*$, $WW^*$, $\tau\tau)$, as given in Ref.~\cite{ATLAS-CONF-2014-009}, are shown on the right panel of Fig.~\ref{fig:subcat2D} (the corresponding CMS results can be found in Fig.~5 of Ref.~\cite{Khachatryan:2014jba}). The solid and dashed contours delinate the 68\% and 95\%~CL allowed regions, respectively.
As the unfolding of the individual measurements is done by the experimental collaborations themselves, all correlations between systematic uncertainties (both experimental and theoretical) are taken into account for a given decay mode $Y$, and are encompassed in the correlation between $\mu({\rm ggH+ttH}, Y)$ and $\mu({\rm VBF+VH}, Y)$. (Other 2-dimensional planes can be relevant, depending on the sensitivity of the searches.)
This is a very significant improvement over the naive combination of categories of Eq.~\eqref{eq:likesimpleprod}, in which all measurements are assumed to be independent.
Moreover, in this approach no approximation needs to be made because of missing information on the signal efficiencies or signal strengths category per category.
For these reasons, we use the results in terms of signal strengths in the theory plane as the primary experimental input in \lilith.

A remark is in order regarding the grouping of the five production modes into just two. First of all, grouping together VBF, WH and ZH is unproblematic for testing the vast majority of the new physics models because custodial symmetry requires that the couplings of the Higgs to $W$ and $Z$ bosons scale in the same way. Probing models that violate custodial symmetry based on this input and on the inclusive breaking into the individual production modes $\VBF, \WH$, and $\ZH$, as will be explicited in Eq.~\eqref{eq:VHeff}, may lead to results that deviate significantly from the ones using the full likelihood, as will be shown in Section~\ref{subsec:comparison}.
The combination of the ggH and ttH production modes is more problematic at first sight. While gluon fusion is dominated by the top-quark contribution in the SM, this can be modified drastically if BSM colored particles are present. However, for all decay modes except $H \to b\bar b$ (where gluon fusion-initiated production of the Higgs is not accessible) the ttH production mode is currently constrained with much poorer precision than ggH because of its small cross section (being 150 times smaller than ggH at $\sqrt{s}=8$~TeV~\cite{Heinemeyer:2013tqa}). Therefore, with the current data it is justified to take $\mu({\rm ggH+ttH}, Y) = \mu({\rm ggH}, Y)$ for all channels except $H \to b\bar b$, and $\mu({\rm ggH+ttH}, b\bar b) = \mu({\rm ttH}, b\bar b)$.\footnote{Constraints on the ttH production mode for decay modes other than $b\bar b$ are taken into account independently in {\tt Lilith}, see Table~\ref{tab:defaultdatabase}.}

Finally, note that all results given in terms of signal strengths are derived assuming the current theoretical uncertainties in the SM predictions. Hence, constraining a scenario with different (usually larger) uncertainties from a fit to the signal strength measurements is a delicate task. This issue will also be discussed, alongside with a possible solution, in Section~\ref{sec:prospects}.

\subsection{Statistical procedure}
\label{sec:statproc}

\begin{figure}
\centering \includegraphics[scale=0.34]{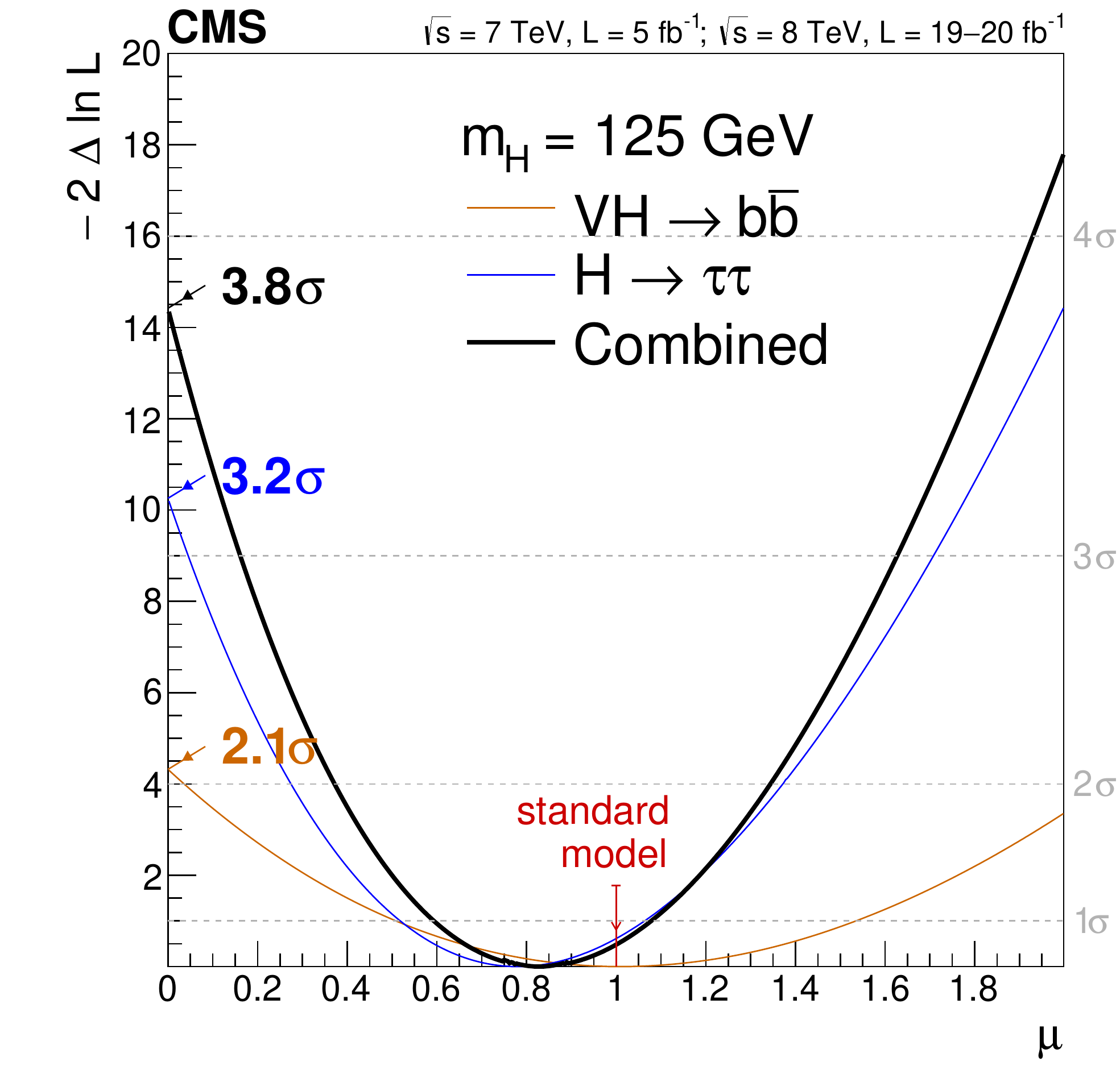} \includegraphics[scale=0.39]{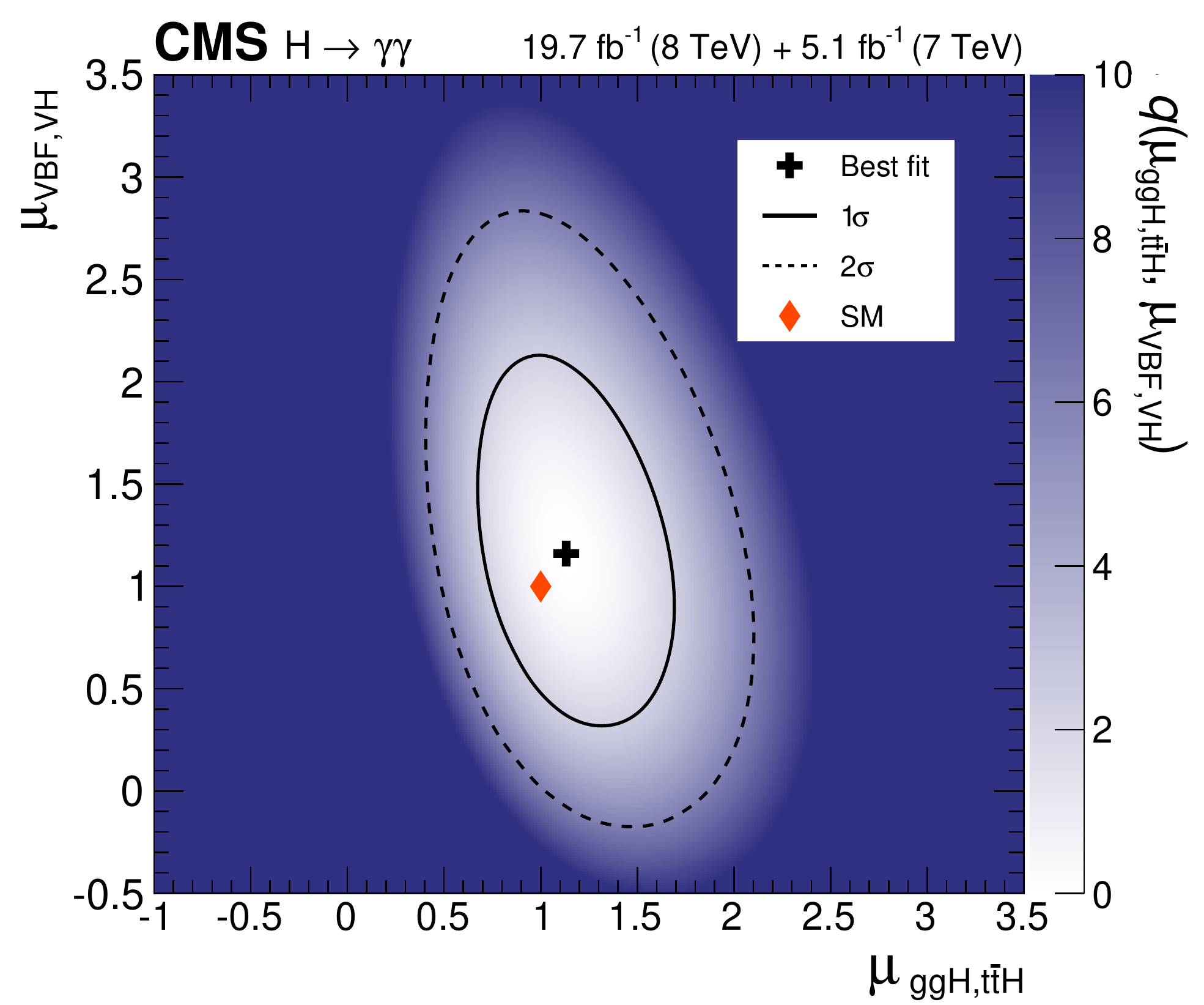}
\caption{Signal strength results from the CMS collaboration: 1D likelihood for VH, $H\to b\bar{b}$ (red curve)~\cite{Chatrchyan:2014vua} (left),
and temperature plot in the plane $(\mu({\rm ggH+ttH}, \gamma\gamma), \mu({\rm VBF+VH}, \gamma\gamma))$~\cite{Khachatryan:2014ira} (right).}
\label{fig:dataformats}
\end{figure}

We use signal strengths for pure production and decay modes  
as basic ingredients for the construction of the Higgs likelihood in \lilith.
However, the full likelihood in the $\mu(X,Y)$ basis is not accessible as only 1- and 2-dimensional (1D and 2D) results are provided by the experimental collaborations; therefore some of the correlations are necessarily missing.
%
In the currently available 1D and 2D results, the full likelihood is provided in some cases in addition to contours of constant likelihood.
This is extremely helpful since the transmission of the result between the collaboration and the reader does not cause any loss of information.
Two examples from the CMS collaboration are given in Fig.~\ref{fig:dataformats}. The 1D likelihood as a function of $\mu({\rm VH}, b\bar b)$~\cite{Chatrchyan:2014vua} is shown on the left panel.\footnote{Note that 2D results in the plane $(\mu(\WH,b\bar b),\mu(\ZH, b\bar b))$ also exist for this analysis~\cite{Chatrchyan:2013zna}. Both results are present in the database of {\tt Lilith}; by default we use the 1D results shown on the left panel of Fig.~\ref{fig:dataformats}.} On the right panel, the full likelihood in the 2D plane $(\mu({\rm ggH+ttH}, \gamma\gamma), \mu({\rm VBF+VH}, \gamma\gamma))$~\cite{Khachatryan:2014ira} is shown as a ``temperature plot''.
Moreover, likelihood grids have been provided by ATLAS in numerical format in the 2D plane $(\mu({\rm ggH+ttH}, Y), \mu({\rm VBF+VH}, Y))$ for $Y\in(\gamma\gamma, ZZ^*, WW^*)$~\cite{HEPDATA1,HEPDATA2,HEPDATA3}.

Any result given in terms of signal strengths can be used in \lilith. Whenever available, we take into account the full likelihood information. The provision of numerical grids for the di-boson final states by the ATLAS collaboration was an important step forward in the communication of the likelihood. Unfortunately, they were derived with previous versions of the analyses, and the same information has not (yet) been given for the corresponding final Run~I results~\cite{Aad:2014eha,Aad:2014eva,ATLAS:2014aga}.
Moreover, in the CMS $H \to \gamma\gamma$ result shown on the right panel of Fig.~\ref{fig:dataformats}, the Higgs boson mass has been profiled over instead of being fixed to a given value, making the interpretation of the result very difficult. Limitations of current way of presenting signal strength results, as well as possible improvements, will be discussed in Section~\ref{sec:prospects}.

If only contours of constant likelihood (the 68\%~CL interval in 1D, 68\% and 95\%~CL contours in 2D) are present, assumptions about the shape of the likelihood have to be made in order to reconstruct it in the full plane.
The 1D case was already discussed above, and resulted in the likelihood of Eq.~\eqref{eq:Lchi2}.
In the 2D case, a natural choice is to use a bivariate normal (Gaussian) distribution.
For two (combination of) production and decay processes $(X,Y)$ and $(X',Y')$, we obtain the following likelihood:
\beq
- 2 \log L(\boldsymbol{\mu}) =
(\boldsymbol{\mu} - \hat{\boldsymbol{\mu}})^T
C^{-1}
(\boldsymbol{\mu} - \hat{\boldsymbol{\mu}}) \,, \label{eq:mu2d}
\eeq
where $\boldsymbol{\mu} = \begin{pmatrix} \mu(X, Y) \\ \mu(X', Y') \end{pmatrix}$, and $C^{-1} = \begin{pmatrix} a&b\\ b&c \end{pmatrix}$ is the inverse of the covariance matrix.
Under the bivariate normal approximation, the 68\% and 95\%~CL contours (which are iso-contours of $-2\log L$) 
 are ellipses. The information on a single contour suffices to reconstruct the likelihood in the full plane: the parameters $a$, $b$ and $c$, as well as $\hat\mu(X,Y)$ and $\hat\mu(X',Y')$, can be fitted from points sitting on the 68\%~CL or 95\%~CL contours as they have known values of $- 2 \log L$ ($2.30$ and $5.99$, respectively). 
In the following, unless stated otherwise, we choose to reconstruct the full likelihood from a fit to the 68\%~CL contour provided by the experimental collaboration. 
However, having more than one contour of constant likelihood is very useful for checking the validity of this approximation.
This will be presented in Section~\ref{sec:validation} for the experimental results included in the database of {\tt Lilith}.
Finally, note that generalization of the previous equations is trivial should higher-dimensional signal strength measurements be published by the experimental collaborations.

{\renewcommand{\arraystretch}{1.3}
\begin{table}
\begin{center}
	\begin{tabular}{c|c|c|c}
	Collaboration & Analysis & Type & Reference \\
	\hline
	\multirow{8}{*}{ATLAS} & $H\to\gamma\gamma$ & 2D contour & \cite{Aad:2014eha} \\
	 & $H\to ZZ^*$ & 2D contour & \cite{Aad:2014eva} \\
	 & $H\to WW^*$ & 2D contour & \cite{ATLAS:2014aga} \\
	 & $H\to\tau\tau$ & 2D contour & \cite{Aad:2015vsa} \\
	 & $\VH, H\to b\bar{b}$ & 2D contour & \cite{Aad:2014xzb} \\
	 & $\ZH, H\to\invisible$ & full 1D & \cite{Aad:2014iia} \\
	 & $\ttH, H\to b\bar{b}$ & 1D interval & \cite{ATLAS-CONF-2014-011} \\
	 & $\ttH, H\to \gamma\gamma$ & 1D interval & \cite{Aad:2014eha} \\
	\hline
	\multirow{3}{*}{CMS} & $H\to\gamma\gamma, ZZ^*, WW^*, b\bar{b}, \tau\tau$ & 2D contours & \cite{Khachatryan:2014jba} \\
	 & $\ttH, H\to \gamma\gamma, \tau\tau$ & 1D interval & \cite{Khachatryan:2014jba} \\
	 & $\ttH, H\to$~leptons & 1D interval & \cite{Khachatryan:2014qaa} \\
	 & $\ZH+\VBF, H\to\invisible$ & full 1D & \cite{Chatrchyan:2014tja}\\
	\hline
	CDF \& D0 & $\VH, H\to b\bar{b}$ & 1D interval & \cite{Aaltonen:2013kxa} \\
	\end{tabular}
\end{center}
\caption{Recommended set of experimental results, for the database of {\tt Lilith} version {\tt 15.02}. This set corresponds to the file {\tt latest.list}, and is used by default when running {\tt Lilith}.}
\label{tab:defaultdatabase}
\end{table}

A database of up-to-date experimental results is shipped with {\tt Lilith}, along with recommended sets of results to use for computing the likelihood (in the form of list files; all technical details will be given in Section~\ref{sec:manual}).
The default set of results, {\tt latest.list}, includes the latest measurements from the LHC.
Its content as of February 2015 is displayed in Table~\ref{tab:defaultdatabase}.
All considered 2D results are in the plane $(\mu({\rm ggH}, Y), \mu({\rm VBF+VH}, Y))$ except for $Y=\gamma\gamma$ in ATLAS, where only VBF is considered instead of VBF+VH, and for $Y=b\bar b$ in CMS, which is given in the plane $(\mu({\rm ttH}, b\bar b), \mu({\rm VH}, b\bar b))$. The CMS 2D results are taken from the combination of Ref.~\cite{Khachatryan:2014jba}, but correspond to the results from Refs.~\cite{Khachatryan:2014ira,Chatrchyan:2013mxa,Chatrchyan:2013iaa,Chatrchyan:1643937,Khachatryan:2014qaa}.
We also take into account all available searches on production in association with a top-quark pair, as well as searches for invisible decays of the Higgs boson from both ATLAS and CMS. Note the presence of the CDF and D0 combined result for VH, $H \to b\bar b$~\cite{Aaltonen:2013kxa}; only in this channel the precision 
of the Tevatron result is comparable with the one of the LHC at Run~I.

All considered experimental results are given at a fixed Higgs mass (that can be read from the database, see Section~\ref{sec:expinput}) in the $[125, 125.6]$~GeV range. Variations of the experimental results within this narrow interval are expected to be small, hence limiting the inconsistencies when combining the results. However, it would be desirable to take into account the variation of the results with mass, as we will argue in Section~\ref{sec:prospects}.
The final Higgs likelihood is the product of the individual (1- or 2-dimensional) likelihoods. 
Validation of the Higgs likelihood used in {\tt Lilith} against official LHC results will be presented in Section~\ref{sec:validation}.

\section{Parametrization of new physics}
\label{sec:npparam}

In order to assess the compatibility of a new physics hypothesis with the LHC measurements presented in the previous section, one needs to compute the expected signal strengths $\mu(X,Y)$ (see Eq.~\eqref{eq:signalstr3}) for the relevant production mechanisms $X$ and decay modes $Y$. This can be achieved in a direct way from $\sigma(X)$, $\sigma^{\rm SM}(X)$, $\BR(H\to Y)$, and $\BR^{\rm SM}(H\to Y)$, but is often found to be impracticable. Indeed, in order to have well-defined signal strengths (for which $\mu=1$ corresponds to the SM prediction) one should take the same prescription for computing cross sections and branching fractions in the SM and in the considered new physics scenario~\cite{Heinemeyer:2013tqa}. Concretely, one needs to consider the same order in perturbation theory, the same set of parton density functions, etc.

In most new physics scenarios only leading order (LO) computations are available. Thus, all available next-to-LO (NLO) corrections to the SM predictions should be ignored. While this leads to properly defined signal strengths, $\sigma_{\rm NLO}(X) / \sigma_{\rm NLO}^{\rm SM}(X)$ will typically differ from $\sigma_{\rm LO}(X) / \sigma_{\rm LO}^{\rm SM}(X)$ (and similarly for branching ratios) as soon as one deviates from the SM prediction. This is because the relative contributions of SM particles to the process will be affected by the NLO corrections. For instance, higher-order corrections to the gluon fusion process will change the relative contribution of the top and bottom quark loops. Therefore, considering LO or NLO cross sections will yield different $\mu(\ggH,Y)$ if new physics affects the couplings of the Higgs to top and bottom quarks in a different way.

These two problems come from the parametrization of new physics effects from cross sections and branching ratios. As we will see, they can be alleviated if new physics is parametrized instead using reduced couplings.

\subsection{Scaling factors and reduced couplings}
\label{sec:redc}

The general signal strength expression given in Eq.~\eqref{eq:signalstr2} can be rewritten as
\begin{align}
\mu = \sum_{X,Y} {\rm eff}_{X,Y} \frac{\sigma(X)\BR(H\to Y)}{\sigma^{\rm SM}(X)\BR^{\rm SM}(H\to Y)} &= \sum_{X,Y} {\rm eff}_{X,Y}\times \frac{C_X^2 \sigma^{\SM} (X)}{\sigma^{\SM}(X)}  \times \frac{C_Y^2 \Gamma^{\rm SM}_Y}{\Gamma^{\rm SM}_Y} \times \frac{\Gamma^{\rm SM}_H}{\sum_Y C_Y^2 \Gamma^{\rm SM}_Y} \nonumber \\
&= \frac{1}{\sum_Y C_Y^2 \BR^{\rm SM}(H\to Y)} \sum_{X,Y} {\rm eff}_{X,Y} C_X^2 C_Y^2
\,, \label{eq:signalstr4}
\end{align}
where $\Gamma_H^{\SM}$ is the total decay width of the SM Higgs boson, and the cross section (partial width) for each process $X$ ($Y$) is scaled with a factor $C_X^2$ ($C_Y^2$) compared to the SM expectation.\footnote{The scaling factors $C_i$ are often seen elsewhere as $\kappa_i$.} The term $\sum_Y C_Y^2 \BR^{\rm SM}(H\to Y)$ accounts for the scaling of the total width of the Higgs boson. (We assume that the narrow-width approximation also holds in the new physics scenarios.) Furthermore, we introduce reduced couplings through the following Lagrangian,
\begin{equation}
	\label{eq:lagrangian}
	\begin{split}
		\mathcal{L} &= g \left[ C_W m_W W^\mu W_\mu + C_Z \frac{m_Z}{\cos \theta_W} Z^\mu Z_\mu\right] H \\
		& + g \left[- C_t \frac{m_t}{2m_W} t\bar{t} - C_b \frac{m_b}{2 m_W} b\bar{b} - 
			C_c \frac{m_c}{2 m_W} c\bar{c} - C_\tau \frac{m_\tau}{2 m_W} \tau\bar{\tau} \right] H \,,
	\end{split}
\end{equation}
where $C_{W,Z}$ and $C_{t,b,c,\tau}$ are bosonic and fermionic reduced couplings, respectively. Light fermions are not taken into account as their phenomenological impact on the SM Higgs sector is negligeable. In the limit where all reduced couplings 
go to 1, the SM case is recovered. At leading order in perturbation theory, the scaling factors $C_X$ and $C_Y$ from Eq.~\eqref{eq:signalstr4} can be directly identified with the reduced couplings $C_i$ from Eq.~\eqref{eq:lagrangian} for processes involving just one coupling to the Higgs boson. We obtain
\begin{equation}
		 C_{\WH}^2 = C_W^2 \,, \quad
		 C_{\ZH}^2 = C_Z^2 \,, \quad
		 C_{\ttH}^2 = C_t^2 \,, \quad
		 C_{f\bar{f}}^2 = C_f^2 \,, \quad
		 C_{VV}^2 = C_V^2 \,, \quad
\end{equation}
where $f=b,c,\tau$ and $V=W,Z$.

For the remaining main processes (ggH and VBF production, decay into $gg$, $\gamma\gamma$ and $Z\gamma$), there is no direct identification unless the Higgs couplings to all involved SM particles scale in the same way.
In the general case, the $C_X$ and $C_Y$ for these processes will be given by a combination of reduced couplings $C_i$, weighted according to the contribution of the particle $i$ to the process.
For the production mechanisms, we have
\begin{align} \label{eq:ggHVBF}
	C_{\rm ggH}^2 &= \frac{\sum\limits_{i,j=t,b,c} C_i C_j \, \sigma_{ij}^{\rm SM}(\rm ggH)}{\sum\limits_{i,j=t,b,c}\sigma_{ij}^{\rm SM}({\rm ggH})} \,, \quad
	C_{\rm VBF}^2 = \frac{\sum\limits_{i,j=W,Z} C_i C_j \, \sigma_{ij}^{\rm SM}(\rm VBF)}{\sum\limits_{i,j=W,Z} \sigma_{ij}^{\rm SM}(\rm VBF)} \,,
\end{align}
where the $\sigma_{ij}^{\rm SM}$ are the different contributions to the cross section in the SM. For $i=j$ it corresponds to the cross section from 
the particle $i$ alone, while $i\neq j$ comes from the interference between the particles $i$ and $j$. (We only consider either the term $\sigma_{ij}^{\rm SM}$ or $\sigma_{ji}^{\rm SM}$ in the sum, not both, to avoid double counting the interference terms.)
Similarly, the reduced couplings for the $gg$, $\gamma\gamma$, and $Z\gamma$ loop-induced decay modes are computed as 
\begin{align} \label{eq:Cloop}
     C_{gg}^2 = \frac{\sum\limits_{i,j=t,b,c} C_i C_j \, \Gamma_{ij}^{\rm SM}(H\to gg)}{\sum\limits_{i,j=t,b,c}\Gamma_{ij}^{\rm SM}(H\to gg)}  \,, \quad
     C_{\gamma\gamma,Z\gamma}^2 = \frac{\sum\limits_{i,j=W,t,b,c,\tau} C_i C_j \, \Gamma_{ij}^{\rm SM}(H\to\gamma\gamma,Z\gamma)}{\sum\limits_{i,j=W,t,b,c,\tau}\Gamma_{ij}^{\rm SM}(H\to\gamma\gamma,Z\gamma)}  \,,
\end{align}
where the $\Gamma_{ij}^{\rm SM}$ are the SM partial widths of the process under consideration. In all cases, all relevant SM contributions have been taken into account. Note that the relative sign of the reduced couplings will affect the interference terms, as they are proportional to $C_i C_j$. 

At LO, the various $\sigma_{ij}^{\rm SM}$ and $\Gamma_{ij}^{\rm SM}$ can be obtained from tree-level amplitudes (for VBF) or from the 1-loop amplitudes (for $gg \to H$ and $H \to gg,\gamma\gamma,Z\gamma$).\footnote{At LO, one obtains the same scaling for gluon fusion and for the decay into two gluons, $C_{\rm ggH} = C_{gg}$.} 
It would however be desirable to take into account the NLO corrections to the Higgs cross sections and partial widths as they modify the relations $C_{X,Y}(C_i)$. This can be achieved in a simple way as long as higher-order corrections only rescale the $\sigma_{ij}^{\rm SM}$ and $\Gamma_{ij}^{\rm SM}$ that are already existing in Eqs.~\eqref{eq:ggHVBF}--\eqref{eq:Cloop}, and do not induce new couplings to the Higgs boson. This is the case for the QCD corrections, but not for the electroweak corrections. Thus, as will be explained in Section~\ref{sec:reducedcouplingmode}, the QCD corrections for all five processes of Eqs.~\eqref{eq:ggHVBF}--\eqref{eq:Cloop} will be included in {\tt Lilith}.

One last remark is in order. The signal strength framework requires that the signal in all searches be a sum of processes that exist for the SM Higgs boson. However, new production or decay modes may exist without spoiling the signal strength interpretation as long as they do not yield sizable contribution in the current Higgs searches. 
Two particularly interesting cases are Higgs boson decays into undetected particles, or into invisible particles. In the first case, this new decay is simply missed by current searches (as would, {\it e.g.}, be the case for the decay of the Higgs into light quarks and gluons), while in the second case this new decay mode gives rise to missing energy in the detector. As was shown in Section~\ref{sec:statproc}, invisible decays of the Higgs boson are constrained by current searches which are taken into account in {\tt Lilith}.
In both cases of undetected and invisible decays, the width of the Higgs boson becomes larger and modifies the signal strength predictions of Eq.~\eqref{eq:signalstr4} as
\begin{equation}
	\label{eq:BRBSM}
	\mu(C_X,C_Y)
    \longrightarrow (1 - \BR_{\rm invisible} - \BR_{\rm undetected}) \mu(C_X,C_Y) \,.
\end{equation}
In {\tt Lilith}, arbitrary invisible and/or undetected decays can be specified, as will be presented in Section~\ref{sec:usermodeinput}.

\subsection{CP-violating admixtures}
\label{sec:npparamCPV}

We also consider the case where the observed Higgs boson is a mixture of CP-even and CP-odd states~\cite{Godbole:2004xe,Accomando:2006ga}.
The Higgs coupling to vector bosons has the form 
\begin{eqnarray}
  VVH: \quad C_V\, \frac{gM_V^2}{m_W}\, g^{\mu \nu} \,,
\end{eqnarray}
where as above $C_V$ measures the departure from the SM: $C_V=1$ for a pure scalar $H^+$ (CP-even)
state with SM-like couplings and $C_V=0$ for a pure pseudoscalar $H^-$ (CP-odd) state. 

In the fermion sector, we find the general vector and axial--vector structure of the Higgs coupling to fermions. Concretely, we have 
\begin{eqnarray}
   H f\bar f: \quad   -\bar f (\Ree(C_f)+i\Imm(C_f)\gamma_5) f \, \frac{gm_f}{2m_W} \,,
\label{eq:hfit-CP-Hff}
\end{eqnarray}
where in the SM one has $\Ree(C_f)=1$ and $\Imm(C_f)=0$, while a purely CP-odd Higgs would have $\Ree(C_f)=0$ and $\Imm(C_f)=1$. Since $m_f^2\ll m_H^2$ for $f=b,c,\tau$, the partial decay widths scale as $\Gamma(H\to f\bar{f})\propto \Ree(C_f)^2+\Imm(C_f)^2 = |C_f|^2$ to a very good approximation~\cite{Djouadi:2005gi}. This is what is implement in $\Lilith$.
Effects of CP mixing will mainly show up at loop level, in particular in the $gg\to H$ and $H\to \gamma\gamma$ rates.
A test of the CP properties of the observed Higgs from a global fit to the signal strengths was  
presented in \cite{Djouadi:2013qya,Brooijmans:2014eja}.
Following Ref.~\cite{Djouadi:2013qya}, at leading order the Higgs rates normalized to 
the SM expectations can be written as
\begin{eqnarray}
\frac{\Gamma( H \to \gamma\gamma)}{\Gamma^{\rm SM}( H \to \gamma\gamma)} 
&  \simeq & 
\frac{\left| \frac{1}{4} C_W A^+_1[m_W] + \left(\frac{2}{3}\right)^2 {\rm Re}(C_U)  \right|^2
+ \left| \left(\frac{2}{3}\right)^2 \frac{3}{2} {\rm Im}(C_U) \right|^2} 
{\left| \frac{1}{4} A^+_1[m_W] + \left(\frac{2}{3}\right)^2  \right|^2 } \,, 
\nonumber \\ 
\frac{\sigma( gg \to H)}{\sigma^{\rm SM}( gg \to H)} & = & 
\frac{\Gamma( H \to gg)}{\Gamma^{\rm SM}( H \to gg)}  \simeq 
\left| {\rm Re}(C_U)  \right|^2 + \left|\frac{3}{2} {\rm Im}(C_U)  
\right|^2 \,, 
\label{eq:hfit-widthsCP} 
\end{eqnarray}
with $A^+_1[m_W] \simeq -8.32$ for $m_H = 125$ GeV. For convenience, the contribution from the other 
quarks has been omitted in the above equations but is taken into account in \lilith.

In the case of ttH production, the approximation that we used above for the other fermions does not hold since $m_t>m_H$.
Instead, the cross section scales as $\sigma(\ttH^{+/-}) \propto {\rm Re}(C_f)^2 + {\rm Im}(C_f)^2 \sigma^{\rm SM}(\ttH^-)/\sigma^{\rm SM}(\ttH^+)$. Following Ref.~\cite{Frederix:2011zi}, a factor $\sigma^{\rm SM}(\ttH^-)/\sigma^{\rm SM}(\ttH^+) \approx 1/3$ is considered in {\tt Lilith}.
However, a significant coupling of the CP-odd component of the Higgs boson to top quarks may modify the acceptance times efficiency compared to the SM value in searches for the Higgs boson in association with a pair of top quarks~\cite{Frederix:2011zi, Demartin:2014fia}, {\it i.e.}, $(A \times \varepsilon)_{{\rm ttH},Y} \neq [(A \times \varepsilon)_{{\rm ttH},Y}]^{\rm SM}$. As this cannot be taken into account in {\tt Lilith}, such cases should be interpreted with care. Moreover, only after the end of Run II will the LHC have enough sensitivity to probe CP violating effects in the $H\to\tau\tau$ decays~\cite{Berge:2014sra}, and the product $(A \times \varepsilon)_{X,\tau\tau}$ can thus be approximated by the SM one for now.
Details on how to specify real and imaginary parts for the couplings are given in Section~\ref{sec:reducedcouplingmode}. More precise measurements at Run~II of the LHC will ultimately call for an implementation of CP admixture that includes NLO effects in {\tt Lilith}.

\section{Running $\Lilith$}
\label{sec:manual}

\subsection{Getting started}

\lilith\ is a library written in {\tt Python} for constraining model of new physics against the LHC 
results. The code is distributed under the GNU General Public License v3.0.
The latest version of \lilith\ and of the database of experimental results (as of February 2015, {\tt Lilith~1.1} and database version {\tt 15.02}) as well as all necessary information can be found at 
\begin{center} \url{http://lpsc.in2p3.fr/projects-th/lilith} \end{center}
The archive of \lilith\ can be unpacked in any directory.
It contains a root directory called {\tt Lilith-1.1/} where the following directories can be found:
\begin{itemize}
\item {\tt lilith/}: the {\tt Python} package itself. The {\tt Lilith} application programming interface (API) will be presented in Section~\ref{sec:api}. It also contains the {\tt Python/C} API that will be presented in Section~\ref{sec:cpp}.
\item {\tt data/}: contains the database of experimental results in {\tt XML} format, as well as {\tt *.list} text files for the recommended sets of results. Details are given in Section~\ref{sec:expinput}.
\item {\tt userinput/}: where parametrizations of new physics models, in the {\tt XML} format described in Section~\ref{sec:usermodeinput}, can be stored. Some basic user input files that include extensive comments are provided with the {\tt Lilith} distribution.
\item {\tt examples/}: concrete examples on how to use {\tt Lilith} for constraining new physics. Two of them will be presented in detail in Section~\ref{sec:examples}; an example for using \lilith\ in {\tt C} and {\tt C++}/{\tt ROOT} programs will be discussed in Section~\ref{sec:cpp}.
\item {\tt results/}: empty folder where results from \lilith\ can be stored.
\end{itemize}
The folder {\tt Lilith-1.1/} moreover contains {\tt run\_lilith.py}, the command-line interface (CLI) of {\tt Lilith} that will be presented in Section~\ref{sec:cli}, as well as general information, information on the license, and a changelog in the files {\tt README}, {\tt COPYING}, and {\tt changelog}, respectively.

$\Lilith$ requires {\tt Python~2.6}~\cite{python} or more recent, but not the {\tt 3.X}~series.
The standard {\tt Python} scientific libraries, {\tt SciPy} and {\tt NumPy}~\cite{scipy}, should furthermore be installed. We require {\tt SciPy~0.9.0} or more recent, and {\tt NumPy~1.6.1} or more recent.
{\tt Python}, {\tt SciPy} and {\tt NumPy} are available for the major platforms, including GNU/Linux, Mac OS~X, and Microsoft Windows.

The easiest way to check if all dependencies of {\tt Lilith} are correctly installed is to try to compute the likelihood from an example file. This can be achieved by typing to the shell (with current directory {\tt Lilith-1.1/}) the command
\begin{lstlisting}
 python run_lilith.py userinput/example_couplings.xml
\end{lstlisting}
Everything is correctly installed if basic information as well as the value of the likelihood is printed on the screen.
Note that the version number of {\tt Python} can be obtained by typing the command {\tt python --version} to the shell, while the presence of {\tt SciPy} and {\tt NumPy} and their version numbers can be checked by typing in an interactive session of {\tt Python} (started by typing {\tt python} to the shell) the following commands:
\begin{lstlisting}
 import scipy
 print scipy.__version__
 import numpy
 print numpy.__version__
\end{lstlisting}

Note that every version of {\tt Lilith} is shipped with the latest version for the database of experimental results at the time of release.
However, as new experimental results usually do not require any modification to {\tt Lilith}, we do not release a new version of the code every time new experimental results come out. Instead, we provide separately an update of the database of experimental results. Each release of the database has version number {\tt YY.MM} ({\it e.g.}, {\tt 15.02}), where {\tt YY} and {\tt MM} correspond to the year and to the month, respectively. If two or more updates of the experimental database are provided the same month, from the second release onwards versions will be numbered {\tt YY.MM.n} (with {\tt n} starting from {\tt 1}).
The version number of the database can be found in {\tt data/version} (also accessible via the {\bf readdbversion}() method of the API, see next section, and printed on the screen when using the CLI). When using the recommended sets of experimental results of {\tt Lilith} in a publication, the version number of the database must also be cited in addition to the experimental publications from which results were used.

\subsection{The {\tt Lilith} API}
\label{sec:api}

{\tt Lilith} provides an API from which all tasks (reading the user and the experimental input, compute the likelihood, print the results in a file, etc.) can be performed, using the methods described below.
This is the recommended way of using {\tt Lilith}.
In order to be used in any {\tt Python} code (or in an interactive session of {\tt Python}), the package of {\tt Lilith}, called {\tt lilith}, first needs to be imported. However, {\tt Python} needs to know the location of the {\tt lilith} package. This can be achieved in at least three ways:
\begin{enumerate}
\item create the {\tt Python} script importing {\tt lilith} in the directory {\tt Lilith-1.1/} (or, in an interactive session, having {\tt Lilith-1.1/} as the current directory).
\item adding the path to {\tt Lilith-1.1/} to the environment variable {\tt PYTHONPATH}. This can be done with the command
\begin{lstlisting}
 export PYTHONPATH=/path/to/Lilith-1.1:$PYTHONPATH
\end{lstlisting}
in {\tt bash} shell or
\begin{lstlisting}
 setenv PYTHONPATH /path/to/Lilith-1.1:$PYTHONPATH
\end{lstlisting}
in {\tt csh}/{\tt tcsh} shell.
In order to permanently have the path to {\tt Lilith} in {\tt PYTHONPATH} (not only for the current session) this command should be added in a {\tt .bashrc} (for {\tt bash}f) or {\tt .cshrc} / {\tt .tcshrc} (for {\tt csh}/{\tt tcsh} shell) file located in the home directory of the user.
\item adding the path to {\tt Lilith-1.1/} to the variable {\tt sys.path} by starting the script with
\begin{lstlisting}
 import sys
 sys.path.append('/path/to/Lilith-1.1')
\end{lstlisting}
before importing {\tt lilith}. Note that the path can also be relative.
\end{enumerate}

The {\tt Lilith} library can then be imported to the current script by typing {\tt import lilith} or {\tt from lilith import *}.
In the first case, all classes, methods and attributes defined in the code will be in the namespace {\tt lilith}, in the second case they will be in the global namespace.
We now present all methods and attributes of the API of {\tt Lilith}.
\begin{description}
  \item[{\rm {\it class} {\footnotesize lilith.}{\bf Lilith}({\it verbose=False}, {\it timer=False})}] \hfill \\
  Instanciate the {\bf Lilith} class. The following public attributes are initialized:
  \begin{description}
    \item[verbose] \hfill \\
    if {\it True}, information will be printed on the screen
    \item[timer] \hfill \\
    if {\it True}, each operation will be timed and the results will be printed on the screen
    \item[exp\_mu] \hfill \\
    list of experimental results read from the database
    \item[exp\_ndf] \hfill \\
    number of measurements ($n$-dimensional results count for $n$ measurements)
    \item[dbversion] \hfill \\
    version number for the database of experimental results
    \item[couplings] \hfill \\
    list of reduced couplings for each Higgs particle contributing to the signal as read from the user input
    \item[user\_mu] \hfill \\
    list of signal strengths for each Higgs particle contributing to the signal as read (or derived from) the user input
    \item[user\_mu\_tot] \hfill \\
    signal strengths for the sum of the Higgs particles present in {\bf user\_mu}
    \item[results] \hfill \\
    list of results after computation of the likelihood for each individual measurement
    \item[l] \hfill \\
    value of $-2\log L$
  \end{description}

  \item[{\rm {\it exception} {\footnotesize lilith.}{\bf LilithError}}] \hfill \\
  Base exception of {\tt Lilith}; all other exceptions inherit from it.
  For the definition of all the exceptions, see {\tt lilith/errors.py}.

  \item[{\rm {\footnotesize Lilith.}{\bf readuserinput}({\it userinput})}] \hfill \\
  Read the string in {\tt XML} format given as argument and fill the attribute {\bf couplings} (if the user input is given in terms of reduced couplings) or {\bf user\_mu} and {\bf user\_mu\_tot} (if the user input is given in terms of signal strengths). User input formats are presented in Section~\ref{sec:usermodeinput}.

  \item[{\rm {\footnotesize Lilith.}{\bf readuserinputfile}({\it filepath})}] \hfill \\
  Read the user input located at {\it filepath} and call {\bf readuserinput}().

  \item[{\rm {\footnotesize Lilith.}{\bf computecouplings}()}] \hfill \\
  Compute from {\bf user\_couplings} the following reduced couplings if not already present in {\bf user\_couplings}: 
  $C_{\rm VBF}$, $C_{\rm ggH}$, $C_{gg}$, $C_{\gamma\gamma}$, and $C_{Z\gamma}$.

  \item[{\rm {\footnotesize Lilith.}{\bf computemufromreducedcouplings}()}] \hfill \\
  Compute the signal strengths (stored in {\bf user\_mu} and {\bf user\_mu\_tot}) from the reduced couplings in {\bf user\_couplings}.

  \item[{\rm {\footnotesize Lilith.}{\bf compute\_user\_mu\_tot}()}] \hfill \\
  Add up the signals from all Higgs bosons contributing to the signal in {\bf user\_mu}; store the result in {\bf user\_mu\_tot}.

  \item[{\rm {\footnotesize Lilith.}{\bf readexpinput}({\it filepath=default\_exp\_list})}] \hfill \\
  Read the experimental input specified in a list file and store the results in {\bf exp\_mu} and {\bf exp\_ndf}. By default, the list file is {\tt data/latest.list}. The formats of the experimental results are presented in Section~\ref{sec:expinput}.

  \item[{\rm {\footnotesize Lilith.}{\bf readdbversion}()}] \hfill \\
  Read the version of the database of experimental results from the file {\tt data/version}, and store the information in {\bf dbversion}.

  \item[{\rm {\footnotesize Lilith.}{\bf compute\_exp\_ndf}()}] \hfill \\
  Compute the number of measurements from {\bf exp\_mu} and store the information in {\bf exp\_ndf}.

  \item[{\rm {\footnotesize Lilith.}{\bf computelikelihood}({\it userinput=None}, {\it exp\_filepath=None}, {\it userfilepath=None})}] \hfill \\
  Evaluate the likelihood function from signal strengths derived from the user input ({\bf user\_mu\_tot}) and the experimental results ({\bf exp\_mu}) and store the results in the attribute {\bf results}.
  If the arguments {\it userinput} and {\it userfilepath} are not specified, {\bf user\_mu\_tot} will be assumed to have been filled already. Else, all information will be read from the {\tt XML} input given in {\it userinput}, or from the file located at {\it userfilepath}, and {\bf user\_mu\_tot} will be computed.
  If the {\it exp\_filepath} argument is not specified, the experimental results will be read from the default list file unless {\bf exp\_mu} is already filled. Else, experimental results from {\it exp\_filepath} will be read before computing the likelihood.

  \item[{\rm {\footnotesize Lilith.}{\bf writecouplings}({\it filepath})}] \hfill \\
  Write reduced couplings from the attribute {\bf couplings} in a file located at {\it filepath} in the {\tt XML} format specified in Section~\ref{sec:reducedcouplingmode}.

  \item[{\rm {\footnotesize Lilith.}{\bf writesignalstrengths}({\it filepath}, {\it tot=False})}] \hfill \\
  Write signal strengths from the attribute {\bf user\_mu} (if {\it tot=False}) or {\bf user\_mu\_tot} (if {\it tot=True}) at {\it filepath} in the {\tt XML} format specified in Section~\ref{sec:signalstrengthsmode}.

  \item[{\rm {\footnotesize Lilith.}{\bf writeresults}({\it filepath}, {\it slha=False})}] \hfill \\
  Write the content of the attribute {\bf results} at the location {\it filepath} in the {\tt XML} format (if {\it slha=False}) or the {\tt SUSY Les Houches Accord} ({\tt SLHA})-like format~\cite{Skands:2003cj} specified in Section~\ref{sec:output}.
\end{description}
Note also that the version of {\tt Lilith} is stored in the {\tt lilith.\_\_version\_\_} variable.

A minimal example of use of the API is as follows:
\begin{lstlisting}
 from lilith import *
 lcal = Lilith()
 lcal.readexpinput()
 lcal.readuserinputfile('userinput/example_mu.xml')
 lcal.computelikelihood()
 print '-2log(likelihood) =', lcal.l
\end{lstlisting}
The first two lines import the {\tt Lilith} library into the global namespace and initialize the computations. They are equivalent to
\begin{lstlisting}
 import lilith
 lcal = lilith.Lilith()
\end{lstlisting}
The three following lines successively read the experimental input, read the user input from the file {\tt userinput/example\_mu.xml}, and compute the likelihood. Alternatively, they could be replaced with a single line,
\begin{lstlisting}
 lcal.computelikelihood(userfilepath='userinput/example_mu.xml')
\end{lstlisting}
Finally, the value of $-2\log L$ is printed on the screen on the last line.

In the example above, any error (corresponding to an exception in {\tt Python}) will interrupt the code.
This may not be the desired behavior.
In particular, if several user inputs are successively given to {\tt Lilith} (as in the case of a scan of a parameter space), it may be preferable to store the error and move on to the next user input instead of stopping the execution of the code.
In {\tt Python}, the handling of errors can be achieved with {\tt try ..\ except} blocks. We provide below a simple example.
\begin{lstlisting}
 try:
     lcal.readuserinputfile('userinput/example_mu.xml')
     lcal.computelikelihood()
     print '-2log(likelihood) =', lcal.l
 except LilithError as e:
     print 'the following error occured:', e
\end{lstlisting}
Here, any error raised by {\tt Lilith} (of type {\footnotesize lilith.}{\bf LilithError} or derived from it) will be catched, in which case the error message will be printed on the screen and the script will continue normally.
It is of course also possible to store the error message in a file, or simply replace the last line with the {\tt pass} statement in order to ignore all errors raised by {\tt Lilith} and continue the execution of the script.
For the definition of all exceptions used in {\tt Lilith} (that all derive from {\footnotesize lilith.}{\bf LilithError}), see {\tt lilith/errors.py}. This makes it possible to treat each type of error in a different way.

\subsection{Command-line interface}
\label{sec:cli}

A command-line interface or CLI is also shipped with {\tt Lilith} for a more basic usage of the tool. It corresponds to the file \texttt{run\_lilith.py} located in the directory {\tt Lilith-1.1/}.
The CLI can be called by typing to the shell (with current directory {\tt Lilith-1.1/}) the command
\begin{lstlisting}
 python run_lilith.py user_input_file (experimental_input_file) (options)
\end{lstlisting}
where arguments in parentheses are optional.

The first argument, \texttt{user\_input\_file}, is the path to the user input file in the {\tt XML} format described in Section~\ref{sec:usermodeinput}. New physics can be parametrized in terms of reduced couplings (see Section~\ref{sec:npparam}) or directly in terms of signal strengths. Examples are shipped with {\tt Lilith} in the directory {\tt userinput/}.
The second argument, \texttt{experimental\_input\_file}, is the path to the list of experimental results to be used for the construction of the likelihood. If not given, the latest LHC results will be used ({\tt data/latest.list}; its content is given in Table~\ref{tab:defaultdatabase}). It is the recommended list of experimental results to be used for performing a global fit. All details about the experimental input will be given in Section~\ref{sec:expinput}.

If no option is given, basic information as well as the value of the likelihood and the number of measurements is printed on the screen. A number of options are provided to control the information printed on the screen and to print the results of {\tt Lilith} in output files. They are listed in Table~\ref{tab:options}.

\begin{table}
\begin{center}
\begin{tabular}{c|c}
Option & Meaning\\
\hline
\multirow{2}{*}{{\tt --help}, {\tt -h}} & Display basic usage of \texttt{run\_lilith.py} \\
& and list of options \\
\hline
{\tt --verbose}, {\tt -v} & Display details about the computation\\
\hline
{\tt --timer}, {\tt -t} & Time each operation and display results on the screen \\
\hline
{\tt --silent}, {\tt -s} & Silent mode\\
\hline
{\tt --couplings=output}, & Obtain and print the attribute {\bf couplings}\\
{\tt -c output} & in the file \texttt{output} in \texttt{XML} format\\
\hline
{\tt --mu=output}, & Obtain and print the attribute {\bf user\_mu}\\
{\tt -m output} & in the file \texttt{output} in \texttt{XML} format\\
\hline
{\tt --results=output}, & Obtain and print the attribute {\bf results}\\
{\tt -r output} & in the file \texttt{output} in \texttt{XML} or \texttt{SLHA}-like format \\
\end{tabular}
\end{center}
\caption{Options available when running {\tt run\_lilith.py}.}
\label{tab:options}
\end{table}

The option \texttt{--couplings} / {\tt -c} only works in the reduced couplings mode. In addition to the reduced couplings already present in the input file, it prints scaling factors computed from the input (\textit{i.e.} $C_{\gamma\gamma}, C_{Z\gamma}, C_{gg}, C_{\ggH},$ and $C_{\VBF}$, see Section~\ref{sec:redc}). 
The option \texttt{--mu} / {\tt -m} prints the complete list of signal strengths in a file in {\tt XML} format. More specifically, all signal strengths $\mu(X,Y)$ with $X\in (\ggH$, $\VBF$, $\WH$, $\ZH$, $\ttH)$ and $Y\in (\gamma\gamma$, $ZZ^*$, $WW^*$, $b\bar{b}$, $\tau\tau$, $c\bar{c}$, $Z\gamma$, $gg$, $\invisible)$ are printed. 
Finally, the option \texttt{--results} / {\tt -r} prints the value of $-2\log L$ and the number of measurements in a file. If the extension of the filename is \texttt{.slha} (case-insensitive), a file in {\tt SLHA}-like format is created. Otherwise, an \texttt{XML} file is created, with extra information on the individual contributions to $-2\log L$ from all the experimental results used in the calculation. 
More details on the structure and content of the output files are given in Section~\ref{sec:output}.

Initialization of {\tt Lilith} and reading of the database of experimental results is done at each execution of {\tt run\_lilith.py}. When computing the Higgs constraints in the context of the scan of a model, this only needs to be done once. Therefore, successive calls to the CLI will be much slower than direct calls to the methods of the API (see previous section), even more so as internal information is stored to optimize successive computations of the results when using the API. Whenever performance can be an issue, we highly recommend using the API instead of the CLI.

\subsection{Interface to {\tt C} and {\tt C++}/{\tt ROOT}}
\label{sec:cpp}

Above, we have presented how to use {\tt Lilith} from the API written in \texttt{Python} and from the CLI.
In order to use \lilith\ within a {\tt C}, {\tt C++}, or {\tt ROOT} code, a first possibility is to call the CLI. However, this will suffer from the performance issues explained at the end of the previous section.
Fortunately, {\tt Python} provides a {\tt Python/C} API~\cite{pythoncapi} from which each method of the API of {\tt Lilith} can be called, and each attribute of {\bf Lilith} can be manipulated, in both {\tt C} and {\tt C++}.
However, direct use of the functions of the {\tt Python}/{\tt C} API can be quite tedious.
For that reason, we have also written a {\tt C} API for {\tt Lilith}, consisting in a series of functions using the {\tt Python/C} API for performing all usual tasks, following closely the methods of the API presented in Section~\ref{sec:api}. 

The {\tt C} API for {\tt Lilith} is contained in the directory {\tt lilith/c-api}.
Its use requires the libraries and header files needed for {\tt Python} development.
They can be installed from most package managers under the name \texttt{python-dev} or \texttt{python-devel}, depending on the platform. More information on how to install these libraries and header, platform by platform, can be found at~\cite{lilith}.

We now present all the functions of the \texttt{C} API:
\begin{description}
\item[{\rm {\bf initialize\_lilith}({\it char* experimental\_input})}] \hfill \\
  Import {\tt lilith}, instanciate the class {\bf Lilith}, and read the experimental input file located at \textit{experimental\_input}. The function returns the instance object. If {\it experimental\_input} is an empty string({\tt ""}), the default experimental input file \texttt{data/latest.list} is used.
  
\item[{\rm {\bf lilith\_readuserinput}({\it PyObject* lilithcalc}, {\it char* XMLinputstring})}] \hfill \\
  Read the user input \texttt{XML} string \textit{XMLinputstring} and store the information in the object \textit{lilithcalc}.
  
\item[{\rm {\bf lilith\_readuserinput\_fromfile}({\it PyObject* lilithcalc}, {\it char* XMLinputpath})}] \hfill \\
  Read the user input \texttt{XML} file located at \textit{XMLinputpath} and store the information in the object \textit{lilithcalc}.  
    
\item[{\rm {\bf lilith\_computelikelihood}({\it PyObject* lilithcalc})}] \hfill \\
  Evaluate and return the value of $-2\log L$ from the object \textit{lilithcalc}.
  
\item[{\rm {\bf lilith\_exp\_ndf}({\it PyObject* lilithcalc})}] \hfill \\
  Evaluate and return the number of measurements from the object \textit{lilithcalc}.
  
\item[{\rm {\bf lilith\_likelihood\_output}({\it PyObject* lilithcalc}, {\it char* outputfilepath}, {\it int slha})}] \hfill \\
  Write the content of the attribute \textbf{results} at \textit{outputfilepath} in the \texttt{XML} format (if \textit{slha}=0) or the \texttt{SLHA}-like format (otherwise) specified in Section~\ref{sec:output}.
    
\item[{\rm {\bf lilith\_mu\_output}({\it PyObject* lilithcalc}, {\it char* outputfilepath}, {\it int tot})}] \hfill \\
  Write signal strengths from the attribute \textbf{user\_mu} (of \textit{tot=0}) or \textbf{user\_mu\_tot} (otherwise) at \textit{outputfilepath} in the \texttt{XML} format specified in Section~\ref{sec:signalstrengthsmode}.
  
\item[{\rm {\bf lilith\_couplings\_output}({\it PyObject* lilithcalc}, {\it char* outputfilepath})}] \hfill \\
  Write reduced couplings from the attribute \textbf{couplings} in a file located at \textit{outputfilepath} in the \texttt{XML} format specified in Section~\ref{sec:reducedcouplingmode}.

\end{description}

Initialization of {\tt Lilith} and the reading of the experimental input file can be done just once with the function \textbf{initialize\_lilith}().
Evaluation of the likelihood can then be performed separately.
An example of use of the {\tt C} interface of {\tt Lilith} is shipped with the code. It is available at \texttt{examples/c/lilith\_compute.c}. We now present it step by step.

\begin{lstlisting}
#include <Python.h>
#include "lilith.h"
int main(int argc, char* argv[])
{
    Py_Initialize();
\end{lstlisting}

The \texttt{Python.h} and \texttt{lilith.h} are the \texttt{Python/C} API and \lilith\ header files, respectively. Those are linked during the compilation by the {\tt Makefile} located in the same directory. To start the \texttt{Python/C} API, the function \texttt{Py\_Initialize()} is mandatory.

\begin{lstlisting}[firstnumber=6]
    char experimental_input[] = "../../data/latest.list";
 
    char output_couplings[] = "lilith_couplings_output.xml";
    char output_XML[] = "lilith_likelihood_output.xml";
    char output_SLHA[] = "lilith_likelihood_output.slha";
    char output_mu[] = "lilith_mu_output.xml";
  \end{lstlisting}
Line 6 is the path to the the experimental input file. Note that in this particular example, it is equivalent to \texttt{char experimental\_input[] = "";}. Various output file paths are then defined.

\begin{lstlisting}[firstnumber=12]
    PyObject* lilithcalc = initialize_lilith(experimental_input);
\end{lstlisting}

Line 12 is the initialization of a \lilith\ object \texttt{lilithcalc} from the experimental input file \texttt{experiment\_input}.

\begin{lstlisting}[firstnumber=13]
    char XMLinputstring[6000] = "";
    [...] // Construction of the user XML input string

    lilith_readuserinput(lilithcalc, XMLinputstring);
\end{lstlisting}
On lines 13 and 14, the user \texttt{XML} string input is constructed, see \texttt{examples/c/lilith\_compute.c} for more details.
The function \texttt{lilith\_readuserinput}() is then called to read the user input.

\begin{lstlisting}[firstnumber=17]
    float my_likelihood;
    my_likelihood = lilith_computelikelihood(lilithcalc);
    printf("-2*log(L) = %lf\n", my_likelihood);
  
    int exp_ndf;
    exp_ndf = lilith_exp_ndf(lilithcalc);
    printf("exp_ndf = %i\n", exp_ndf);
  
    lilith_likelihood_output(lilithcalc, output_XML, 0);
    lilith_likelihood_output(lilithcalc, output_SLHA, 1);
    lilith_mu_output(lilithcalc, output_mu, 0);
    lilith_couplings_output(lilithcalc, output_couplings);

    Py_Finalize();
    return 0;
}
\end{lstlisting}

Once the experimental and user input files have been read, computations can be performed. First, $-2\log L$ and the number of measurements are printed on the screen. Various output files are also created. Finally, \texttt{Py\_Finalize()} can be used to free all memory allocated by the {\tt Python} interpreter.

This example, {\tt lilith\_compute.c}, can be compiled and executed by typing to the shell
\begin{lstlisting}
 make
 ./lilith_compute
\end{lstlisting}
while the executable and the intermediate object files can be removed by typing to the shell {\tt make clean}.
In order to use {\tt Lilith} in {\tt C++} or {\tt ROOT} codes, only mininal changes need to be made compared to the case of {\tt C} codes. {\tt C++} users may want to change the compiler ({\tt CC} in the {\tt Makefile}) from {\tt gcc} to {\tt g++}.
{\tt ROOT} users should furthermore link the headers and libraries of {\tt ROOT} by adding the following two lines to the {\tt Makefile}:
\begin{lstlisting}
 CFLAGS += $(shell root-config --cflags)
 LFLAGS += $(shell root-config --glibs)
\end{lstlisting}

\subsection{Experimental input}
\label{sec:expinput}

We have seen that the evaluation of the likelihood in {\tt Lilith} requires the input of a list of experimental results to be considered. It corresponds to a simple text file with a {\tt .list} extension listing the paths to experimental result files in $\XML$ format (each containing a single 1D or 2D signal strength result).
{\tt Lilith} is shipped with the latest LHC Higgs results (plus a Tevatron result), see Table~\ref{tab:defaultdatabase} in Section~\ref{sec:statproc}, in the form of {\tt XML} files present in subdirectories of {\tt data/}. Moreover, several lists of experimental results are provided in {\tt data/}, with {\tt latest.list} being the default list file. This is the one recommended for a global fit to the LHC+Tevatron Higgs data.

The user can also create his/her own list file in the {\tt data/} directory.
For instance, in order to put constraints on new physics using only the latest di-boson results from ATLAS~\cite{Aad:2014eha,Aad:2014eva,ATLAS:2014aga}, one can create a file list that contains
\begin{verbatim}
 # ATLAS di-boson analyses
 ATLAS/Run1/HIGG-2013-08_ggH-VBF_gammagamma_n68.xml
 ATLAS/Run1/HIGG-2013-21_ggH-VVH_ZZ_n68.xml
 ATLAS/Run1/HIGG-2013-13_ggH-VVH_WW_n68.xml
\end{verbatim}
The first line, starting with a {\tt \#}, is a comment and is not read by $\Lilith$.
The three following lines indicate the paths to the $\XML$ files to be considered.
As can be read in the paths, these are published results from the ATLAS collaboration based on Run~I data.
The conventional, though not mandatory, naming scheme for the files is as follows. The identifier of the analysis (HIGG-2013-XX in this case) comes first, followed by the 2D plane in which results are given; finally, {\tt n68} indicates that the likelihood has been reconstructed from the contour at 68\%~CL under the Gaussian approximation.
Note that no consistency check is done by {\tt Lilith}. When creating a new list file, the user should make sure that there is no overlapping between experimental results ({\it e.g.}, that the results of two versions of the same analysis, based on overlapping event sets, are not used at the same time).

\subsubsection{$\XML$ format}

Every single experimental result (1D or 2D) is stored in a different $\XML$ file. In this way,
modifying and updating the database is an easy process. We now present the format of the experimental results files in {\tt Lilith}.

The root tag of each file is {\tt <expmu>}.
It has two mandatory attributes, {\tt dim} and {\tt type}, that specify the type of signal strength result (1D interval, full 1D, 2D contour, or full 2D, see Section~\ref{sec:statproc}).
Possible values for the attributes are given in Table~\ref{tab:xmlattributes}.
In addition, the {\tt <expmu>} tag has two optional attributes: {\tt prod} and {\tt decay}. They can be given a value listed in Table~\ref{tab:xmlattributes} if the analysis under consideration is only sensitive to one production mode ({\it e.g.}, ttH) or to one decay mode ({\it e.g.}, $\gamma\gamma$) of the Higgs boson.
In the general case, the {\tt prod} and {\tt decay} attributes can be skipped.
Indeed, all relevant efficiencies ${\rm eff}_{X,Y}$ (see Section~\ref{sec:mumeasurements}) can be specified in {\tt <eff>} tags.

\begin{table}
	\begin{center}
	\begin{tabular}{c|c|c|c|c}
	 Attribute & 1D interval & full 1D & 2D contour & full 2D \\
	\hline
	{\tt dim} & {\tt "1"} & {\tt "1"} & {\tt "2"} & {\tt "2"} \\
	\hline
	{\tt type} & {\tt "n"} & {\tt "f"} & {\tt "n"} & {\tt "f"}\\
	\hline
	({\tt prod}) & \multicolumn{4}{c}{{\tt "ggH", "ttH", "VBF", "WH", "ZH", "VH", "VVH"}}\\
	\hline
	\multirow{2}{*}{({\tt decay})} & \multicolumn{4}{c}{{\tt "gammagamma", "ZZ", "WW", "Zgamma",}}\\
                   & \multicolumn{4}{c}{{\tt "tautau", "bb", "cc", "invisible"}}\\
	\end{tabular}
	\end{center}
	\caption{Allowed values for the attributes of the {\tt <expmu>} tag, in experimental files in {\tt XML} format.
	The four different formats of experimental data are defined by the mandatory {\tt dim} and {\tt type} attributes.}
	\label{tab:xmlattributes}
\end{table}

Taking for instance the case of a 1D measurement, one can specify
\begin{verbatim}
 <eff prod="ggH">0.5</eff>
 <eff prod="VBF">0.5</eff>
\end{verbatim}
if {\tt decay} is specified in {\tt <expmu>}. Similarly, one can specify
\begin{verbatim}
 <eff decay="WW">0.5</eff>
 <eff decay="tautau">0.5</eff>
\end{verbatim}
if {\tt prod} is specified in {\tt <expmu>}. If none of them is present, one could specify efficiencies in the following way
\begin{verbatim}
 <eff prod="ggH" decay="WW">0.25</eff>
 <eff prod="VBF" decay="WW">0.25</eff>
 <eff prod="ggH" decay="tautau">0.25</eff>
 <eff prod="VBF" decay="tautau">0.25</eff>
\end{verbatim}
where it is required that the sum of all efficiencies is 1.
In the case of 2D signal strengths ({\it i.e.} if {\tt dim="2"} in {\tt <expmu>}) the efficiencies should be given for both dimensions (and separately add up to 1). In this case, the attribute {\tt axis} should be provided in every {\tt <eff>} tag, with possible values being {\tt "x"} or {\tt "y"} for the first and second dimension of the results.

Before turning to the syntax case by case, we comment on the optional information that can be provided in the experimental files. The following tags can be given:
\begin{itemize}
\item {\tt <experiment>}, referring to the experiment that has produced the current result.
\item {\tt <source>}, that contains the name of the analysis, and has attribute {\tt type} that contains the status of the analysis (published or preliminary).
\item {\tt <sqrts>} contains the collider center-of-mass energy.
\item {\tt <CL>}: when the likelihood has been extrapolated from a 2D contour, this can be used to indicate the CL of the contour thas has been used to extract the covariance matrix (usually the 68\% or 95\%~CL contour).
\item {\tt <mass>} contains the Higgs boson mass considered in the results. If not given, $m_H = 125$~GeV is assumed.

\end{itemize}

Note that, if {\tt prod="VH"} or {\tt prod="VVH"} is given as attribute to the {\tt <expmu>} tag or to an {\tt <eff>} tag, the relative contributions of WH and ZH (for VH) and of WH, ZH and VBF (for VVH) will be computed internally assuming an inclusive search, {\it i.e.}, for VVH,
\begin{equation}
	\label{eq:VHeff}
	{\rm eff}(X,Y) = \frac{\sigma^{\rm SM}(X)}{\sum_{X={\rm VBF}, {\rm WH}, {\rm ZH}} \sigma^{\rm SM}(X)}
    \quad {\rm for}\ X\in({\rm VBF}, {\rm WH}, {\rm ZH}) \,,
\end{equation}
where the cross sections are evaluated at the Higgs mass given in the {\tt <mass>} tag using 
the LHC Higgs Cross Section Working Group (HXSWG) results for the 8~TeV LHC~\cite{Heinemeyer:2013tqa}.

An explicit example of well-formed experimental input is
\begin{verbatim}
 <expmu dim="2" type="n" decay="ZZ">
   <experiment>ATLAS</experiment>
   <source type="published">HIGG-2013-21</source>
   <sqrts>7+8</sqrts>
   <mass>125.36</mass>
   <CL>68%</CL>

   <eff axis="x" prod="ggH">1.0</eff>
   <eff axis="y" prod="VVH">1.0</eff>
  
   <!-- (...) -->
 </expmu>
\end{verbatim}
for the results of the ATLAS $H \to ZZ^*$ analysis~\cite{Aad:2014eva}.
The comment {\tt <!-- (...) -->} indicates where the likelihood information should be placed.
We now present explicitly the different possibilities for specifying the likelihood in itself.
\begin{description}
\item[1D interval] \hfill \\
We consider as an example the $H\to b\bar{b}$ Tevatron search 
\cite{Aaltonen:2013kxa}. The following signal strength is provided: $\mu(\VH,b\bar{b})=1.59^{+0.69}_{-0.72}$.
\begin{verbatim}
 <bestfit>1.59</bestfit>
 <param>
   <uncertainty side="left">-0.72</uncertainty>
   <uncertainty side="right">0.69</uncertainty>
 </param>
\end{verbatim}
The {\tt <bestfit>} tag contains the best-fit value for the signal strength.
The {\tt <uncertainty>} tag contains the left (negative) and right (positive) 
	$1\sigma$ errors. If the left and right errors are equal in magnitude, the {\tt side} attribute is not 
	necessary and can be omitted.

\item[full 1D] \hfill \\
We consider as an example the $H\to b\bar{b}$ CMS search~\cite{Chatrchyan:2014vua}. 
The 1D profile likelihood as a function of $\mu(\VH,b\bar{b})$ is provided in Ref.~\cite{Chatrchyan:2014vua}, see Fig.~\ref{fig:dataformats}.
\begin{verbatim}
 <grid>
   0.00217269   4.30368
   0.00617181   4.26767
   .........
   1.99356   3.32783
   1.99756   3.35332
 </grid>
\end{verbatim}
The digitized likelihood information is stored in the tag {\tt <grid>}. The first column is the signal strength (whose nature is determined by the {\tt <eff>} tags) while the second column is the value of $-2\log L$.

\item[2D contour] \hfill \\
We consider as an example the CMS search $H\to b\bar{b}$~\cite{Chatrchyan:2013zna}. 
The 68\% and 95\%~CL contours are provided in the plane $(\mu(\WH,b\bar b),\mu(\ZH,b\bar b))$. 
As was explained in Section~\ref{sec:statproc}, we start by fitting the 68\%~CL contour assuming that the likelihood follows a bivariate normal distribution, and we extract the experimental best-fit point and the inverse of the covariance matrix $C^{-1} = \begin{pmatrix} a&b\\ b&c \end{pmatrix}$.
\begin{verbatim}
 <bestfit>
   <x>1.123</x>
   <y>0.997</y>
 </bestfit>

 <param>
   <a>1.393</a>
   <b>0.190</b>
   <c>2.217</c>
 </param>
\end{verbatim}
The tag {\tt <bestfit>} specifies the location of the best-fit point in the ({\tt x},{\tt y}) plane.
The tag {\tt <param>} contains the sub-tags {\tt <a>}, {\tt <b>}, and {\tt <c>}, that parametrize the inverse of the covariance matrix in the ({\tt x},{\tt y}) plane.

\item[full 2D] \hfill \\
We consider as an example the $H\to\gamma\gamma$ ATLAS search~\cite{Aad:2013wqa} for 
which the full likelihood information is given in the plane $(\mu({\rm ggH+ttH},\gamma\gamma,{\rm VBF+VH},\gamma\gamma))$~\cite{HEPDATA1}.
\begin{verbatim}
 <grid>
   4.26000000e-01	-4.50000000e-01	4.45215260e+01
   4.78000000e-01	-4.50000000e-01	4.19894480e+01	
   5.30000000e-01	-4.50000000e-01	3.94115060e+01
   .........
   2.87000000e+00	4.45000000e+00	1.84899120e+01	
   2.92200000e+00	4.45000000e+00	1.90217220e+01	
   2.97400000e+00	4.45000000e+00	1.95605440e+01
 </grid>
\end{verbatim}
The tag {\tt <grid>} contains the grid provided by the experimental collaboration.
	The first and second columns are defined by the {\tt axis="x"} and {\tt axis="y"} attributes
	of the {\tt <eff>} tag, respectively. The third column is the value of $-2\log L$.
\end{description}

\subsection{User model input}
\label{sec:usermodeinput}

The user model input, parametrizing the new physics model under consideration, can be given either in terms of signal strengths $\mu(X,Y)$ directly (defined as in Eq.~\eqref{eq:signalstr3}), or in terms of reduced couplings and scale factors (see Section~\ref{sec:npparam}). In the latter case, scale factors that might be missing in the input are computed, and signal strengths are derived from the scale factors.

The user model input has {\tt XML} syntax and can be provided as a string or in the form of a file (see the methods \texttt{readuserinput}() and \texttt{readuserinputfile}() in Section~\ref{sec:api}). In this section, we present the format that is used in {\tt Lilith}.

\subsubsection{{\tt XML} format for signal strengths}
\label{sec:signalstrengthsmode}

In the signal strengths mode, the basic inputs are the signal strengths defined as in Eq.~\eqref{eq:signalstr3}. An example of $\XML$ 
input file for the signal strengths mode is now presented.
\begin{verbatim}
 <lilithinput>
   <signalstrengths>
     <mass>125</mass>

     <mu prod="ggH" decay="gammagamma">1.0</mu>
     <mu prod="ggH" decay="VV">1.0</mu>
     <mu prod="ggH" decay="bb">1.0</mu>
     <mu prod="ggH" decay="tautau">1.0</mu>

     <mu prod="VVH" decay="gammagamma">1.0</mu>
     <mu prod="VVH" decay="VV">1.0</mu>
     <mu prod="VVH" decay="bb">1.0</mu>
     <mu prod="VVH" decay="tautau">1.0</mu>
  
     <mu prod="ttH" decay="gammagamma">1.0</mu>
     <mu prod="ttH" decay="VV">1.0</mu>
     <mu prod="ttH" decay="bb">1.0</mu>
     <mu prod="ttH" decay="tautau">1.0</mu>
  
     <!-- optionnal: if not given, no decay into invisible -->
     <redxsBR prod="ZH" decay="invisible">0.0</redxsBR>
     <redxsBR prod="VBF" decay="invisible">0.0</redxsBR>
   </signalstrengths>
 </lilithinput>
\end{verbatim}
\begin{itemize}
	\item {\tt <lilithinput>} is the root tag of the $\XML$ file, it defines a $\Lilith$ input file.
	\item The {\tt <signalstrengths>} tag indicates that the user input is given in terms of signal strengths.
	\item The {\tt <mass>} tag defines the Higgs boson mass at which the likelihood should be computed. It should be in the $[123,128]$~GeV range. This information is not used in the calculations with the current experimental input, where results are only given for a fixed Higgs mass.
	\item The signal strengths themselves are defined in {\tt <mu>} tags. Two mandatory arguments should be given:
\begin{itemize}
	\item The {\tt prod} attribute can be {\tt ggH}, {\tt WH}, {\tt ZH}, {\tt VBF}, {\tt ttH}. 
	For convenience, multi-particle attributes have been defined. They are listed in Table~\ref{tab:mumultidecayprod}.
	\item The {\tt decay} attribute can be {\tt gammagamma}, {\tt Zgamma}, {\tt WW}, {\tt ZZ}, {\tt bb}, 
	{\tt cc}, {\tt tautau}. As for the {\tt prod} attribute, multi-particle labels have been defined and are listed in Table~\ref{tab:mumultidecayprod}.
\end{itemize}
Note that every {\tt <mu>} tag can be omitted; in such a case the SM value will be assumed. A warning will furthermore be issued in case of missing $\mu(X,Y)$ for $Y\in(\gamma\gamma, ZZ^*, WW^*, b\bar b, \tau\tau)$ (after resolving multi-particle labels).
	\item Finally, there is the possibility to specify an invisible branching ratio in the {\tt <redxsBR>} tag. This is defined as 
	\begin{equation}
		{\rm redxsBR}(X,\invisible) =
        \frac{\sigma(X)}{\sigma^{\rm SM}(X)}\BR_{\invisible} \,.
	\end{equation}
As the branching fraction of the SM Higgs boson into invisible particles is very small ($\BR^{\rm SM}(H\to4\nu) = 0.11\%$ at $m_H = 125$~GeV~\cite{Heinemeyer:2013tqa}) and cannot be probed at the LHC, one usually does not express the results of invisible Higgs searches in terms of signal strengths.
Invisible decays of the Higgs boson are currently constrained in association with two jets from VBF, and in association with a $Z$ boson from ZH production, see Table~\ref{tab:defaultdatabase}.
\end{itemize}

\begin{table}
	\begin{center}
	\begin{tabular}{c|c|c}
	 \multicolumn{2}{c|}{Attribute} & shortcut for... \\
	\hline
	\multirow{2}{*}{{\tt prod}} & {\tt "VVH"} & {\tt "VBF"}, {\tt "WH"}, {\tt "ZH"} \\
	& {\tt "VH"} & {\tt "WH"}, {\tt "ZH"} \\
	\hline
	\multirow{4}{*}{{\tt decay}} & {\tt "VV"} & {\tt "ZZ"}, {\tt "WW"} \\
    & {\tt "ff"} & {\tt "cc"}, {\tt "bb"}, {\tt "tautau"} \\
    & {\tt "uu"} & {\tt "cc"} \\
    & {\tt "dd"} & {\tt "bb"}, {\tt "tautau"} \\
	\end{tabular}
	\end{center}
	\caption{Possible multi-particle attributes for the tag {\tt <mu>} in the signal strengths mode.}
	\label{tab:mumultidecayprod}
\end{table}

Note that the signal strengths for several Higgs states contributing to the signal can be defined by specifying an arbitrary number of {\tt <signalstrengths> ..\ </signalstrenths>} tags in the input. After reading the input, the signal strengths from each individual state contributing to the signal will be stored in the attribute {\bf user\_mu}, and the sum of the signal from the different particules (signal strength per signal strength) will be stored in the attribute {\bf user\_mu\_tot} (for more details, see Section~\ref{sec:api}). We neglect possible interferences between the different states.
It can be useful to provide an identifier for each particle. This can be achieved with a {\tt part} attribute to the {\tt <signalstrengths>} tag.
An example of user input in terms of signal strengths is stored in {\tt userinput/example\_mu.xml} for the case of a single Higgs boson contributing to the signal, and in {\tt userinput/example\_mu\_multiH.xml} for the case of two or more particles.

\subsubsection{{\tt XML} format for reduced couplings}
\label{sec:reducedcouplingmode}

New physics can be parametrized in terms of scaling factors that can be identified as (or derived from) reduced couplings, as was presented in Section~\ref{sec:npparam}. In this section we present the user input in terms of reduced couplings.
Before turning to the format of the user input, that also has {\tt XML} syntax, we comment on the computation of couplings and of signal strengths.
First of all, as we have seen in Section~\ref{sec:redc}, predictions for the Higgs boson can be obtained from the reduced couplings $C_W$, $C_Z$, $C_t$, $C_b$, $C_c$, and $C_\tau$ appearing in Eq.~\eqref{eq:lagrangian}.
Scaling factors for VBF production and loop-induced processes are function of the $C_i$ and can be expressed in as Eq.~\eqref{eq:ggHVBF}--\eqref{eq:Cloop}.
In the following, we will consider two possible cases: that these scaling factors are obtained from leading-order calculations ({\it i.e.} tree-level results for VBF and one-loop analytical expressions for other processes), or including NLO QCD corrections. The former case will be denoted as {\tt LO}, the latter one as {\tt BEST-QCD}.
We comment on the computations currently implemented in \lilith:
\begin{description}
\item[VBF] \hfill \\
The contribution from the $W$ boson, the one from the $Z$ boson, and the interference between them have been obtained from {\tt VBFNLO-2.6.3}~\cite{Arnold:2008rz} for Higgs masses in the $[123,128]$~GeV range with (for {\tt BEST-QCD} mode) and without (for {\tt LO} mode) NLO QCD corrections at the LHC 8~TeV, using the MSTW2008 parton distribution functions~\cite{Martin:2009iq}. The results for $\sigma^{\rm SM}_{WW}({\rm VBF})$, $\sigma^{\rm SM}_{ZZ}({\rm VBF})$ and $\sigma^{\rm SM}_{WZ}({\rm VBF})$ as a function of the Higgs mass were stored in text files shipped with \lilith\ and read internally when using {\bf computereducedcouplings}() (see Section~\ref{sec:api}).
\item[ggH] \hfill \\
The contributions from the three heaviest quarks ($t$, $b$, $c$) to the SM cross section are taken into account.
In the {\tt LO} mode, we use analytical expressions~\cite{Djouadi:2005gi}. In the {\tt BEST-QCD} mode, those have been generated in the $[123,128]$~GeV range with {\tt HIGLU}~\cite{Spira:1995mt} at the LHC 8~TeV with the MSTW2008 parton distribution functions.
\item[$H \to gg,\gamma\gamma,Z\gamma$] \hfill \\
The relevant SM partial widths of these processes (taking into account particles listed in Eq.~\eqref{eq:Cloop}) are obtained from  analytical expressions~\cite{Djouadi:2005gi} in the {\tt LO} mode.
In the {\tt BEST-QCD} mode, those have 
been generated in the $[123,128]$~GeV range with {\tt HDECAY}~\cite{Djouadi:1997yw} including the available QCD corrections.
\end{description}

However, the Lagrangian defined in Eq.~\eqref{eq:lagrangian} does not exhaust the possibilities for new physics affecting the properties of the Higgs processes.
One particularlity interesting case is that BSM particles enter the loop-induced processes, such as $gg \to H$ and $H \to \gamma\gamma$. An explicit example will be given in Section~\ref{sec:stau}.
To account for these cases, we allow direct definition of scaling factors for the four main loop-induced processes ($gg \to H$ and $H \to gg,\gamma\gamma,Z\gamma$), {\it i.e.} direct definition of $C_{\rm ggH}$ and $C_{gg,\gamma\gamma,Z\gamma}$.
If some or all of the scaling factors are missing from the input, they will be computed internally using Eq.~\eqref{eq:ggHVBF}--\eqref{eq:Cloop}, {\it i.e.} assuming that only SM particles are involved.
Finally, note that we use the SM branching ratios provided by the LHC HXSWG~\cite{Heinemeyer:2013tqa}, at the Higgs mass given in the user input, when computing the signal strengths (see Eq.~\eqref{eq:signalstr4}).

The user input file for the $\Lilith$ reduced couplings mode has the following structure.

\begin{verbatim}
 <lilithinput>
   <reducedcouplings>
     <mass>125</mass>

     <C to="ZZ">1.0</C>
     <C to="WW">1.0</C>
     <C to="tt">1.0</C>
     <C to="cc">1.0</C>
     <C to="bb">1.0</C>
     <C to="tautau">1.0</C>

     <C to="gammagamma">1.0</C>
     <C to="Zgamma">1.0</C>
     <C to="gg">1.0</C>
     <C to="VBF">1.0</C>

     <precision>BEST-QCD</precision>

     <extraBR>
       <BR to="invisible">0.0</BR>
       <BR to="undetected">0.0</BR>
     </extraBR>
   </reducedcouplings>
 </lilithinput>
\end{verbatim}
\begin{itemize}
	
	\item {\tt <lilithinput>} is the root tag of the $\XML$ file, it defines a $\Lilith$ input file.
	
	\item The {\tt <reducedcouplings>} tag is specific to the reduced couplings mode. This is where the 
	reduced couplings are specified. The correspondence between the $\XML$ notation and Eq.~\eqref{eq:lagrangian} 
	is given in Table~\ref{tab:correspondence}. Note the possibility to define common couplings for the up-type 
	fermions, down-type fermions, all fermions, and electroweak gauge bosons.

	\item The tag {\tt <mass>} defines the Higgs boson mass at which the likelihood should be computed.
	 The allowed range is $[123,128]$~GeV. This affects the computation of the SM branching ratios and partial
	  cross sections and widths as explained above. If it is not given, a Higgs mass of 
	  $125$~GeV is assumed.
	
	\begin{table}
	\begin{center}
	\begin{tabular}{c|c|c|c|c|c|c}
	{\tt to} attribute & {\tt "tt"} & {\tt "bb"} & {\tt "cc"} & {\tt "tautau"} & {\tt "WW"} & {\tt "ZZ"} \\
    \hline
    corresponds to & $C_t$ & $C_b$ & $C_c$ & $C_\tau$ & $C_W$ & $C_Z$ \\
    \end{tabular}

    \vspace{5mm}

	\begin{tabular}{c|c|c|c|c|c}
	\multirow{2}{*}{{\tt to} attribute} & \multirow{2}{*}{{\tt "gammagamma"}} &
    \multirow{2}{*}{{\tt "Zgamma"}} & \multicolumn{2}{|c|}{{\tt "gg"}} & \multirow{2}{*}{{\tt "VBF"}} \\
    & & & {\tt for="prod"} & {\tt for="decay"} & \\
    \hline
    corresponds to & $C_{\gamma\gamma}$ & $C_{Z\gamma}$ & $C_{\rm ggH}$ & $C_{gg}$ & $C_{\rm VBF}$ \\
    \end{tabular}

    \vspace{5mm}

	\begin{tabular}{c|c}
	{\tt to} attribute & shortcut for... \\
	\hline
	{\tt "VV"} & {\tt "WW"}, {\tt "ZZ"} \\
	{\tt "ff"} & {\tt "tt"}, {\tt "cc"}, {\tt "bb"}, {\tt "tautau"} \\
	{\tt "uu"} & {\tt "tt"}, {\tt "cc"} \\
	{\tt "dd"} & {\tt "bb"}, {\tt "tautau"} \\
	\end{tabular}
	\end{center}
	\caption{Upper and middle tables: correspondence between the $\XML$ notation and that of Eq.~\eqref{eq:lagrangian}. Bottom table: common 
	reduced couplings definitions.}
	\label{tab:correspondence}
	\end{table}
	
	\item Regarding the effective coupling to a pair of gluons, NLO corrections affect gluon fusion (ggH) and the decay into two gluons ($H\to gg$) in a different way. Therefore, scaling factors $C_{\rm ggH}$ and $C_{gg}$ can be specified separately as
	\begin{verbatim}
 <C to="gg" for="prod">1.0</C>
 <C to="gg" for="decay">1.05</C>
	\end{verbatim}
	If {\tt for="all"} is specified, the same coupling is assigned to the production and decay modes. 
	This is the default behavior if the {\tt for} attribute is missing.
	
	\item CP violation was presented in Section~\ref{sec:npparamCPV}. In the {\tt LO} mode, the fermionic couplings $C_t, C_b, C_c, C_\tau$ can be given a real and an imaginary component.
	For the top quark for instance, this can be specified as
	\begin{verbatim}
 <C to="tt" part="re">0.8</C>
 <C to="tt" part="im">0.2</C>
	\end{verbatim}
	If {\tt part="re|im"} is not specified, the coupling is assumed to be purely real. In the {\tt BEST-QCD} mode, 
	only the real part of the coupling is taken into account.
	
	\item The {\tt <precision>} tag contains either {\tt BEST-QCD} or {\tt LO}. If not specified, or wrongly 
	spelled, the {\tt BEST-QCD} mode is the default mode.
	
	\item The {\tt <extraBR>} tag contains the declaration of the invisible or undetected branching ratios (see Section~\ref{sec:npparam}).
    		
\end{itemize}
As in the case of input in terms of signal strengths, several tags {\tt <reducedcouplings> ..\ </reducedcouplings>} can be defined, corresponding to the case where several Higgs states contribute to the observed signal around 125~GeV.  These particles can also be given a name with a {\tt part} attribute to the {\tt <reducedcouplings>} tag.
An example of user input in terms of reduced couplings is stored in {\tt userinput/example\_couplings.xml} for the case of a single Higgs boson contributing to the signal, and in {\tt userinput/example\_couplings\_multiH.xml} for the case of two or more Higgs states.

\subsection{Output}
\label{sec:output}

When using the API, all relevant information can be accessed from the public attributes of the class {\tt Lilith} presented in Section~\ref{sec:api}. The user can manipulate and store this information in any way he/she wants.
However, having standardized output formats is important in order to interface {\tt Lilith} with other programs.
We provide three output methods: {\bf writecouplings}(), {\bf writesignalstrengths}(), and {\bf writeresults}() (for details on how to call these methods, see Section~\ref{sec:api}).
The first two methods write the reduced coupling information (if available) and the signal strength information, respectively, in a file. The formats specified in Section~\ref{sec:signalstrengthsmode} and \ref{sec:reducedcouplingmode} above are used, such that these output files can also be used as input to a subsequent call to {\tt Lilith}. In the command-line interface, these two methods can be called with the options {\tt -c} or {\tt --couplings}, and {\tt -m} or {\tt --mu}, respectively (for more details, see Section~\ref{sec:cli}).

The last method, {\bf writeresults}(), is used to store the results after evaluation of the likelihood.
Depending on the second argument, the output will be written in {\tt XML} format (if {\it slha=False}) or in {\tt SLHA}-like format (if {\it slha=True}). By default the output file is in {\tt XML} format for consistency with what is used otherwise in {\tt Lilith}; an {\tt SLHA}-like output was added as it is widely used in BSM phenomenology.
This method can be called with the option {\tt -r} or {\tt --results} in the command-line interface.

We start with the description of the {\tt XML} format. Its root tag is {\tt <lilithresults>}.
The results from each analysis are given within an {\tt <analysis>} tag having two attributes: {\tt experiment} and {\tt source}, corresponding to the value of the tags {\tt <experiment>} and {\tt <source>} read from the experimental input, if present; otherwise the attribute is given an empty value {\tt ""}. Each {\tt <analysis>} tag contains a {\tt <l>} tag, whose value is $-2\log L$ from this experimental result alone, and an {\tt <expmu>} tag that follows the syntax of the experimental input (see Section~\ref{sec:expinput}) except that it only provides information on the efficiencies (in {\tt <eff>} tags).

Finally, {\tt <lilithresults>} contains tags summarizing the information: {\tt <ltot>}, whose value is the sum of the $-2\log L$ values from all experimental results, and {\tt <exp\_ndf>}, the number of measurements. We also provide the version of {\tt Lilith} in a {\tt <lilithversion>} tag and the version of the experimental database of results in a {\tt <dbversion>} tag. To summarize, we give a complete example of output considering only two experimental results.
\begin{verbatim}
 <lilithresults>
   <lilithversion>1.1</lilithversion>
   <dbversion>15.02</dbversion>

   <analysis experiment="ATLAS" source="HIGG-2013-08">
     <expmu decay="gammagamma" dim="2" type="n">
       <eff axis="x" prod="ggH">1.0</eff>
       <eff axis="y" prod="VBF">1.0</eff>
      </expmu>
      <l>0.9</l>
   </analysis>

   <analysis experiment="CMS" source="CMS-HIG-13-030">
    <expmu decay="invisible" dim="1" type="f">
      <eff prod="VBF">0.79</eff>
      <eff prod="ZH">0.21</eff>
    </expmu>
    <l>1.1</l>
  </analysis>

  <ltot>2.0</ltot>
  <exp_ndf>3</exp_ndf>
 </lilithresults>
\end{verbatim}

The output in {\tt SLHA}-like format is much more basic. Currently, we only provide the total $-2\log L$ value as well as the number of measurements. Taking the same example as above, the output would read:
\begin{verbatim}
BLOCK         LilithResults
  0           2.0               # -2*LogL
  1           3                 # exp_ndf
\end{verbatim}
The {\tt SLHA} structure, where each element of a block is identified by a set of numbers, makes it more complicated to integrate all necessary information in a well-structured way. For this reason we recommend the {\tt XML} format. Extensions of the {\tt SLHA} format will be considered in the future depending on the needs of the users of {\tt Lilith}.

\section{Validation}
\label{sec:validation}

Having explained how to use \lilith\ in the previous section, we now turn to the validation of the likelihood derived from the experimental input shipped with the code.
We begin by discussing the validity of the bivariate normal distribution as an approximation to the 2D likelihood functions in the signal strength planes $(\mu(X,Y), \mu(X',Y'))$. The use of this approximation is necessary whenever only contours of constant likelihood are provided instead of the full information.
Several coupling fits from the ATLAS and CMS collaboration are then reproduced. The results from \lilith\ are compared to the official ones to assess the validity of the likelihood used in \lilith.

\subsection{Reconstruction of the experimental likelihoods}
\label{sec:2dmurec}

In a signal strength plane $(\mu(X,Y), \mu(X',Y'))$, an approximation to the likelihood function can be obtained assuming that the measurements follow a bivariate normal distribution, as explained in Section~\ref{sec:statproc}.
Using the 68\%~CL contour provided by the experimental collaboration, we reconstruct the shape of the likelihood and compare the location of the best-fit point as well as the 68\% and 95\%~CL contours with what is provided by ATLAS or CMS.

\begin{figure}
\centering \includegraphics[scale=0.38] {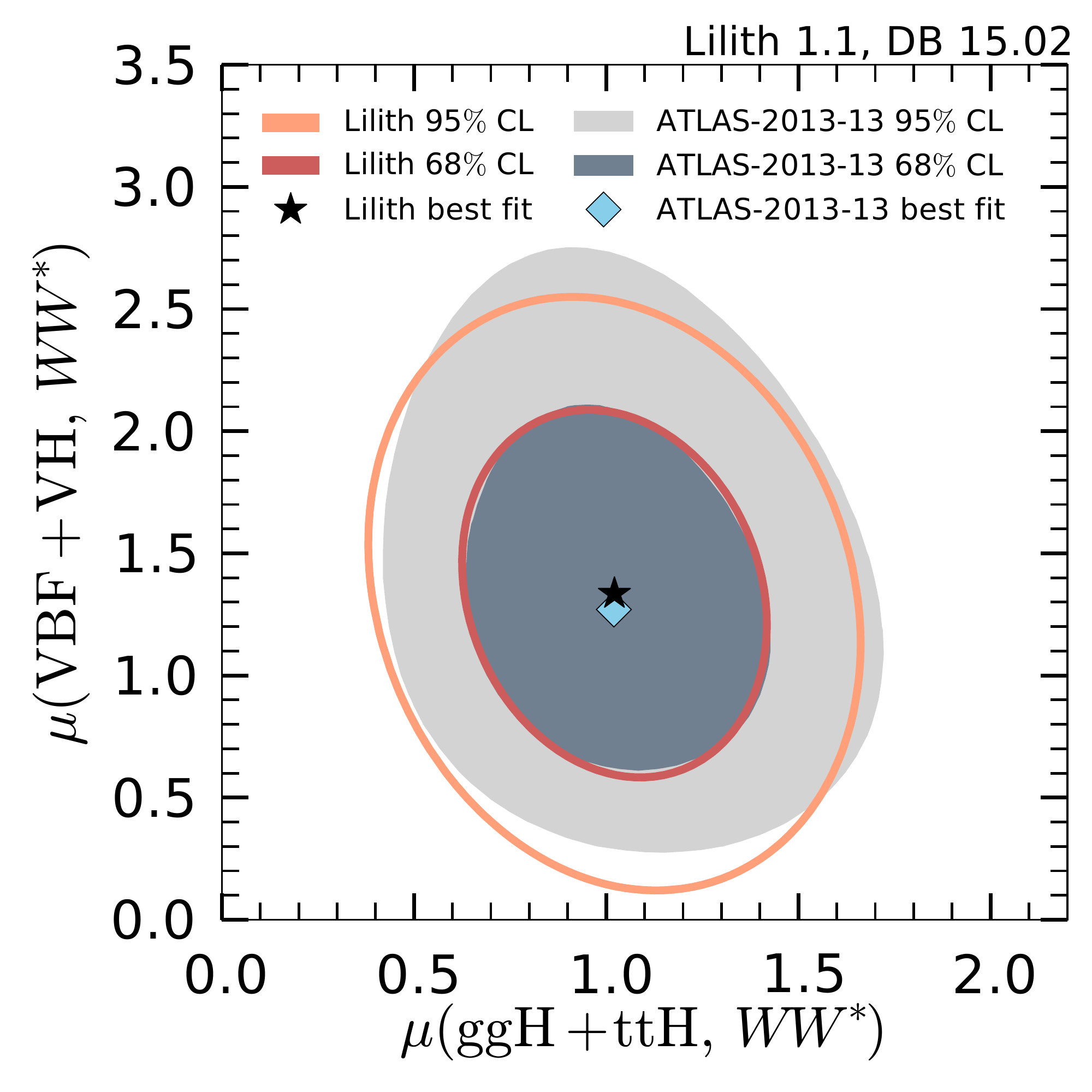}  \includegraphics[scale=0.38] {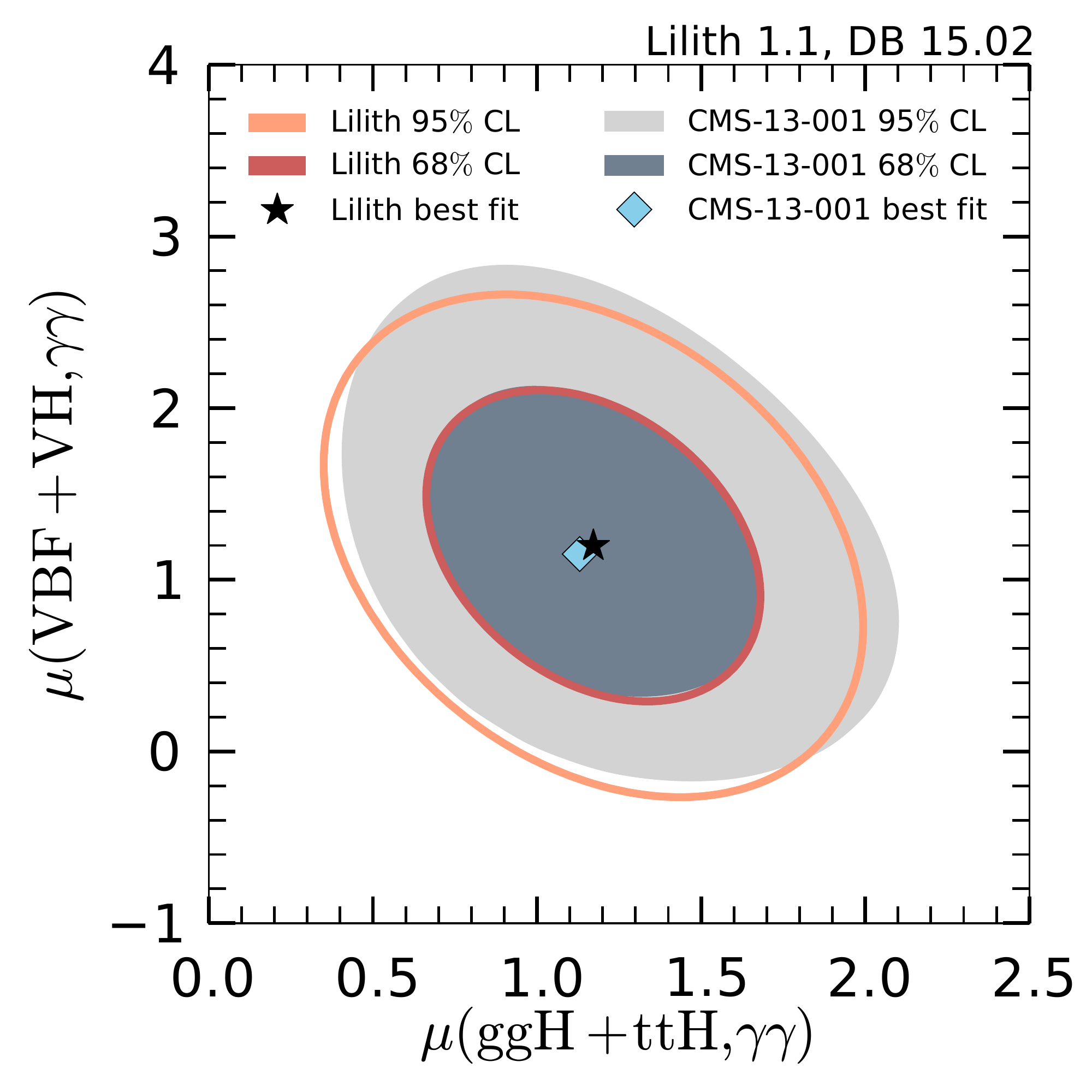}
\caption{Reconstruction of the experimental likelihood from a bivariate normal approximation for the ATLAS $WW^*$ search~\cite{ATLAS:2014aga} (left)
 CMS $\gamma\gamma$ search~\cite{Khachatryan:2014ira} (right). The filled dark and light grey contours show the 68\% and 95\%~CL experimental contours while the red and orange solid lines show the reconstructed likelihood contours. The blue diamond and the black star indicates the experimental and reconstructed best-fit points, respectively.}
\label{WWgammagammacomparison}
\end{figure}

Two examples are shown in Fig.~\ref{WWgammagammacomparison}: the reconstruction of the likelihood for the ATLAS $WW^*$~\cite{ATLAS:2014aga} and the CMS $\gamma\gamma$~\cite{Khachatryan:2014ira} final states.\footnote{Additional validation materials for the reconstruction of the likelihood can be found at~\cite{lilith}.} In both cases we observe an excellent agreement between the reconstructed likelihood and the official result. The 68\%~CL regions are perfectly reproduced and the reconstructed best-fit points are very close to the experimental ones. The extrapolation towards the 95\%~CL regions also shows very good agreement. We find equally good agreements with all other decay modes (with the exception of $H \to ZZ^*$), and we conclude that the Gaussian distribution is a very good approximation to the true distribution.

The largest deviations from the normal approximation are expected to occur for final states with low statistics since the counting of the events, that follows the Poisson distribution, has not yet entered the Gaussian regime. In particular, this is the case for the $ZZ^*$ channel. In Fig.~\ref{ZZcomparison}, we show the comparison between the \lilith\ reconstructed likelihood in the $ZZ^*$ final state and the corresponding ATLAS and CMS ones.

\begin{figure}
\centering  \includegraphics[scale=0.38]{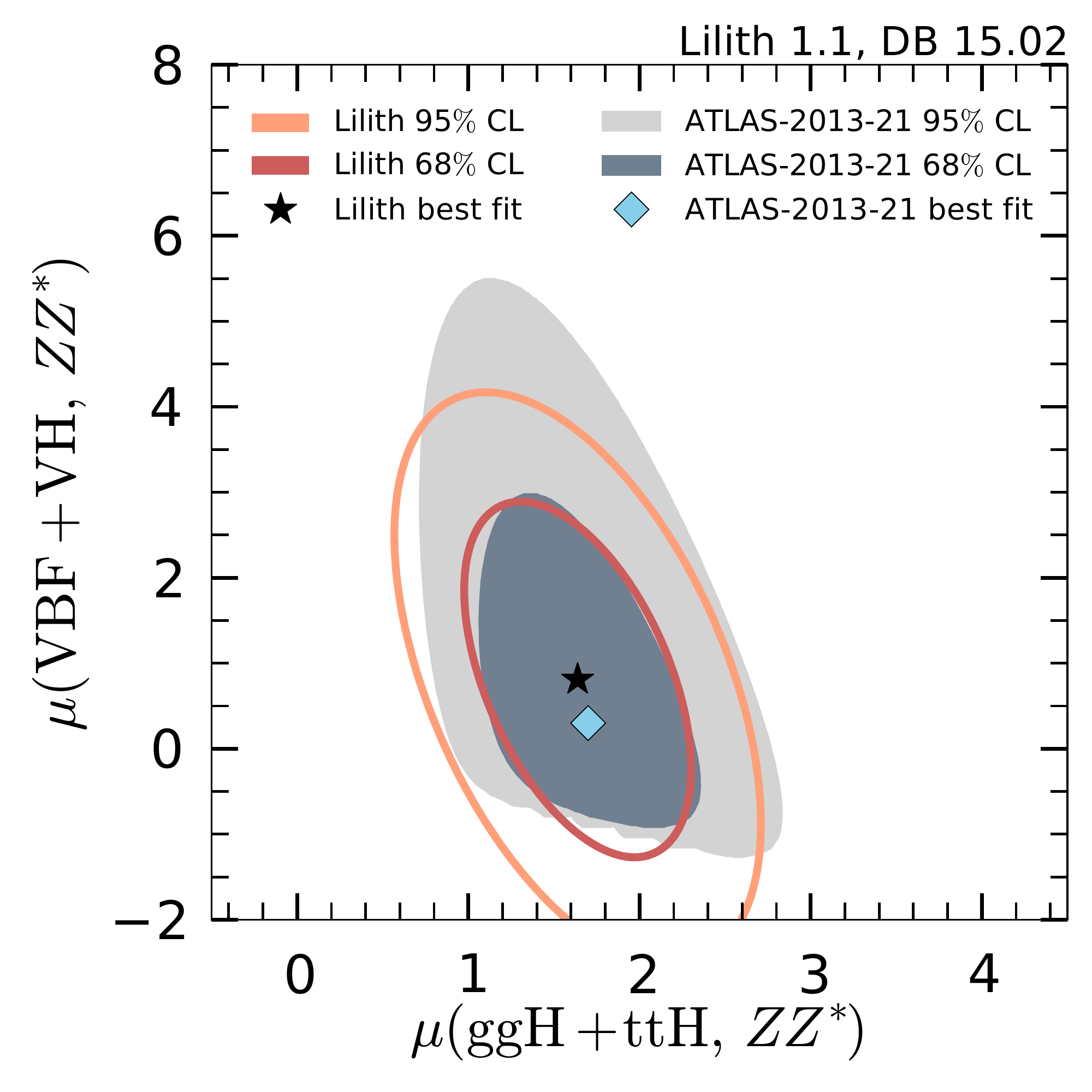}  \hspace{-0.3cm} \includegraphics[scale=0.38] {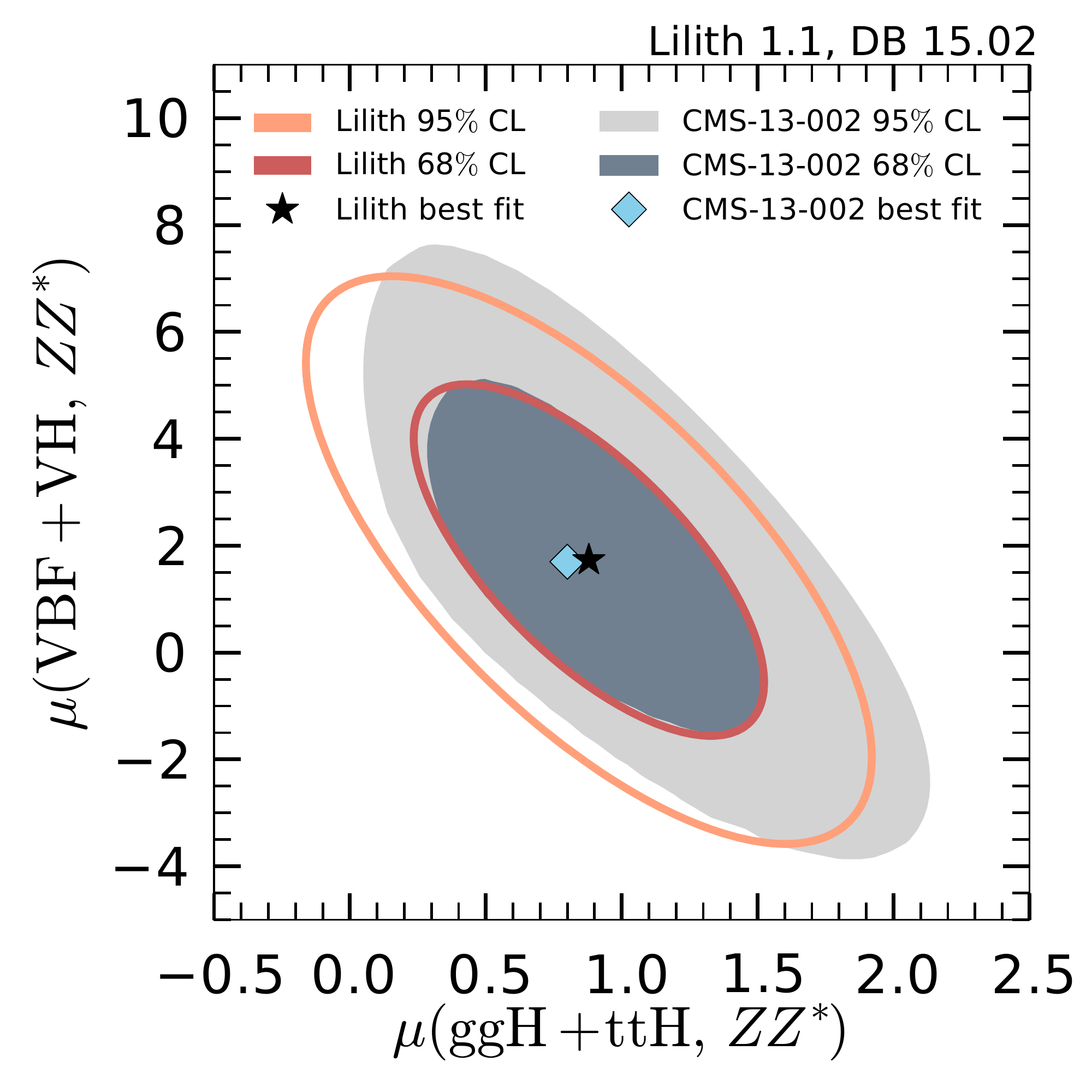}
\caption{Reconstruction of the experimental likelihood from a bivariate normal approximation for the ATLAS~\cite{Aad:2014eva} (left) and CMS~\cite{Chatrchyan:2013mxa} (right) $H\to ZZ^*$ searches. The filled dark and light grey contours show the 68\% and 95\%~CL experimental contours while the red and orange solid lines show the reconstructed likelihood contours. The blue diamond and the black star indicates the experimental and reconstructed best-fit points, respectively.}
\label{ZZcomparison}
\end{figure}

As can be seen, the deviation of the ATLAS likelihood from the bivariate normal approximation can be substantial. In the positive region of the plane (the one that is relevant), the approximation holds well near the best-fit point. However, going away from it the reconstructed shape fails to reproduce the ATLAS 95\%~CL contour at large $\mu(\VBFVH,ZZ^*)$. Due to non-Gaussian effects, the reconstructed best-fit point is quite distant to the experimental one.
For the CMS case, the approximation holds to a better approximation. The reconstructed best-fit point is very close to the experimental one and the shape of the 95\%~CL contour is very well reproduced although a small shift in the $\mu(\ggHttH,ZZ^*)$ direction is observed.
As will be argued in Section~\ref{sec:prospects}, provision of the full likelihood information would 
yield a significant improvement over the normal approximation in such cases.

\subsection{Comparison to Higgs coupling fits from ATLAS and CMS}
\label{subsec:comparison}

In order to validate the approximate Higgs likelihood used in \lilith, we attempt to reproduce coupling fit results from combination notes from ATLAS~\cite{ATLAS-CONF-2014-009} and CMS~\cite{Khachatryan:2014jba}.
Note that while the CMS combination~\cite{Khachatryan:2014jba} makes use of the final Run~I results, a number of analyses considered in the combination of the ATLAS results given in Ref.~\cite{ATLAS-CONF-2014-009} have been updated since then. The final, legacy combination of the Higgs measurements from ATLAS at Run~I has not yet been released. For this reason, the lists of recom\-mended experimental results \texttt{data/CMS-HIG-14-009.list} and \texttt{data/latestCMS.list} are identical, as of February 2015, while \texttt{data/ATLAS-CONF-2014-009.list} and \texttt{latestATLAS.list} differ.

First, results for two benchmark scenarios proposed by the LHC HXSWG in~\cite{LHCHiggsCrossSectionWorkingGroup:2012nn} are presented.
In the first scenario, SM-like tree-level couplings are assumed (\textit{i.e.}, all $C_i=1$ in Eq.~\eqref{eq:lagrangian}) but two scaling factors are introduced: $C_\gamma \equiv C_{\gamma\gamma}$ (scaling $H\to\gamma\gamma$), and $C_g \equiv C_{\rm ggH} = C_{gg}$ (scaling ggH production and $H\to gg$).
In the second benchmark scenario, two reduced couplings are introduced: $C_V \equiv C_W = C_Z$, for the coupling of the Higgs boson to a pair massive vector bosons, and $C_F \equiv C_t = C_b = C_c = C_\tau$, a universal coupling to fermions. In this case, the effective coupling to gluons is simply $C_F$, while $C_{\gamma\gamma}$ is a function of both $C_V$ and $C_F$ that was obtained taking into account QCD corrections (see Section~\ref{sec:reducedcouplingmode}).

\begin{figure}
\centering \includegraphics[scale=0.38]{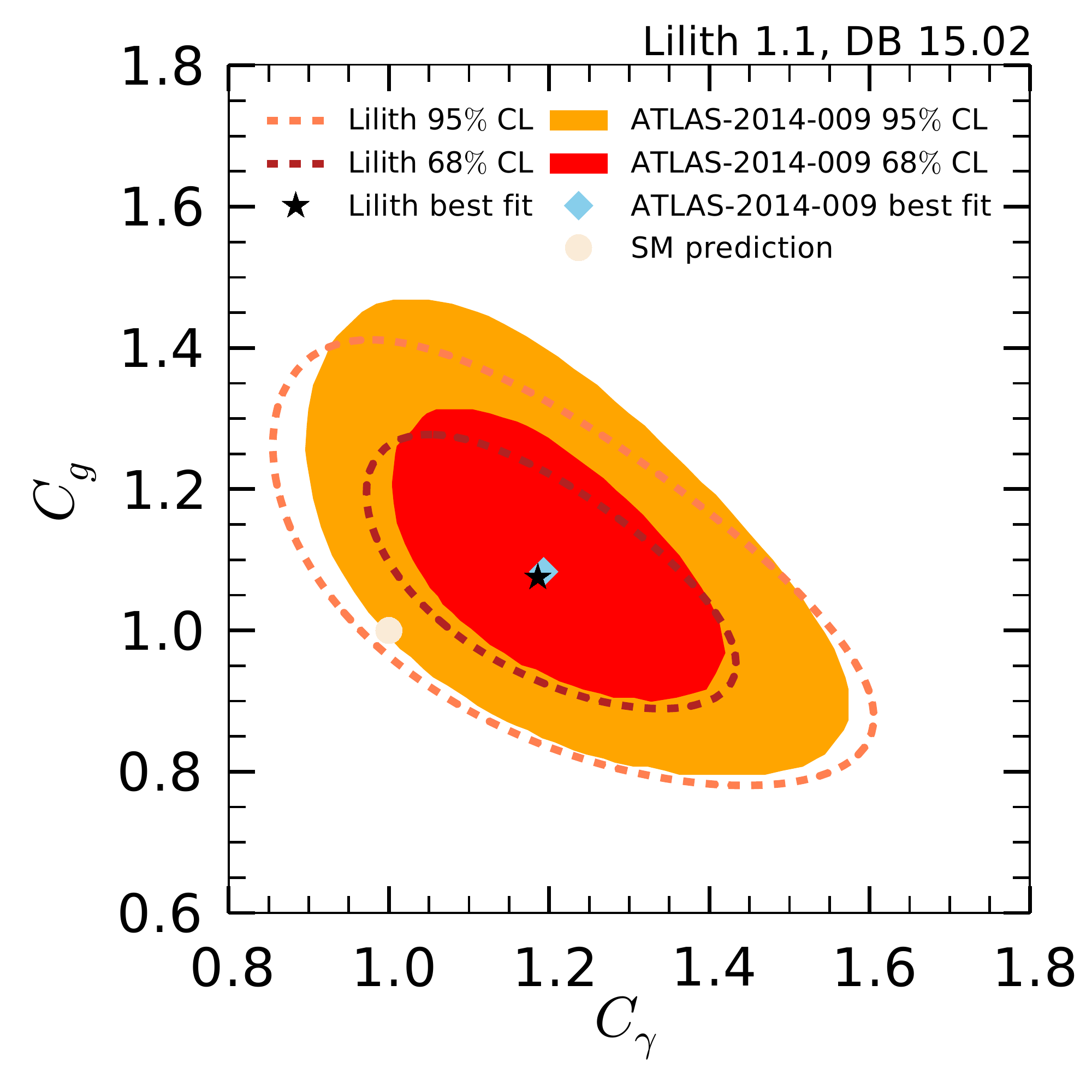} \includegraphics[scale=0.38]{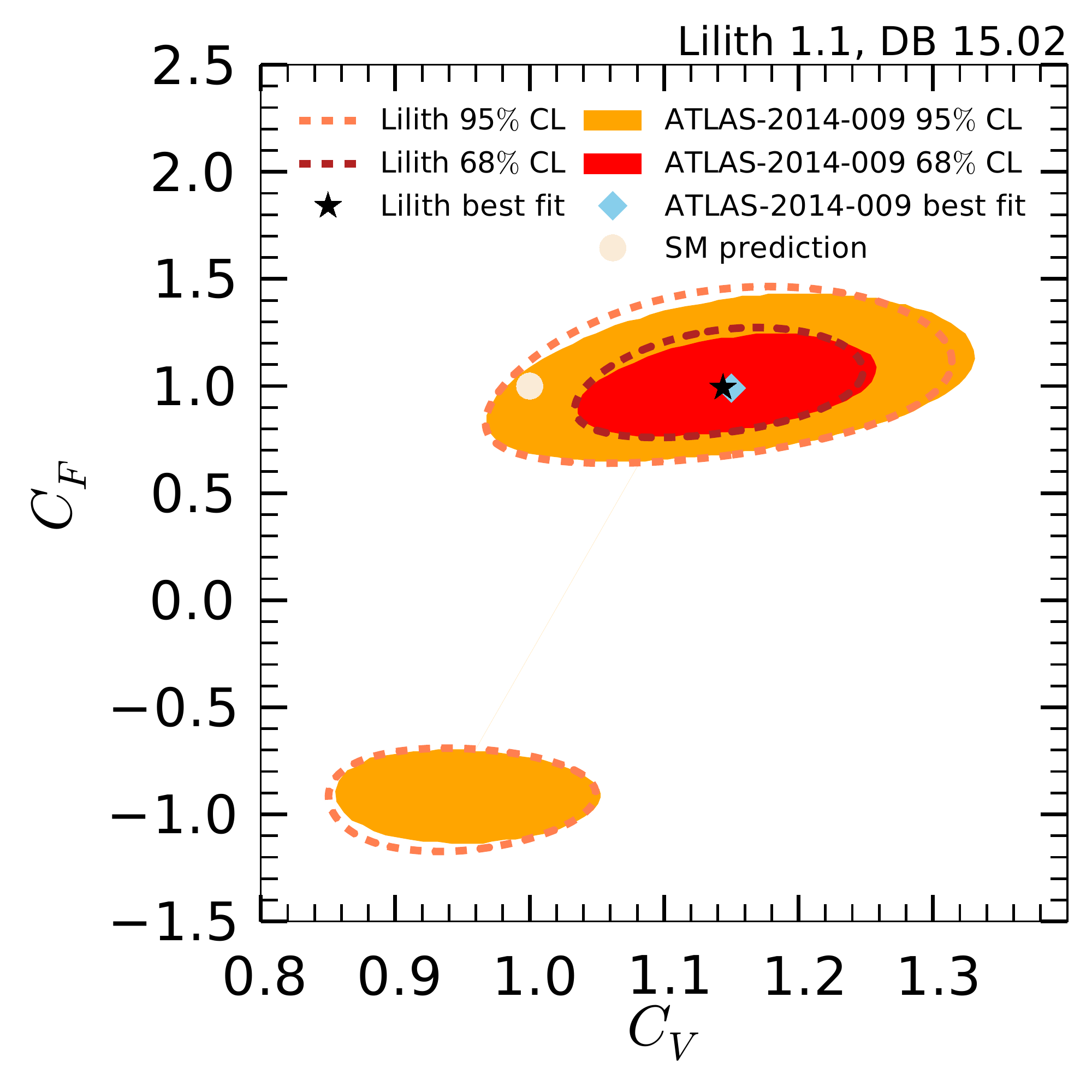}
\caption{$(C_\gamma, C_g)$ (left) and $(C_V,C_F)$ (right) fits using data from the ATLAS combination~\cite{ATLAS-CONF-2014-009}. The red and orange filled surfaces correspond to the 68\% and 95\%~CL regions obtained by the ATLAS collaboration while the corresponding dashed lines show the {\tt Lilith} results. The black star indicates the position of the $\Lilith$ best-fit point, the blue diamond is the ATLAS best-fit point and the white circle shows the SM prediction.}
\label{ATLASfit}
\end{figure}

Let us first discuss the results from ATLAS, obtained using the list of experimental results \texttt{data/ATLAS-CONF-2014-009.list}.
Results of the two fits are presented in Fig.~\ref{ATLASfit}. In both scenarios, very good agreement is observed between the results from ATLAS and the ones obtained with \lilith. Both the reconstructed best-fit point and contours reproduce very well the ATLAS results. The most significant deviation is a slight deformation of the 95\%~CL region in the $(C_\gamma,C_g)$ plane.
The corresponding results for CMS are shown in Fig.~\ref{CMSfit}. CMS results are well reproduced with {\tt Lilith}, even for the contour at 99.7\%~CL. Slight shifts of the best-fit points and minor deformations of the contours are observed. The overall agreement is nevertheless very good.

\begin{figure}
\centering  \includegraphics[scale=0.38]{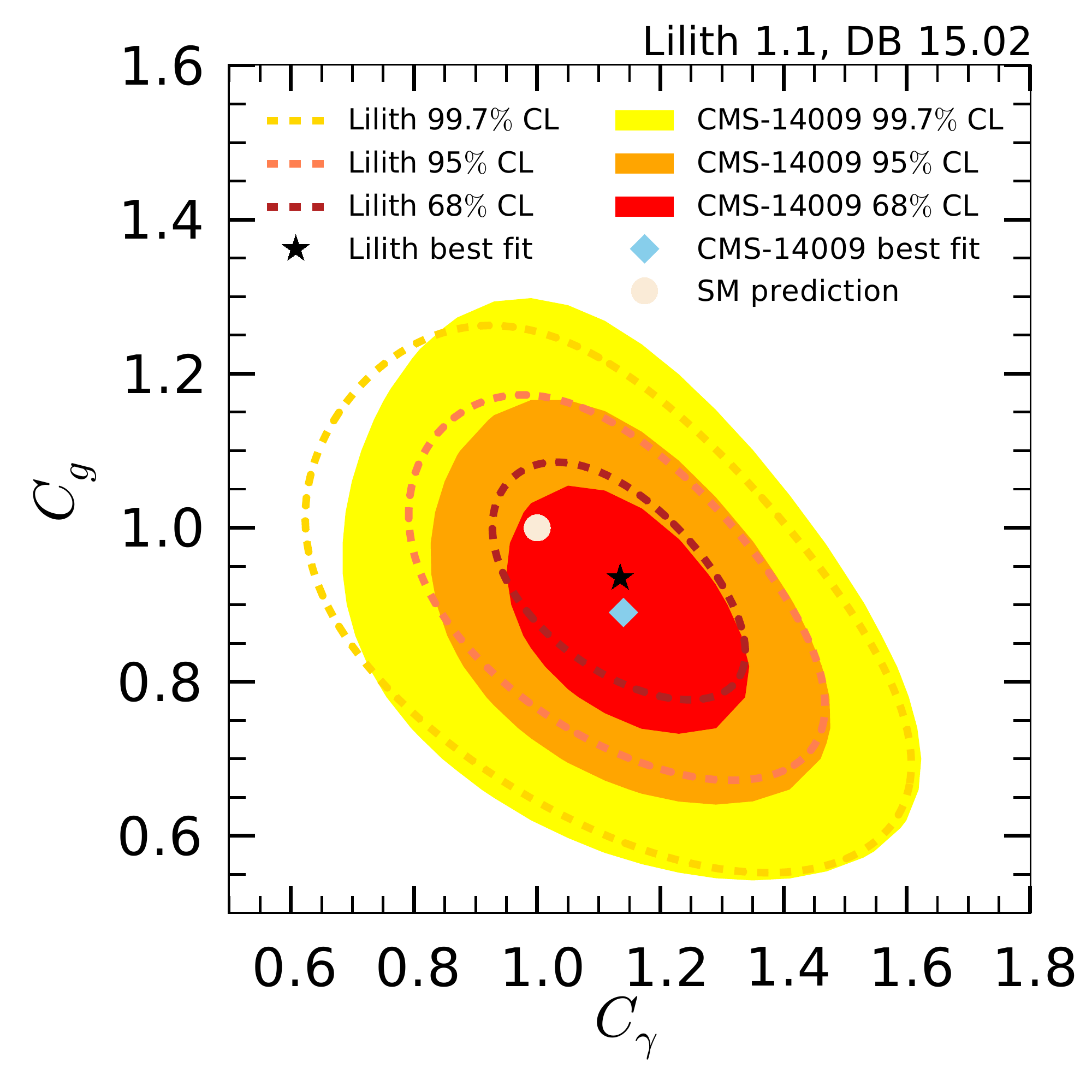} \includegraphics[scale=0.38]{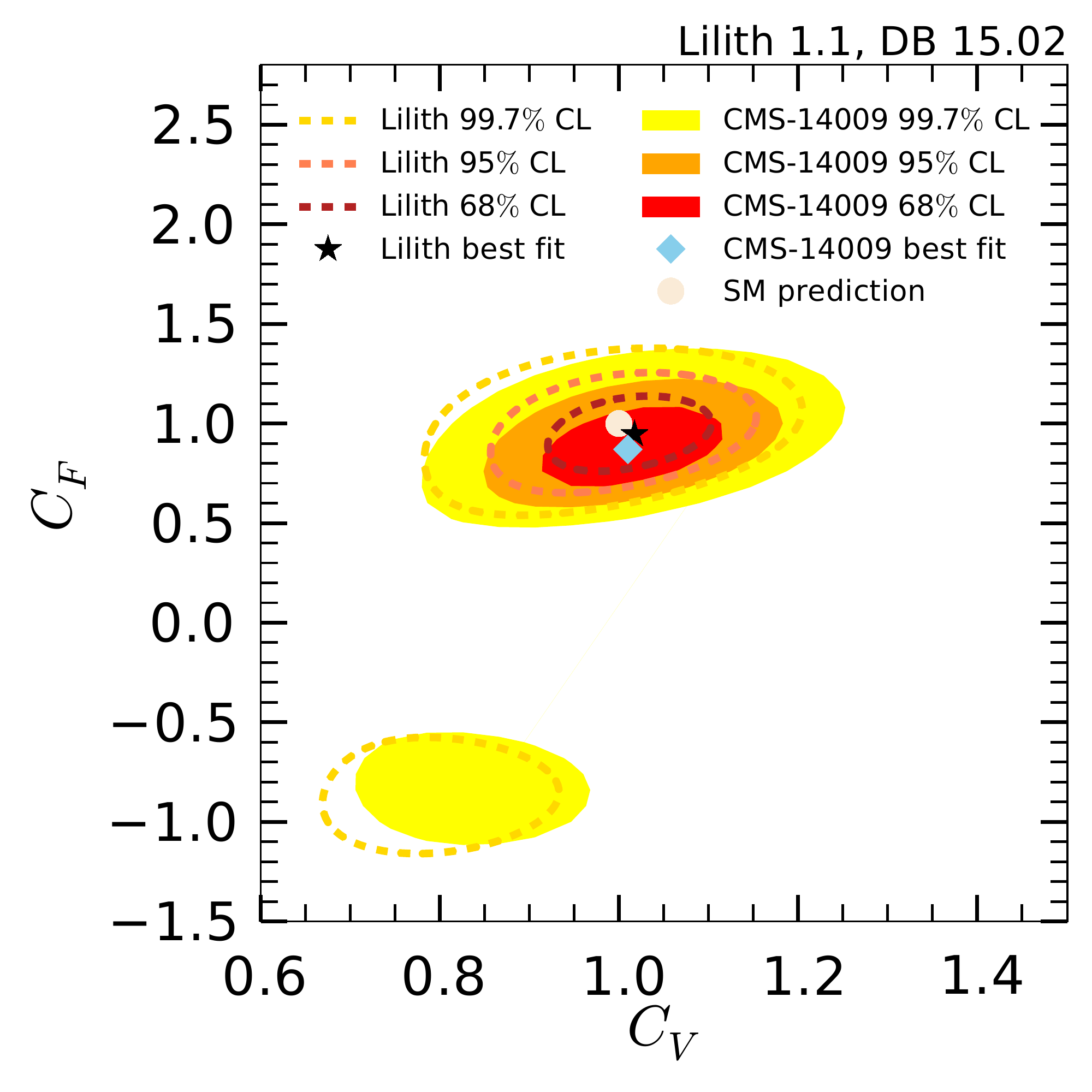} 
\caption{$(C_\gamma, C_g)$ (left) and $(C_V,C_F)$ (right) fits using data from the CMS combination~\cite{Khachatryan:2014jba}. The red, orange and yellow filled surfaces correspond to the 68\%, 95\% and 99.7\%~CL regions obtained by the CMS collaboration while the corresponding dashed lines show the {\tt Lilith} results. The black star indicates the position of the $\Lilith$ best-fit point, the blue diamond is the CMS best-fit point and the white circle shows the SM prediction.}
\label{CMSfit}
\end{figure}

Let us move on to the 3-parameter fit $(C_W, C_Z, C_F)$. As in the $(C_V,C_F)$ benchmark scenario discussed above a universal coupling to fermions is introduced, but instead of a single coupling to vector boson one defines separately the reduced coupling to $W$ bosons, $C_W$, and to $Z$ bosons, $C_Z$
Defining $C_{WZ} \equiv C_W/C_Z$, a direct test of custodial symmetry can be performed using the Higgs measurements alone. The 1-dimensional likelihood profile for $C_{WZ}$ is shown in Fig.~\ref{CWZ_ATLAS_CMS} for both the ATLAS and CMS combination.

\begin{figure}
\centering  \includegraphics[scale=0.38]{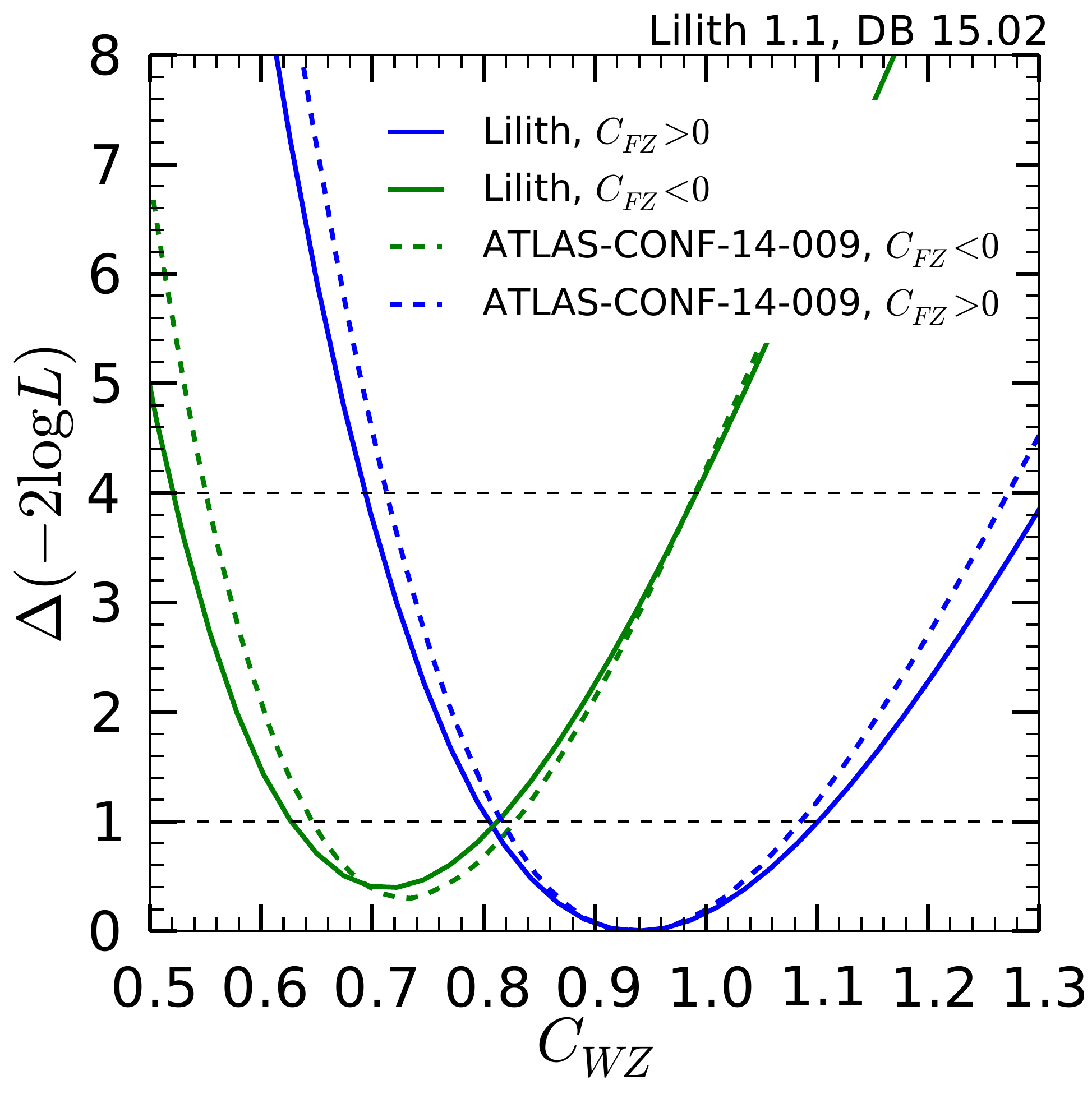} \includegraphics[scale=0.38]{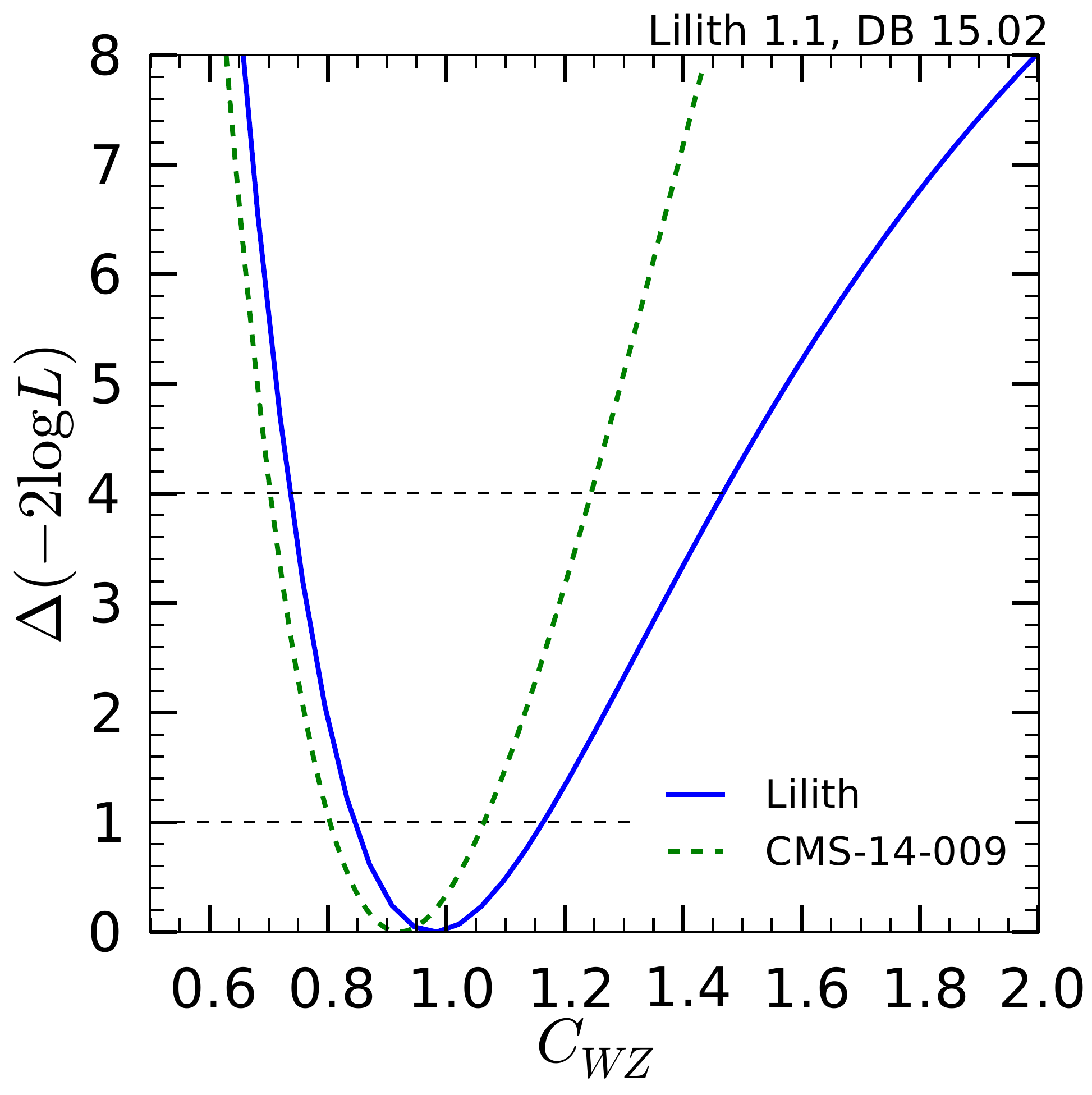}
\caption{1D likelihood profiles of $C_{WZ}\equiv C_W/C_Z$ from a $(C_{W}, C_{Z}, C_F)$ fit to the ATLAS~\cite{ATLAS-CONF-2014-009} (left) and CMS~\cite{Khachatryan:2014jba} (right) data and comparison to the official results. The ATLAS fit considers both signs for the Higgs--fermion--fermion coupling and furthermore defines $C_{FZ}\equiv C_F/C_Z$. The results are given for both signs of $C_{FZ}$.}
\label{CWZ_ATLAS_CMS}
\end{figure}

Although the ATLAS result is almost perfectly reproduced, a significant discrepancy is observed in the case of CMS for $C_{WZ} > 1$.
This does not come as a surprise: several experimental results were considered in the $(\mu({\rm ggH+ttH}, Y), \mu({\rm VBF+VH}, Y))$ plane.
The breaking of $\VBFVH$ into the individual production modes $\VBF$, $\WH$ and $\ZH$ (assumed to be inclusive, see Eq.~\eqref{eq:VHeff}) becomes relevant for $C_W \neq C_Z$.
Moreover, ATLAS results make use of the full numerical likelihood grids that were provided in Refs.~\cite{HEPDATA1,HEPDATA2,HEPDATA3} while the bivariate normal approximation is used in the case of CMS.
Thus, constraints on models in which $C_W \neq C_Z$ should be interpreted with care given the experimental information being used as input to {\tt Lilith}.

\begin{figure}
\centering \includegraphics[scale=0.38]{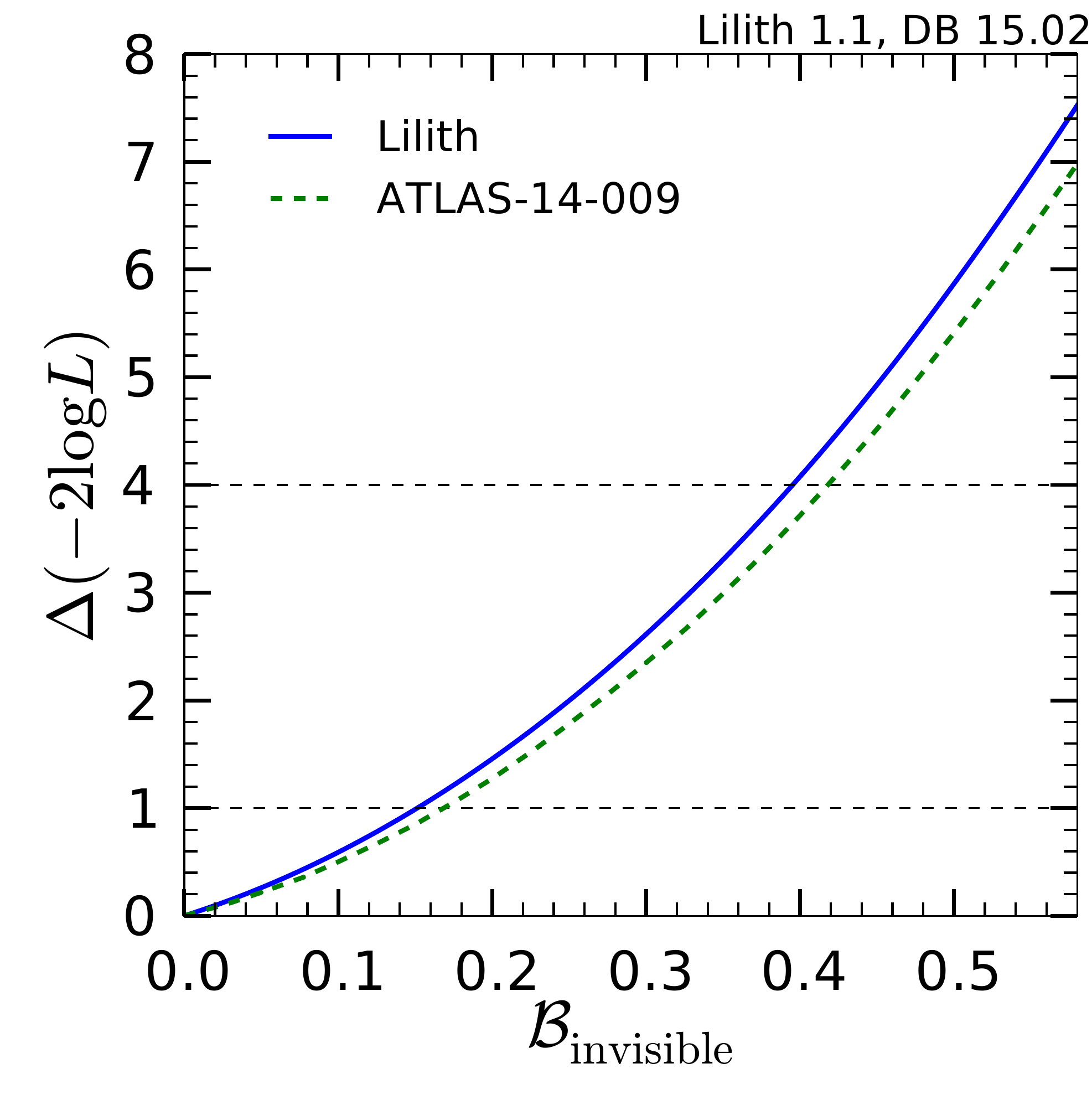} \includegraphics[scale=0.38]{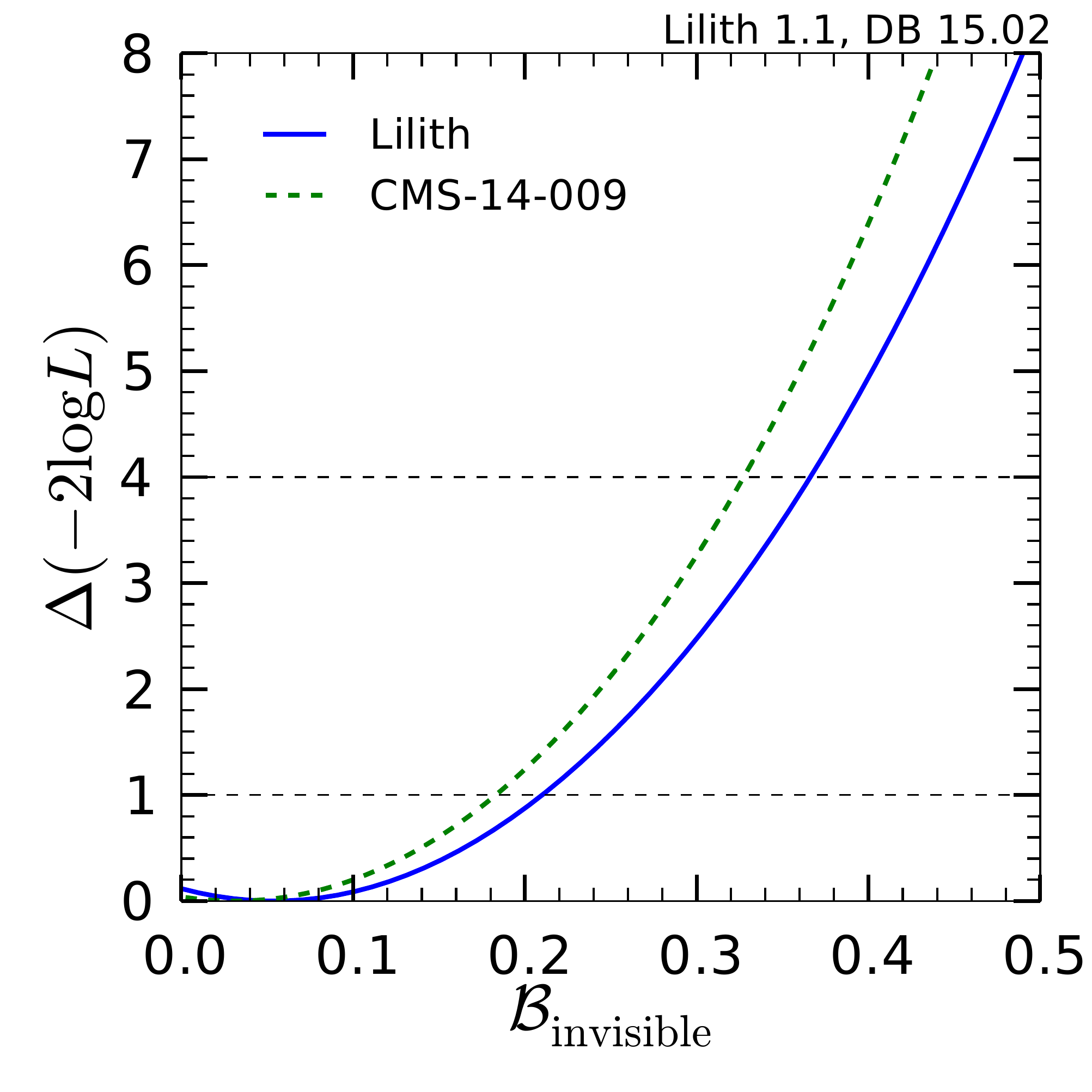}
\caption{1D likelihood profiles of $\mathcal{B}_{\rm invisible}$ from a $(C_\gamma,C_g,\mathcal{B}_{\rm invisible})$ fit and comparison to the ATLAS~\cite{ATLAS-CONF-2014-009} (left) and CMS~\cite{Khachatryan:2014jba} (right) results.}
\label{BRinv}
\end{figure}

Finally, we present the result of a 3-parameter fit $(C_\gamma, C_g, \mathcal{B}_{\rm invisible})$ in terms of the 1D profile likelihood of $\mathcal{B}_{\rm invisible}$ in Fig.~\ref{BRinv}.
A very good agreement is observed in ATLAS, and in CMS for moderate values of $\mathcal{B}_{\rm invisible}$.
As explained in Section~\ref{sec:redc}, the presence of a branching ratio into invisible particles is constrained by direct searches for invisible decays of the Higgs boson, and also by every Higgs search since it modifies the total Higgs width and therefore scales all signal strengths collectively.

\section{Examples of applications}
\label{sec:examples}

Having validated the Higgs likelihood in \lilith\ from results obtained by the ATLAS and CMS collaborations, we now turn to deriving constraints on specific new physics scenarios using the latest LHC results (as of February 2015) present in \texttt{data/latest.list}. The \texttt{Python} routines used to obtain these results are available in the folder \texttt{examples/python} and will be described shortly.

\subsection{Reduced coupling determination}

As a first illustration of the use of \lilith, constraints on the benchmark scenario $(C_V, C_F)$ introduced in Section~\ref{subsec:comparison} are derived.
The right panels of Figs.~\ref{ATLASfit}~and~\ref{CMSfit} show results on this scenario in the 2D plane $(C_V, C_F)$ using only ATLAS or CMS results. Here we combine the latest ATLAS and CMS results and derive 1D profile likelihood constraints on $C_V$ and $C_F$. Results are shown in Fig.~\ref{CVCFglobalfit}.

\begin{figure}
\centering \includegraphics[scale=0.38]{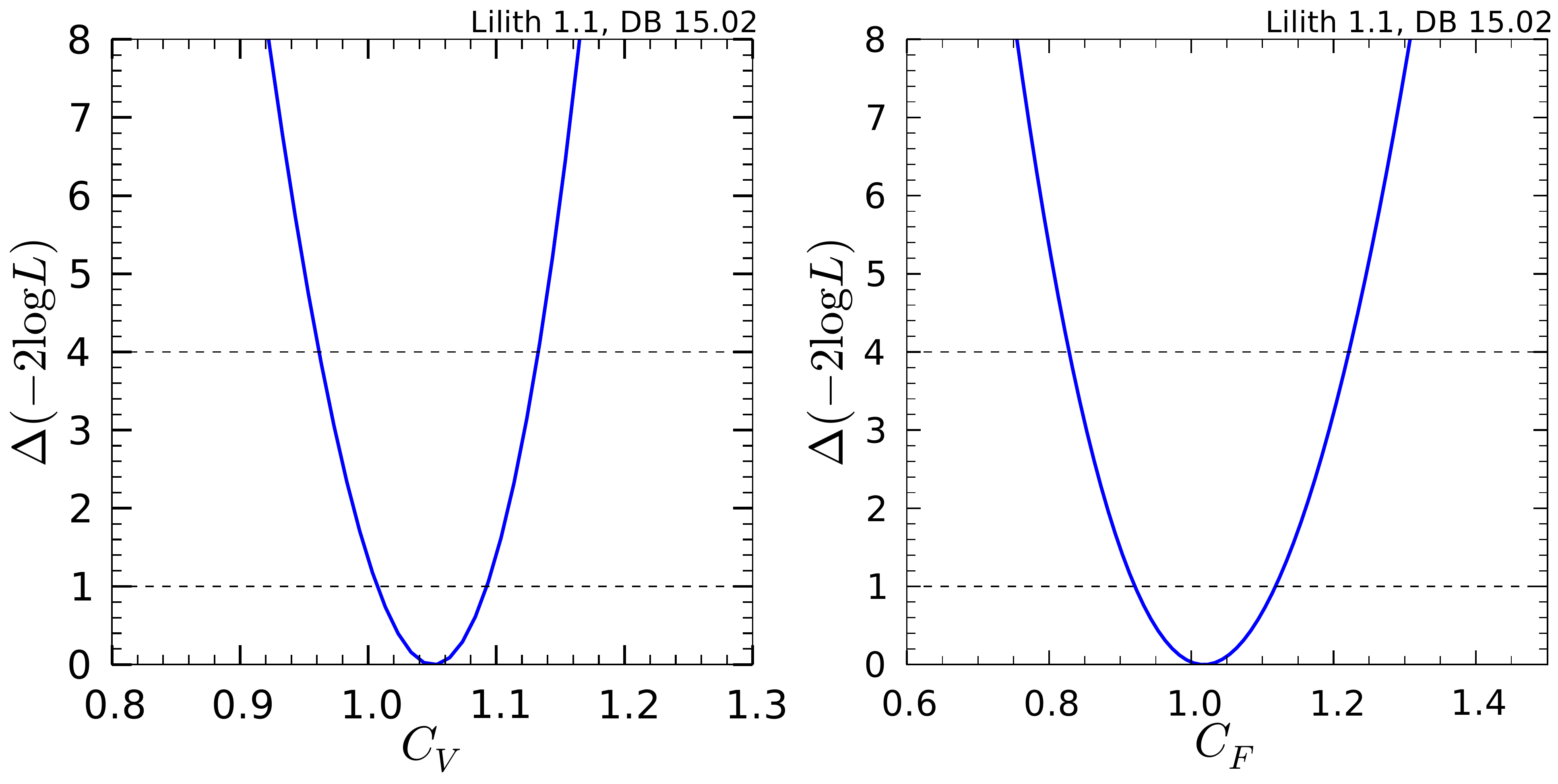} 
\caption{1-dimensional likelihood profiles of $C_V$~(left) and $C_F$~(right) from a global fit of the benchmark scenario $(C_V,C_F)$.}
\label{CVCFglobalfit}
\end{figure}

The \texttt{Python} routine used to obtain this result is \texttt{CVCF\_1dprofile.py}. It can be executed from the \texttt{Lilith-1.1/} folder with the command line
\begin{lstlisting}
python examples/python/CVCF_1dprofile.py
\end{lstlisting}

This example uses the class \texttt{Minuit} of the library \texttt{iminuit}~\cite{iminuit}, a \texttt{Python} implementation of the \texttt{MINUIT}~\cite{James:1975dr} minimization library, in order to minimize $-2\log L$ and derive the 1D profile around the minimum. Moreover, \texttt{matplotlib}~\cite{Hunter:2007} is used to produce the resulting figures. Below, we describe parts of the routine.

After having instantiated the \lilith\ class and read the experimental data with
\begin{lstlisting}
lilithcalc = lilith.Lilith(verbose, timer)
lilithcalc.readexpinput(myexpinput)
\end{lstlisting} 
a function \texttt{getL} returning $-2\log L$ for a given $(C_V,C_F)$ point is defined
\begin{lstlisting}[firstnumber=3]
def getL(CV, CF):
    myXML_user_input = usrXMLinput(mh=mh, CV=CV, CF=CF, precision=precision)
    lilithcalc.computelikelihood(userinput=myXML_user_input)
    return lilithcalc.l
\end{lstlisting}
where the function \texttt{usrXMLinput} creates a \texttt{XML} user input string from $C_V$ and $C_F$, for a given precision mode \texttt{precision}.

An object \texttt{m} of the class \texttt{Minuit} is then created
\begin{lstlisting}[firstnumber=7]
m = Minuit(getL, CV=1, limit_CV=(0,3), CF=1, limit_CF=(0,3))
\end{lstlisting}
where the initial point of the minimization and the range of parameters are defined. 
The function \texttt{m.mnprofile} is then called
\begin{lstlisting}[firstnumber=8]
xV, yV, rV = m.mnprofile("CV", bins=300, bound=(0., 2), subtract_min=True)
\end{lstlisting}
and returns the 1D likelihood profile $\Delta(-2\log L(C_V))\equiv -2\log (L(C_V)/L(\text{best fit}))$ for a given range and number of points.\footnote{In general for a function $-2\log L(\{C_i\}, \{C_j\})$, where $\{C_i\}, \{C_j\}$ can be sets of parameters, the profile likelihood $-2\log L(\{C_i\})$ is obtained by minimizing the full function with respect to $\{C_j\}$ for a given $\{C_i\}$.} Without the option \texttt{substract\_min=True}, the ``absolute'' likelihood $-2\log L(C_V)$ would be returned instead.

The parameter range in which $\Delta(-2\log L(C_V))<1\ (4)$ defines the 68\% (95\%) CL intervals of $C_V$. The constraints on $C_F$ are derived in the same way, and all results are plotted and stored in \texttt{results/CVCF\_1dprofile.pdf}.
In this scenario, the best-fit point is obtained for $C_V=1.05$ and $C_F = 1.02$. In 1D, \textit{i.e.} profiling over the other parameter, the 68\% (95\%) CL intervals read $C_V=[1.00, 1.09]\ ([0.96, 1.13])$ and $C_F=[0.92, 1.12]\ ([0.83, 1.22])$.

\begin{figure}
\centering \includegraphics[scale=0.38]{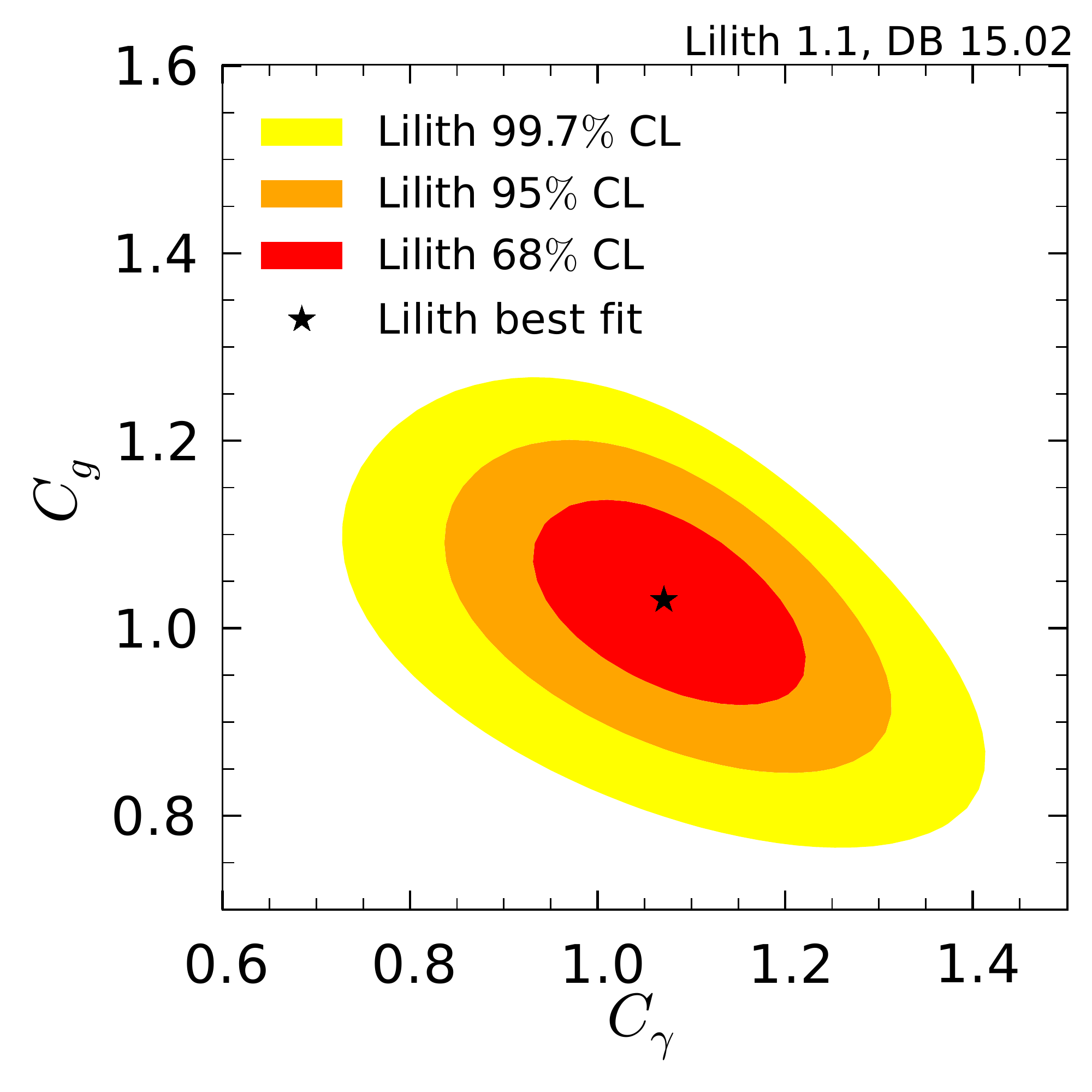} \includegraphics[scale=0.38]{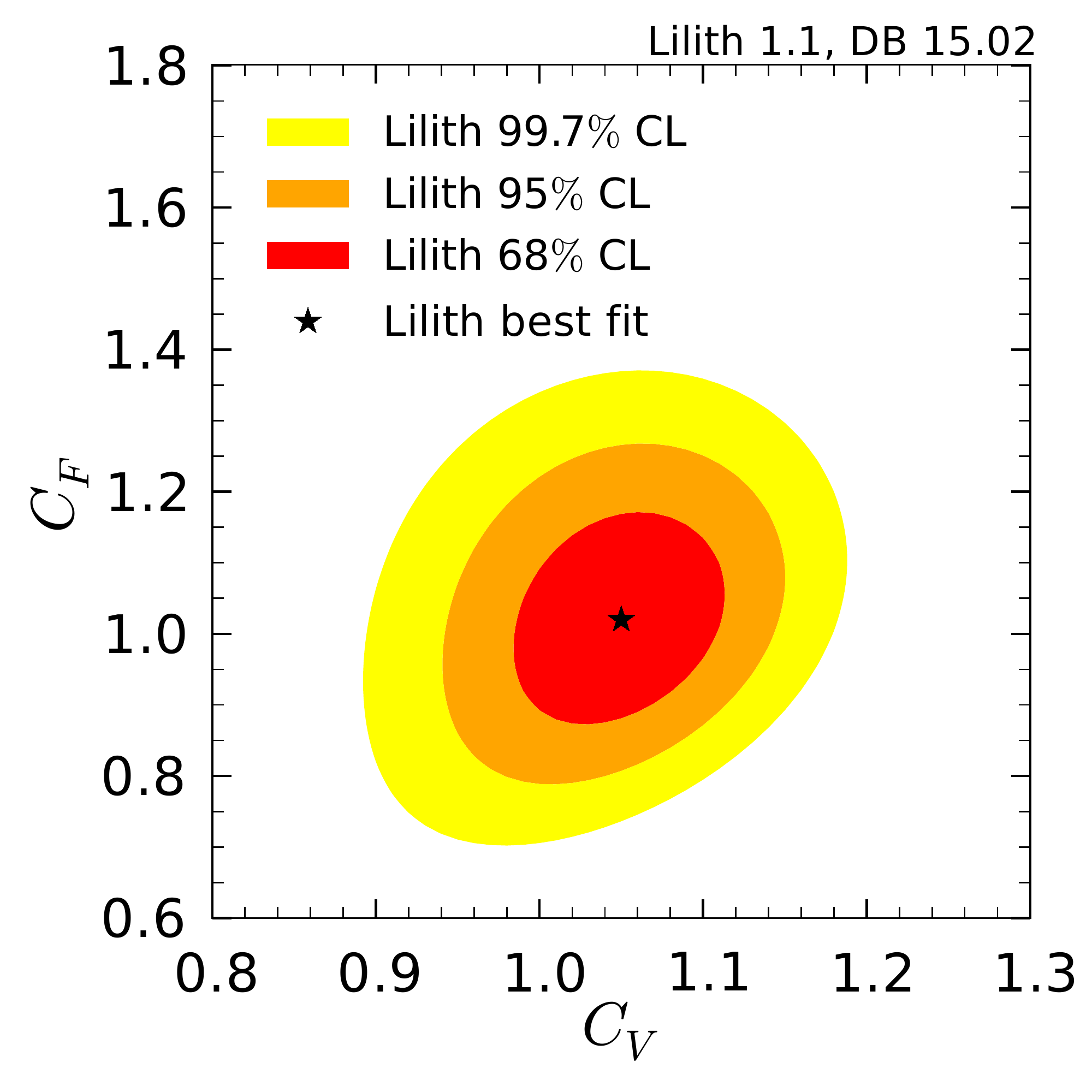}
\caption{Contraints on $(C_\gamma,C_g)$ (left) and $(C_V, C_F)$ (right) from a global fit to the Higgs data. The red, orange and yellow filled surfaces correspond to the 68\%, 95\% and 99.7\%~CL regions. The black star shows the position of the best-fit point.}
\label{CGaCgCVCF}
\end{figure}

We also provide example on how to derive constraints and produce figures for a 2D parameter space.
The left panel of Fig.~\ref{CGaCgCVCF} presents the 2D contraints obtained from a global fit of the $(C_\gamma, C_g)$ model presented above.
The corresponding \texttt{Python} routine is \texttt{CgammaCg\_2d.py}. It can be executed from the \texttt{Lilith-1.1/} folder with the command line
\begin{lstlisting}
python examples/python/CgammaCg_2d.py
\end{lstlisting}
A scan of the $(C_\gamma, C_g)$ parameter space is performed, and results are stored in the file \texttt{results/CgammaCg\_2d.out} in the form
\begin{verbatim}
 0.04040    0.00000    119.05462     
 0.04040    0.02020    119.00658     
 0.04040    0.04040    118.86261     
 0.04040    0.06061    118.62314     
 ......     ......       ......
\end{verbatim}
where the first, second and third columns contain the values of $C_\gamma$, $C_g$ and $-2\log L(C_\gamma,C_g)$, respectively.
The 68\%, 95\%, 99.7\%~CL regions in the $(C_\gamma,C_g)$ plane then corresponds to $\Delta(-2\log L(C_\gamma,C_g))<2.3, 5.99, 11.83$, respectively. This identification is performed by \texttt{matplotlib} with
\begin{lstlisting}
ax.contour(xi,yi,Z,[2.3,5.99,11.83])
\end{lstlisting}
where \texttt{xi}, \texttt{yi} and \texttt{Z} are list of points defining the grid in the $C_\gamma$ and $C_g$ directions, and the corresponding $\Delta(-2\log L(C_\gamma,C_g))$ value, respectively. The results are displayed and stored in \texttt{results/CgammaCg\_2d.pdf}.
For completeness, the 2D constraints on the $(C_V,C_F)$ benchmark scenario, using the latest LHC measurements, are also presented in the right panel of Fig.~\ref{CGaCgCVCF}. They have been derived in the same way.

\subsection{Higgs constraints on superpartners of the tau lepton}
\label{sec:stau}

Supersymmetric scalar partners of the tau leptons, known as staus, can have substantial contribution to the $H\to\gamma\gamma$ decay rate if they are light and have a large mixing~\cite{Carena:2012gp, Endo:2014pja}. Constraints on the parameters controlling this new contribution can therefore be obtained from the Higgs precision measurements. Here, we consider the Minimal Supersymmetric Standard Model (MSSM) and assume that the only deviation from a SM-like Higgs behavior comes from the contribution of staus to the loop-induced process $H\to\gamma\gamma$. More precisely, it is assumed that the supersymmetric partners of the Higgs boson and of the remaining fermions are decoupled, that the second Higgs doublet is phenomenologically irrelevant and that a Higgs mass of 125 GeV can be obtained for any point of the analysis. In this case, the contribution from staus to the $H\to\gamma\gamma$ decay width is parametrized by the two physical masses $m_{\stau_1}$ and $m_{\stau_2}$ (with $m_{\stau_1}<m_{\stau_2}$), the mixing angle $\theta_{\stau}$ and the ratio of vacuum expectation values for the two Higgs doublets, $\tan\beta$. The corresponding amplitude at LO reads~\cite{Djouadi:2005gi,Djouadi:2005gj}
\begin{equation}
\mathcal{M}_{H\gamma\gamma}^{\stau} = \sum_{i=1,2} \frac{g_{h\stau_i\stau_i}(m_{\stau_1},m_{\stau_2},\theta_{\stau},\tan\beta)}{m_{\stau_i}^2} A_0^H\left((m_H/(2 m_{\stau_i}))^2\right)
\end{equation}
where the sum runs over the two stau mass-eigenstates, $A_0^H$ is a loop form factor and  $g_{H\stau_i\stau_i}$ is the Higgs--stau--stau coupling.

The effective Higgs--$\gamma$--$\gamma$ reduced coupling can therefore be expressed as
\begin{equation}
C_\gamma(m_{\stau_1},m_{\stau_2},\theta_{\stau},\tan\beta) = \frac{\left|\mathcal{M}_{H\gamma\gamma}^{\SM}+\mathcal{M}_{H\gamma\gamma}^{\stau}(m_{\stau_1},m_{\stau_2},\theta_{\stau},\tan\beta)\right|}{\left|\mathcal{M}_{H\gamma\gamma}^{\SM}\right|}
\label{stauCGamma}
\end{equation}
Note that the SM amplitude $\mathcal{M}^{\SM}_{H\gamma\gamma}$ appears both in the numerator and denominator of Eq.~\eqref{stauCGamma} since SM tree-level couplings are assumed.

\begin{figure}
\centering \includegraphics[scale=0.38]{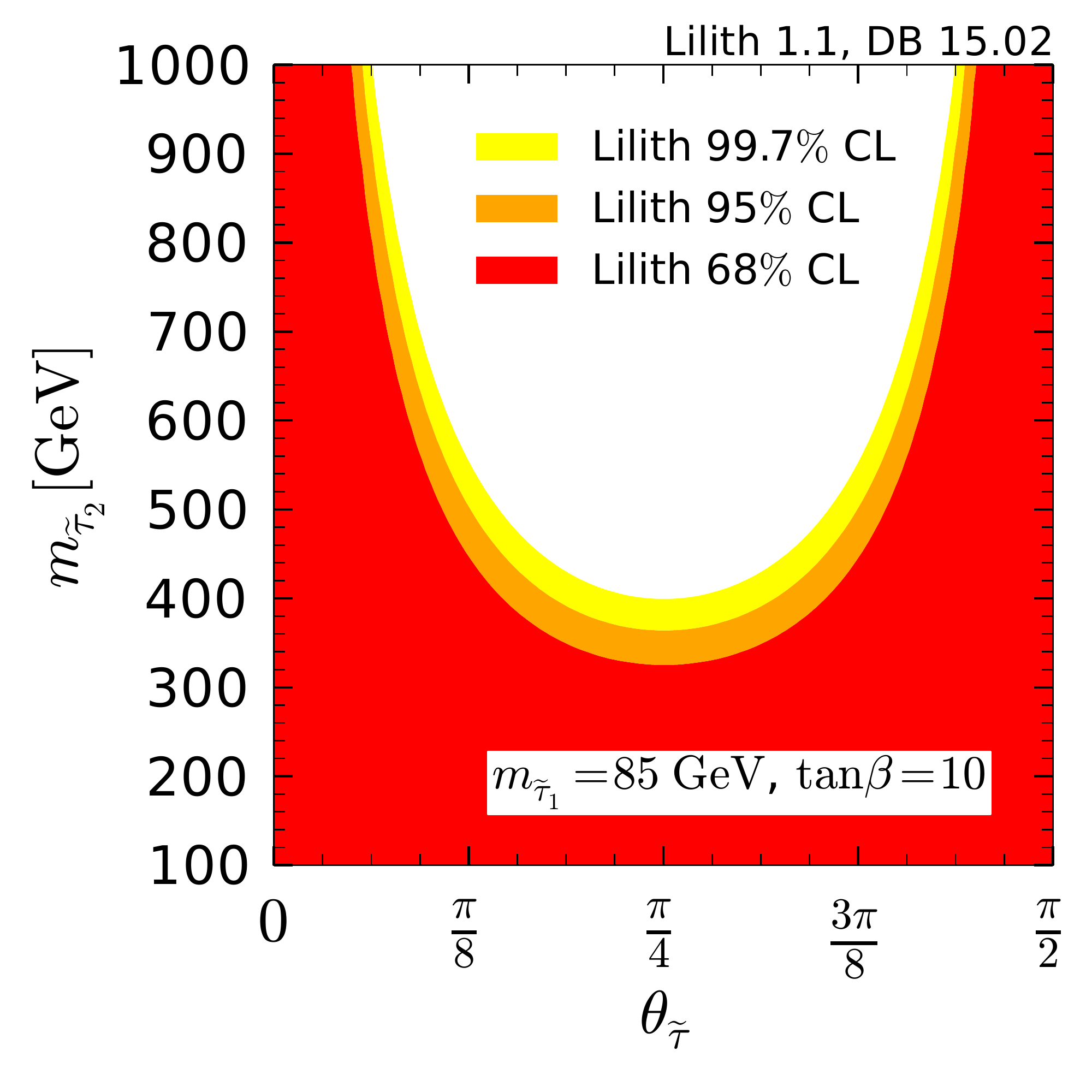} \includegraphics[scale=0.38]{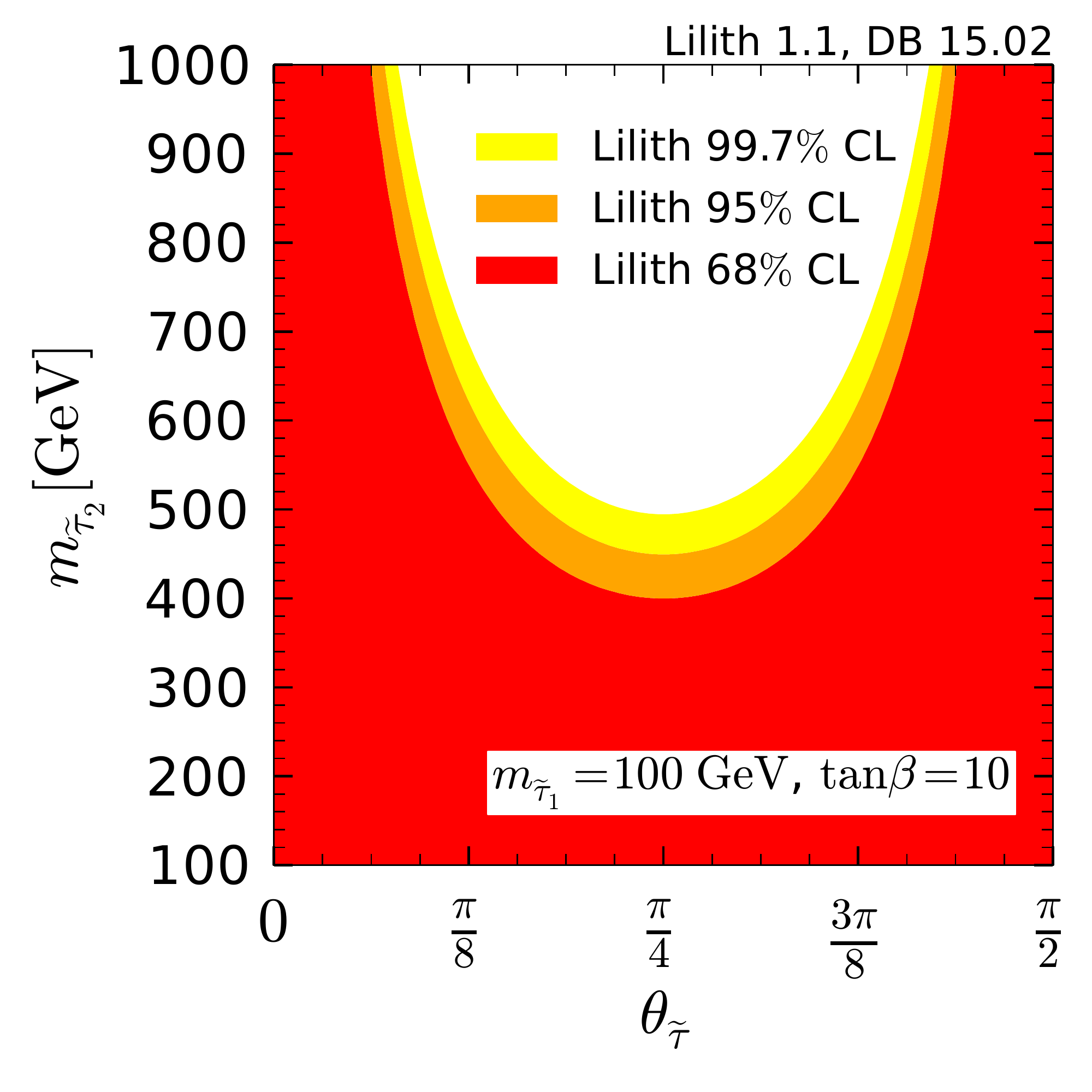}
\caption{Constraints on the staus masses and mixing angle in the $(m_{\widetilde{\tau}_2},\theta_{\widetilde{\tau}})$ plane for $m_{\widetilde{\tau}_1}=85$~GeV~(left) and $m_{\widetilde{\tau}_1}=100$~GeV~(right) and $\tan\beta=10$. The red, orange and yellow filled surfaces correspond to the allowed 68\%, 95\% and 99.7\%~CL regions, respectively. }
\label{staus}
\end{figure}

Fixing $\tan\beta=10$ and the mass of the lightest stau $m_{\stau_1}=85, 100$ GeV, we show constraints in the plane $(m_{\stau_2},\theta_{\stau})$ in Fig.~\ref{staus}.
For $\theta_{\stau}=\pi/4$, the 2-dimensional 95\%~CL upper limit on $m_{\stau_2}$ reads $m_{\stau_2}< 360$ $(460)$~GeV for $m_{\stau_1}=85$ $(100)$ GeV. More generally, the upper limit on $m_{\stau_2}$ becomes weaker as $m_{\stau_1}$ is increased.

The corresponding \texttt{Python} code is \texttt{stau\_gammagamma.py}. It can be executed by typing the following command line to the shell from the \texttt{Lilith-1.1/} folder:
\begin{lstlisting}
python examples/python/stau_gammagamma.py
\end{lstlisting}
The routine works as follows. Functions returning $C_\gamma$ according to Eq.~\eqref{stauCGamma} and $-2\log L(C_\gamma)$ are defined. Since $\tan\beta$ and $m_{\stau_1}$ are fixed, a 2-dimensional grid scan is then performed over the two remaining parameters: for each couple $(m_{\stau_2}, \theta_{\stau})$, the corresponding $\Delta(-2\log L)$ is obtained. The 2-dimensional 68\%, 95\%, 99.7\%~CL regions in the plane $(m_{\stau_2}, \theta_{\stau})$ are obtained with $\Delta(-2\log L)<2.3,5.99,11.83$, respectively.

Note that direct searches from LEP~\cite{lep} and vacuum metastability condition~\cite{Kitahara:2013lfa} impose further constraints on this scenario. Moreover, this simplified SUSY scenario could easily be generalized, {\it e.g.} by taking into account $H \to \tilde\chi^0_1\tilde\chi^0_1$. Light staus are especially relevant in the case where $\tilde\chi^0_1$ is light in order to have a viable neutralino dark matter candidate in the MSSM (see, {\it e.g.}, Ref.~\cite{Belanger:2013pna} and references therein).

\section{Prospects for Run~II of the LHC}
\label{sec:prospects}

As we discussed in Section~\ref{sec:expresults}, approximations necessarily need to be made when combining signal strength results from several categories or several searches, making it necessary to validate the approach.
In Section~\ref{sec:validation}, we have shown that we reproduce well the results of coupling fits from ATLAS and CMS (separately).
However, it is clear that the situation will change as more statistics will be collected at Run~II of the LHC. Indeed, systematic uncertainties will then dominate over statistical uncertainties in the majority of the channels.
Missing correlations between systematic uncertainties (both theoretical and experimental) will thus become a more pressing issue.
Moreover, more combinations for production and decay of the Higgs boson, $(X,Y)$, will be determined with a good precision. This will spoil the simple interpretation we have for a number of the results we currently use, in particular for results given in the plane $(\mu({\rm ggH+ttH}, Y), \mu({\rm VBF+VH}, Y))$.
In this section we recall the main limitations when using the information currently provided by the ATLAS and CMS collaborations for constructing a likelihood. We also discuss new ways of presenting the LHC Higgs results in order to be able to construct a good approximation to the Higgs likelihood at Run~II of the LHC. This section is partly based on the note ``On the presentation of the LHC Higgs Results''~\cite{Boudjema:2013qla} that was put forth by a collaboration of theorists and experimentalists with the aim to maximize the impact of the LHC Higgs results and their utility to the whole high-energy physics community.

First of all, in most cases only contours of constant likelihood (at least the 68\%~CL interval or contour, sometimes contours at 95\%~CL) are provided by ATLAS and CMS. This makes it necessary to extrapolate the likelihood assuming, most naturally, a Gaussian shape. When using a given contour to extrapolate the likelihood, the validity of this approximation can be tested from a comparison of the position of the best-fit point and from contours provided by the experimental collaboration. This was done in Section~\ref{sec:2dmurec}, where we concluded that the reconstruction is generally very good, although in some cases asymmetrical effects are washed out (see, {\it e.g.}, Fig.~\ref{ZZcomparison}). However, in all cases it induces an unnecessary source of error.
The ATLAS and CMS collaborations initiated some efforts during Run~I to provide the full likelihood information in 1D and 2D planes (see the right panel of Fig.~\ref{fig:dataformats} and Refs.~\cite{HEPDATA1,HEPDATA2,HEPDATA3}). 
We strongly hope that this will become standard practice during Run~II of the LHC, and that the information will systematically be provided in numerical form.

Another issue is the dependence of the results on the assumed Higgs boson mass $m_H$. Currently, we use results given at a fixed Higgs mass. As not all results are provided at the same $m_H$, a slight inconsistency is introduced in the combination of the different results (the assumed Higgs mass varies within a few hundreds of MeV).
Official combination notes allow us to get rid of this inconsistency, as all results are therein given at the same Higgs mass.
However, the dependence of the experimental results on the Higgs mass can be very important for the high-resolution channels, that target decay of the Higgs boson into charged leptons and photons (such as $H \to ZZ^* \to 4\ell$ and $H \to \gamma\gamma$). Thus, it would be highly desirable to have access to mass-dependent likelihood results.

Current results are presented in 1- or 2-dimensional projections, often corresponding to the combination of production modes (in 2D, typically $\ggHttH$ and $\VBFVH$). As we discussed above, this becomes a limitation as measurements get more precise, in which case we would like to investigate deviations in all of the five production modes separately. 
For such reasons, a total breakdown of the signal strength measurements in terms of the five Higgs production modes ($\ggH$, $\VBF$, $\WH$, $\ZH$, $\ttH$) would be a considerable step forward regarding the interpretation of the LHC Higgs results.
We would therefore like to advocate the experimental collaborations to provide the likelihood as a function of the Higgs mass and a full set of production modes, that is to say, in the 
\beq
 \left(m_H, \mu(\ggH,Y),\mu(\ttH,Y),\mu(\VBF,Y),\mu(\ZH,Y),\mu(\WH,Y)\right) 
\eeq
parameter space for each final state $Y$. For some final states, all five production modes are certainly not constrained by the experimental searches and only lower dimensional projections of this space would be relevant.

This would solve most of the limitations currently faced, with the notable exception of correlations between the measurements of different decay modes. For instance, theoretical uncertainties on gluon fusion production affect both the $\gamma\gamma$ and $ZZ^*$ final states. 
Recently, an interesting proposal was made in this direction in Ref.~\cite{Cranmer:2013hia}. Provided experimental collaborations publish likelihoods that are not profiled over a set of theoretical nuisance parameters of interest, but instead given for a fixed scenario, it is possible to build a ``recoupled'' likelihood incorporating these uncertainties at the later stage. This has the advantage of not being restricted to the Gaussian approximation. It would certainly be of great interest if the information in the 2D plane $(\mu({\rm ggH+ttH}, Y), \mu({\rm VBF+VH}, Y))$, or even better in the possibly 6D plane discussed above, could be given without profiling over the theoretical uncertainties on the Higgs signal. With the method presented in Ref.~\cite{Cranmer:2013hia}, one could then fully correlate the theoretical uncertainties between the different channels and experiments, and modify these uncertainties compared to what is done in ATLAS and CMS if desired.

\section{Conclusions}
\label{conclusions}

Crucial information on the origin of the electroweak symmetry breaking---and, more generally, on the presence of BSM physics around the electroweak scale---can be obtained from the study of the properties of the Higgs boson with mass around 125~GeV at the LHC. The presentation of the experimental results in terms of signal strengths makes it possible to combine all measurements and perform a global fit to the properties of the Higgs boson. The results of such fits can be used to discriminate between models where the structure of the couplings to SM particles is as in the SM.

However, using all available information from the ATLAS and CMS collaborations to construct a likelihood is a non-trivial task. Indeed, there is a wealth of experimental searches, from which the necessary information often needs to be extracted and put in numerical form. Care should also be taken in order to include all available correlations between systematic uncertainties.
To this aim, we provide a new public tool, {\tt Lilith}.
{\tt Lilith} is a library written in {\tt Python}, and for which we provide an API as well as a command-line interface and a basic interface to {\tt C} and {\tt C++}/{\tt ROOT}.
The experimental results are read from a database in {\tt XML} format that is shipped with the code and which is easy to modify and extend. {\tt Lilith} uses as a primary input results in which the fundamental production and decay modes are unfolded from experimental categories.

New physics can be parametrized in terms of reduced couplings, or signal strengths directly, which are given as input to {\tt Lilith} in {\tt XML} format. If needed, scaling factors for the loop-induced processes and VBF production are computed taking into account QCD corrections. CP-violating Higgs couplings can also be given as input to {\tt Lilith}.
The likelihood is evaluated from a set of experimental results and given as output; detailed results can moreover be stored in {\tt XML} or {\tt SLHA}-like format. For convenience, \lilith\ is provided with several applications of the code where constraints on effective or explicit models of new physics are derived. They include extensive comments.

The Higgs likelihood of \lilith\ obtained from the latest measurements at the LHC has been thoroughly validated against ATLAS and CMS results and can be used to constrain new physics.
Future measurements at Run~II of the LHC will, however, call for new ways of presenting results in order to derive a good approximation to the Higgs likelihood.
In particular, further disentanglement of the different production and decay modes will become necessary. Moreover, correlations between systematic uncertainties, and in particular the treatment of theoretical uncertainties, will become a more pressing issue. 
The structure of the code is such that {\tt Lilith} can easily be adapted to handle extended signal strength information.

\section*{Acknowledgments}

We are deeply indebted to Sabine Kraml for her support at all stages of the project and for comments on the manuscript.
We would also like to thank Genevi\`eve B\'elanger, Ulrich Ellwanger, John F.\ Gunion and Sabine Kraml for the fruitful collaboration on Higgs coupling fits in 2012 and 2013, from which the idea of {\tt Lilith} originated. 
This work was supported in part by the ANR project DMAstroLHC, the
``Investissements d'avenir, Labex ENIGMASS'', and by the IBS under
Project Code IBS-R018-D1. JB thanks the CTPU-IBS for hospitality and
financial support
for a research stay during which this work was finished.


\bibliographystyle{utphys}
\bibliography{biblio}

\end{document}